# Sagnac interference in integrated photonics for reflection mirrors, gyroscopes, filters, and wavelength interleavers

*David J. Moss*

Optical Sciences Centre, Swinburne University of Technology, Hawthorn, VIC 3122, Australia

E-mail: *dmoss@swin.edu.au*

**Abstract**

As a fundamental optical approach to interferometry, Sagnac interference has been widely used for reflection manipulation, precision measurements, and spectral engineering in optical systems. Compared to other interferometry configurations, it offers attractive advantages by yielding a reduced system complexity without the need for phase control between different pathways, thus offering a high degree of stability against external disturbance and a low wavelength dependence. The advance of integration fabrication techniques has enabled chip-scale Sagnac interferometers with greatly reduced footprint and improved scalability compared to more conventional approaches implemented by spatial light or optical fiber devices. This facilitates a variety of integrated photonic devices with bidirectional light propagation, showing new features and capabilities compared to unidirectional-light-propagation devices such as Mach-Zehnder interferometers (MZIs) and ring resonators (RRs). Here, we present our latest results for functional integrated photonic devices based on Sagnac interference. We outline the theory of integrated Sagnac interference devices with comparisons to other integrated photonic building blocks such as MZIs, RRs, photonic crystal cavities, and Bragg gratings. We present our latest results for Sagnac interference devices realized in integrated photonic chips, including reflection mirrors, optical gyroscopes, basic filters, wavelength (de)interleavers, and optical analogues of quantum physics.

## I. Introduction

Optical interferometers are of fundamental importance for precision measurements in modern age, underpinning research and industrial applications in a variety of fields such as astronomy [1-3], remote detection [4, 5] , surface profiling [6, 7], optical communications [8], quantum optics [9, 10], biosensing [11, 12], fluid dynamics [13, 14], optometry [15], and holographic imaging [16, 17]. Generally, an optical interferometer starts with a light input, then splits it into several beams, exposes parts of them to external effects (e.g., changes in length or refractive index), and finally recombines for superposition. Hence, the power or spatial form of the recombined beam can be utilized to extract relevant physical quantity such as refractive index, distance, surface irregularity, and mechanical stress.

The earliest reported optical interferometer can be traced back to 1801 by British scientist Thomas Young in his famous two-slit interference experiment [18, 19]. Since then, a wide range of optical interferometers have been developed [20, 21], which can be classified into two main categories depending on whether they operate based on wavefront or amplitude splitting. Wavefront splitting interferometers, represented by Young's two slits [18, 19], Lloyd's mirror [22], and Rayleigh interferometers [23], are mainly implemented in spatial light devices to split input light wavefront emerging from a point or a narrow slit. In contrast, amplitude splitting interferometers split the amplitude of the input light into directional paths, which can be realized in both spatial light and waveguide devices. **Figure 1** shows schematic configurations of typical amplitude splitting interferometers, including Fizeau, Michelson, Mach–Zehnder, Fabry–Pérot, Twyman–Green, and Sagnac interferometers [20, 23].

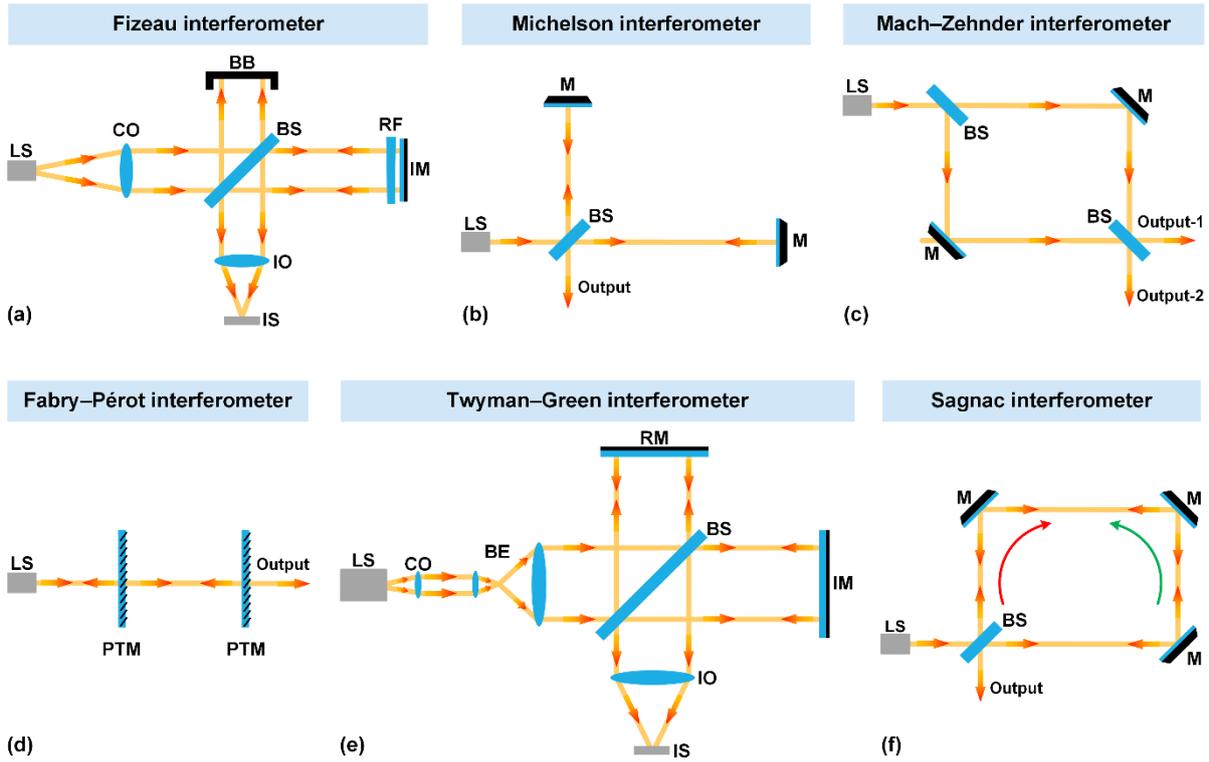

**Figure 1.** Schematic configurations of typical optical interferometers including (a) Fizeau, (b) Michelson, (c) Mach–Zehnder, (d) Fabry–Pérot, (e) Twyman–Green, and (f) Sagnac interferometers. LS: light source. CO: collimating optics. BB: beam block. BS: beam splitter. IO: imaging optics. IS: image sensor. RF: reference flat. IM: inspected mirror. M: mirror. PTM: partially transmissive mirror. BE: beam expander. RM: reference mirror.

Sagnac interferometers [24, 25], which were named after French scientist Georges Sagnac (**Figure 2(a)**), were first demonstrated for rotation sensing in 1913 (**Figure 2(b)**). After that, several milestones in their development history greatly broadened their capabilities and applications. The first Sagnac ring gas laser (**Figure 2(c)**) was proposed in 1962 by Rosenthal [26], which was subsequently implemented in 1963 by Macek and Davis to detect the rotation rate [27]. After that, Sagnac ring laser gyroscopes (**Figure 2(d)**) were also used for detecting general relativity and geodesic phenomena [28-31]. Sagnac fiber ring interferometers, which have achieved many successes in sensing and optical communication applications [32-34], were first suggested by Brown in 1968 in a study of inertial rate sensing [35], and blossomed along with the development of optical fiber technologies [36-40]. The first fiber-optic gyroscope (**Figure 2(e)**) was demonstrated by Vali and Shorthill in 1976 [41], and experienced rapid progress after 1980s (**Figure 2(f)**) [42, 43]. Nowadays, Sagnac interferometers have been used in extensive applications such as inertial navigation [44, 45], optical communication [46], lasering [47, 48], and sensing [49, 50] (**Figures 2(g) – (k)**).

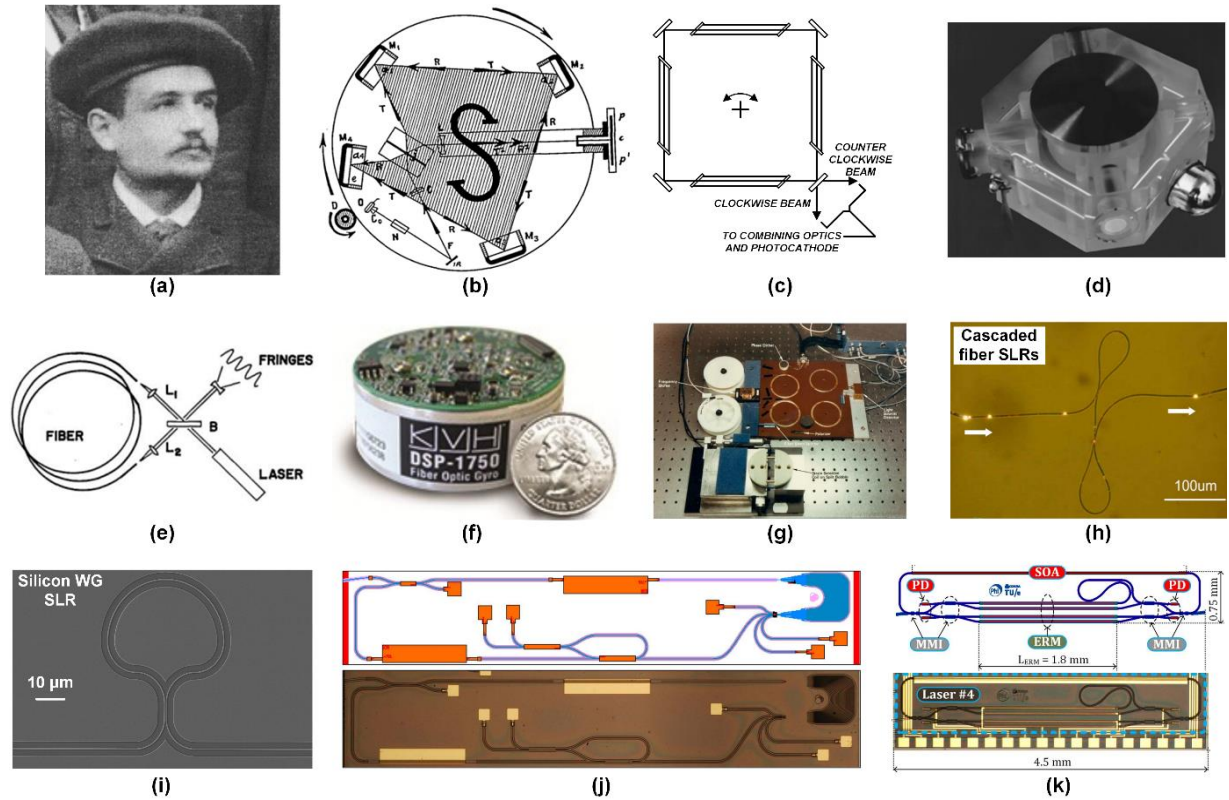

**Figure 2.** The development of Sagnac interferometers. (a) A photo for French physicist Georges Sagnac (1869 – 1928). (b) Schematics of Sagnac's original interferometer in 1913. (c) Schematic diagram of the first Sagnac ring laser demonstrated by Macek and Davis in 1963. (d) A Sagnac ring laser gyroscope. (e) Schematics of the first fiber-optic gyroscope built by Vali and Shorthill in 1976. (f) A fiber optic gyroscope based on Sagnac effect. (g) A strain sensor based on Sagnac effect in optical fibers. (h) Microscope image of a Fabry–Pérot resonator assembled using microfiber Sagnac interferometers. (i) A scanning electron microscope (SEM) image of an integrated Sagnac interferometer. (j) Mask layout (up) and a microscope image (down) of an integrated Sagnac ring laser gyroscope. (k) Mask layout (up) and a microscope image (down) of an integrated tunable laser including a Sagnac interferometer. SOA: a semiconductor optical amplifier. MMI: multi-mode interference couplers. ERM: electro-refractive modulators. PD: photodiodes. (a) Adapted with permission [51]. Copyright 2014, Pleiades Publishing, Ltd. (b) Adapted with permission [26]. Copyright 1962, Optica Publishing Group. (c) Adapted with permission [27]. Copyright 1963, American Institute of Physics. (d) Adapted with permission [52]. Copyright 2011, IEEE. (e) Adapted with permission [41], Copyright 1976, Optica Publishing Group. (f) Adapted with permission [53]. Copyright 2016, SPIE. (g) Adapted with permission [54]. Copyright 2017, SPIE. (h) Adapted with permission [55]. Copyright 2013, De Gruyter. (i) Adapted with permission [56]. Copyright 2018, AIP Publishing LLC. (j) Adapted with permission [57]. Copyright 2018, IEEE. (k) Adapted with permission [58]. Copyright 2016, Optica Publishing Group.

In a Sagnac interferometer, two split light beams travel in opposite directions and share a common optical pathway. Compared to Michelson and Mach–Zehnder interferometers that have different pathways for the split beams, Sagnac interferometers are free of phase control between different pathways, thus providing high stability against external disturbance such as vibration [20, 25]. Moreover, Sagnac interferometers, with their output intensity being determined only by the power split ratio of the beam splitter, exhibit much lower wavelength dependence than the Michelson, Mach–Zehnder, and Fabry–Pérot interferometers.

The hardware implementation of Sagnac interferometers has been realized based on a wide range of device platforms, including spatial light systems (e.g., **Figures 2(b) – (d)**), optical fibers (**Figures 2(e) – (h)**), and photonic integrated circuits (PICs, **Figures 2(i) – (k)**). Compared to discrete off-chip devices that suffer from limitations in system complexity and production scale, integrated Sagnac interferometers fabricated via mature complementary metal-oxide-semiconductor (CMOS) technologies provide competitive advantages in achieving compact device footprint, low power consumption, and low-cost manufacturing. More importantly, the high stability and scalability of integrated devices enable the design and engineering of functional on-chip systems with Sagnac interferometers as building blocks – similar to the ring resonator systems that have achieved many success [59, 60]. In contrast to ring resonators where light only propagates in one direction, Sagnac interferometers involve light waves propagating in two opposite directions as well as mutual interaction between them. This offers additional degree of freedom in engineering the mode interference in resonators formed by Sagnac interferometers and hence more versatile spectral responses.

Here, we present our latest results for integrated photonic devices based on Sagnac interference, which have a wide range of applications to reflection manipulation, spectral engineering, and precision measurements. We highlight the role of integrated Sagnac interferometers as fundamental building blocks in PICs, as well as their comparison and synergy with other building blocks such as Mach-Zehnder interferometers, ring resonators, photonic crystal cavities, and Bragg gratings.

This paper is structured as follows. In **Section II**, the fundamentals of integrated Sagnac interference devices are briefly introduced, including modeling method, basic properties, and comparisons with other building blocks in PICs. Next, we present our latest results for integrated Sagnac interference devices in **Section III**, including reflection mirrors, optical gyroscopes, basic filters, wavelength (de)interleavers, optical analogues of quantum physics, and others. Some outstanding challenges are discussed in **Section IV**, followed by conclusions in **Section V**.

## II. Fundamentals of integrated Sagnac interference devices

In this section, we briefly introduce the fundamentals of integrated Sagnac interference devices, highlighting their comparison with other building blocks in PICs such as Mach-Zehnder interferometers (MZIs), ring resonators (RRs), photonic crystal (PhC) cavities, and Bragg gratings. Note that the integrated photonic devices discussed in this section delineate a wide range of on-chip devices that can be implemented based on different material platforms such as silicon, silicon nitride (SiN), doped silica, III-V materials, and chalcogenide glasses [61-65]. For linear optical devices, the basic principles introduced in this section are universal for all material platforms.

### 1. Waveguide Sagnac interferometers and other integrated building blocks

Directional couplers are basic elements that comprise MZIs, RRs, and Sagnac interferometers, all of which are building blocks of PICs. A directional coupler is formed by two closely placed waveguides with mutual energy coupling (**Figure 3(a)**), which can split a guided optical wave into two physically separated coherent components and vice versa. The universal relation between the input and output of the directional coupler in **Figure 3(a)** can be given by [66]:

$$\begin{bmatrix} E_{out-1} \\ E_{out-2} \end{bmatrix} = \begin{bmatrix} t & j\kappa \\ j\kappa & t \end{bmatrix} \begin{bmatrix} E_{in-1} \\ E_{in-2} \end{bmatrix} \quad (1)$$

where $j = \sqrt{-1}$, $E_{in-1}$, $E_{in-2}$, $E_{out-1}$, and $E_{out-2}$ are the input and output optical fields right before and after the coupling region, $t$ and $\kappa$ are the self-coupling and cross-coupling coefficients, satisfying the relation $t^2 + \kappa^2 = 1$ when assuming lossless coupling. When $t = \kappa = \frac{1}{2}$, the directional coupler works as a 3-dB coupler for equal power split.

A MZI (**Figure 3(b)**) can be realized by cascading two directional couplers in **Figure 3(a)**, its field transfer function is expressed as [67]:

$$\begin{bmatrix} E_{out-1} \\ E_{out-2} \end{bmatrix} = \begin{bmatrix} t^2 e^{jk\Delta L/2} - \kappa^2 e^{-jk\Delta L/2} & j\kappa t(e^{jk\Delta L/2} + e^{-jk\Delta L/2}) \\ j\kappa t(e^{jk\Delta L/2} + e^{-jk\Delta L/2}) & t^2 e^{-jk\Delta L/2} - \kappa^2 e^{jk\Delta L/2} \end{bmatrix} \begin{bmatrix} E_{in-1} \\ E_{in-2} \end{bmatrix} \quad (2)$$

where $\Delta L$ is the length difference between the two arms, and $k = 2\pi n_g/\lambda$, with $n_g$ denoting the group index and $\lambda$ the wavelength. When assuming 3-dB coupling for the two directional couplers, **Eq. (2)** can be simplified as:

$$\begin{bmatrix} E_{out-1} \\ E_{out-2} \end{bmatrix} = j \begin{bmatrix} \sin(k\Delta L/2) & \cos(k\Delta L/2) \\ \cos(k\Delta L/2) & -\sin(k\Delta L/2) \end{bmatrix} \begin{bmatrix} E_{in-1} \\ E_{in-2} \end{bmatrix} \quad (3)$$

By connecting two ports of the same waveguide in a directional coupler to form a closed loop, a RR (**Figure 3(c)**) can be obtained, its field transfer function is given by [59, 60]:

$$T_{RR} = \frac{E_{out-1}}{E_{in}} = \frac{t - ae^{-j\varphi}}{1 - tae^{-j\varphi}} \quad (4)$$

where $a = e^{-\alpha L/2}$ is the round-trip transmission factor, with $\alpha$ denoting the power propagation loss factor and $L$ the loop circumference. In **Eq. (4)**, $\varphi = 2\pi n_g L/\lambda$ is the round-trip phase shift.

A waveguide Sagnac interferometer (**Figure 3(d)**) can be formed by connecting two ports of different waveguides on the same side of a directional coupler. In contrast to RRs that only allow for unidirectional light propagation, it supports bidirectional light propagation as well as mutual coupling between the counter-propagated light waves. Its field transmission and reflection functions are:

$$T_{SI} = \frac{E_{out-1}}{E_{in}} = (t^2 - \kappa^2)ae^{-j\varphi} \quad (5)$$

$$R_{SI} = \frac{E_{out-2}}{E_{in}} = 2jtkae^{-j\varphi} \quad (6)$$

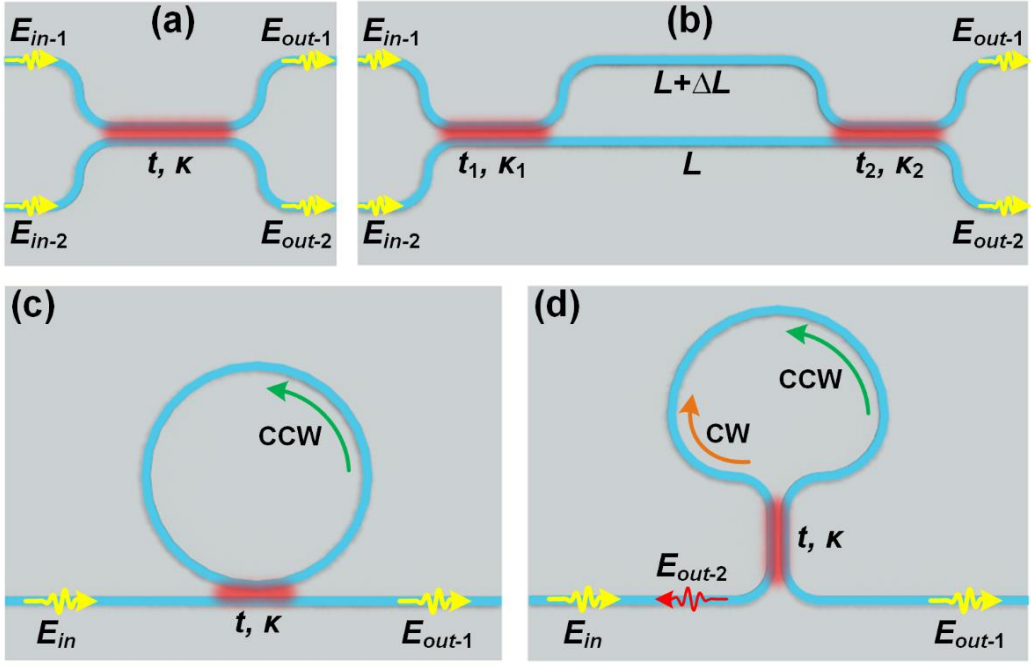

**Figure 3.** (a) Schematic of a directional coupler. (b) – (d) Schematics of building blocks in PICs formed by directional couplers, including (b) a MZI, (c) an all-pass RR, and (d) a waveguide Sagnac interferometer. CW: clockwise. CCW: counter-clockwise.

As can be seen from **Eqs. (5)** and (**6**), for constant $t$ and $\kappa$, the output light intensities (i.e., $|T_{SI}|^2$ or $|R_{SI}|^2$) are constants without any wavelength dependence. This is because the clockwise (CW) and counter-clockwise (CCW) light waves that interfere at the output ports share a common optical pathway and hence experience the same phase shift. By changing $t$ and $\kappa$, the output light intensities can be varied, with the wavelength independence being maintained. This forms the basis of implementing complex photonic systems consisting of Sagnac interferometers, particularly for integrated photonic devices with high stability and scalability. When $t = \kappa$, the Sagnac interferometer operate as a total reflection mirror with a zero transmission, i.e., $T_{SI}$ in **Eq. (5)** equals zero.

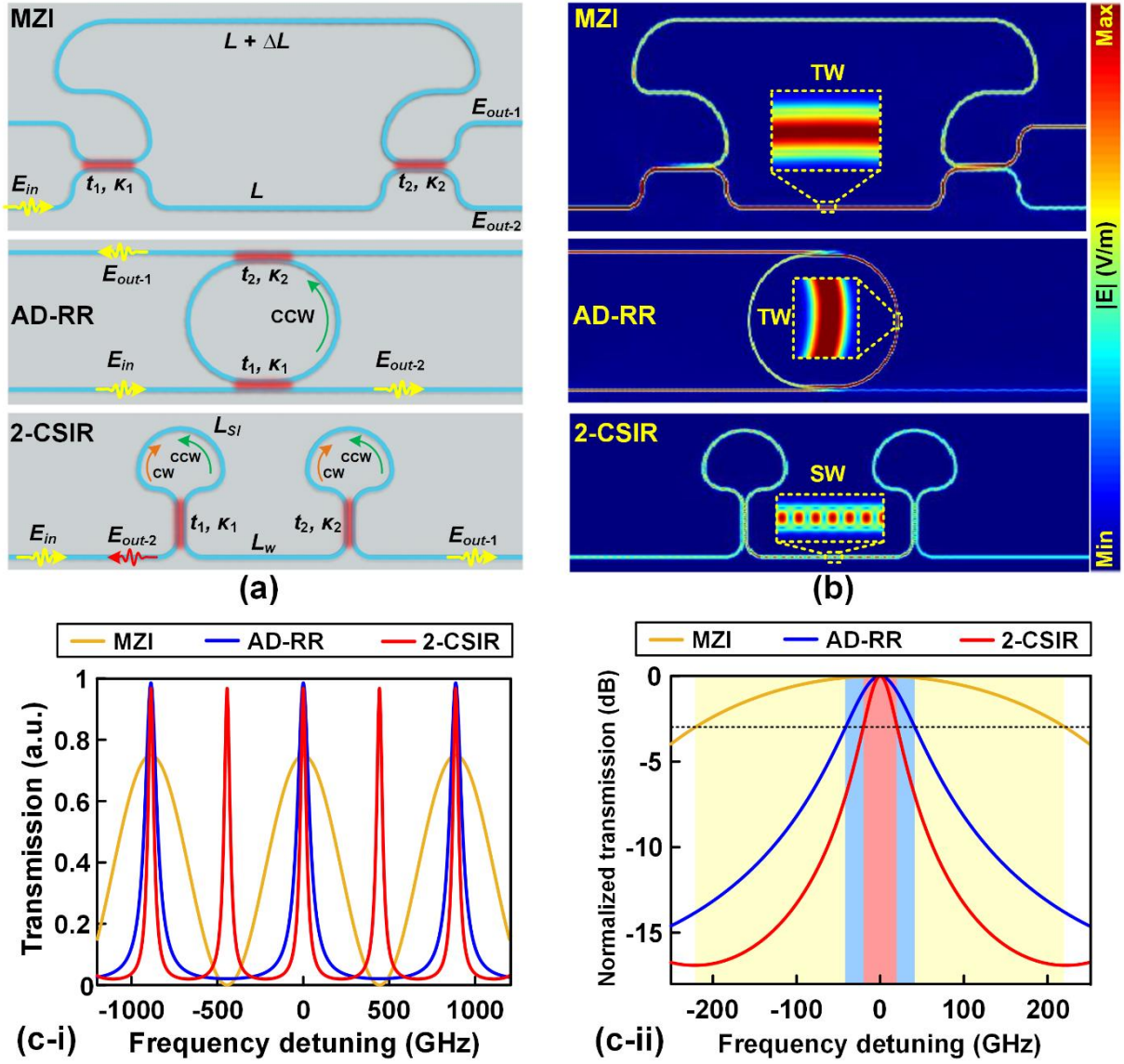

**Figure 4.** Comparison of a MZI, an add-drop RR (AD-RR), and a resonator formed by two cascaded Sagnac interferometers (2-CSIR). (a), (b), and (c) shows the device schematics, electric field distribution profiles, and power transmission spectra, respectively. In (c), (i) shows the transmission spectra corresponding to the outputs at port $E_{out-1}$ for all the three devices, and (ii) compares the 3-dB bandwidths of the transmission peaks in (i). For comparison, the maximum transmission and center frequencies of the transmission peaks in (c-ii) are normalized.

By cascading waveguide Sagnac interferometers, Fabry–Pérot (FP) cavities can be formed. where the Sagnac interferometers perform as reflection mirrors similar to those in FP laser diodes [68, 69]. In **Figure 4**, we compare the performance of a MZI, an add-drop RR (AD-RR), and a resonator formed by two cascaded Sagnac interferometers (2-CSIR). **Figure 4(a)** shows the device schematics. For comparison, all three devices are designed based on the silicon-on-insulator (SOI) platform with the same $n_g = 4.3350$, $\alpha = 55$ m$^{-1}$ (i.e., 2.4 dB/cm), and $t_{1,2} =$ 0.865. In addition, the length difference between the two arms of the MZI (i.e., $\Delta L$), the circumference of the AD-RR, and the cavity length of the 2-CSIR (i.e., $L_{SI} + L_w$, with $L_{SI}$ and $L_w$ denoting the lengths of the Sagnac loop and the connecting waveguide, respectively) are assumed to be the same. **Figure 4(b)** shows the electric field distribution profiles obtained via three-dimensional finite-difference time-domain (3D-FDTD) simulations. As can be seen, the MZI and AD-MRR show a travelling-wave (TW) interference pattern, whereas the 2-CSIR shows a standing-wave (SW) interference pattern. **Figure 4(c)** shows the power transmission spectra calculated based on the scattering matrix methods [56, 70]. In **Figure 4(c-i),** the free spectral range (FSR) of the 2-CSIR is half of those of the AD-RR and MZI – a result of its SW resonator nature. This indicates that the 2-CSIR has a cavity length that is half that of an AD-RR with the same FSR, thus allowing for a more compact device footprint. In **Figure 4(c-ii)**, the MZI has the lowest quality (Q) factor (defined as the ratio of the resonance wavelength to the resonance 3-dB bandwidth). This arises from its finite-impulse-response (FIR) filter nature, in contrast to that of infinite-impulse-response (IIR) resonators such as the AD-RR and the 2-CSIR. The Q factor of the 2-CSIR is about twice that of the AD-RR, making it attractive for applications requiring high Q factors [71-74].

The ability to control the energy coupled into and out of the resonant cavities is crucial for practical devices. Depending on the difference between the energy coupled inside the resonant cavities and their intrinsic loss, the resonator can be classified into three coupling regimes – under coupled, critically coupled, and over coupled [60, 66, 75]. For MRRs, the three coupling regimes with distinctive intensity, phase, and group delay responses have been widely exploited for a range of signal processing applications such as fast/slow light [76-78], analog computing [79-81], and advanced optical modulation formats [82-84]. In **Figure 5**, we compare the intensity, phase, and group delay responses of the AD-RR and 2-CSIR in **Figure 4** for various $t_1$ but constant $t_2 = 0.865$. As can be seen, the AD-RR show typical responses corresponding to under coupled, critically coupled, and over coupled regimes at the through port when $t_1 = 0.83$, 0.865, and 0.9, respectively. Similarly, the 2-CSIR show the responses corresponding to the three coupling regimes at the reflection port, indicating that the three coupling regimes can also be achieved in a SW resonator like the 2-CSIR.

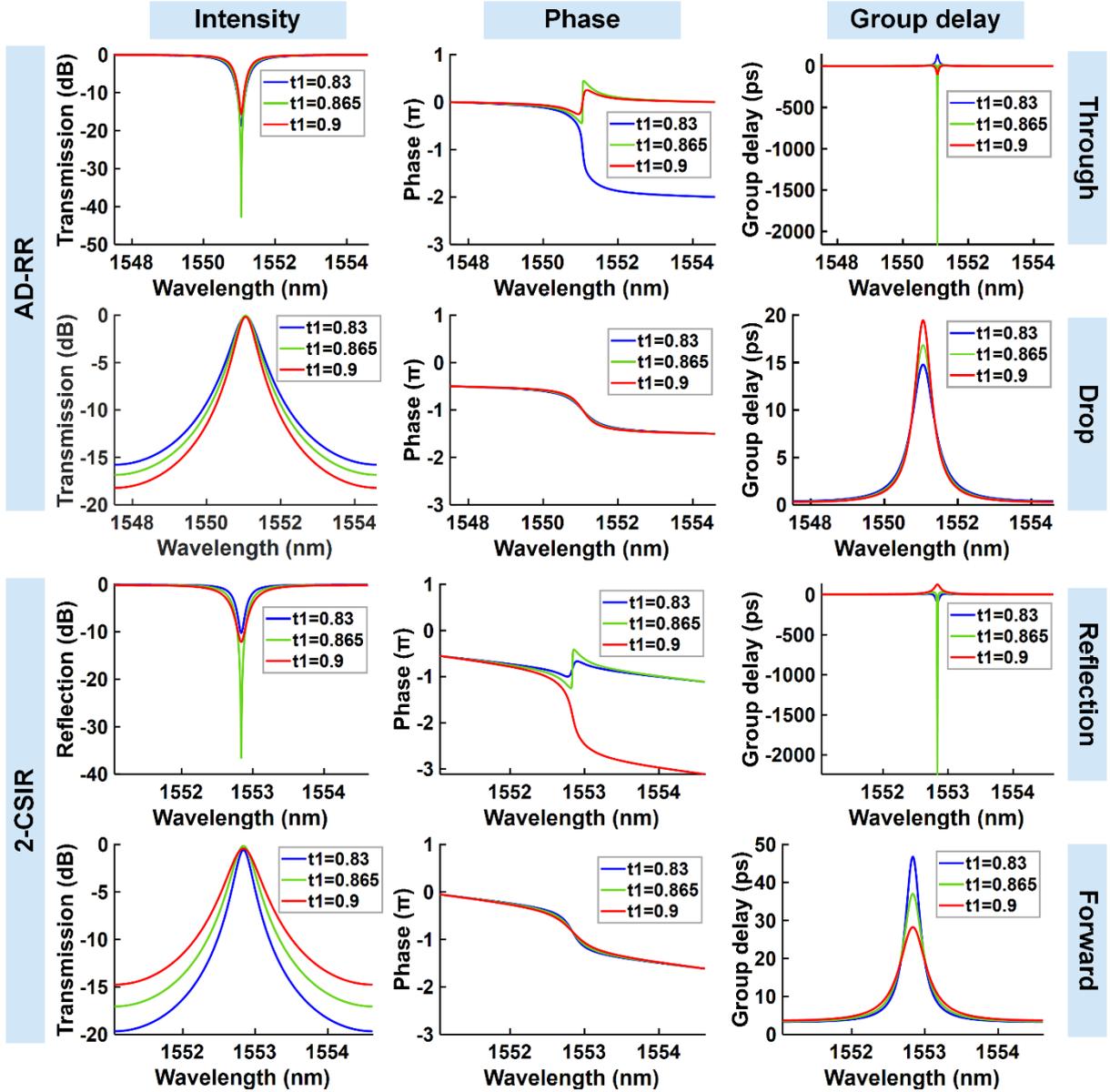

**Figure 5.** Comparison of intensity, phase, and group delay responses of the AD-RR and 2-CSIR in **Figure 3** for various $t_1$ but constant $t_2 = 0.865$.

More versatile spectral responses can be obtained by cascading more Sagnac interferometers. For a resonator formed by multiple cascaded Sagnac interferometers, each Sagnac interferometer acts as a reflection/transmission element and contributes to the overall output spectra, which is similar to other SW resonators such as PhC cavities [85-87] and Bragg gratings [88-91]. In **Figure 6**, we compare the performance of a one-dimensional PhC (1D-PhC) resonant cavity, Bragg gratings, and a resonator formed by eight cascaded Sagnac interferometers (8-CSIR). For comparison, all of the three devices are designed based on the SOI platform. In each device, all the reflection/transmission elements are assumed to be identical except that an additional phase shift of π/2 is introduced to the central element. As can be seen, all the three SW resonators show similar transmission spectra, with a transmission peaking appearing in the stop band. This is induced by enhanced light trapping in the central elements resulting from the additional π/2 phase shift. Compared to the 1D-PhC cavity and the Bragg gratings that have sub-wavelength cavity lengths, the 8-CSIR with a longer cavity length shows a smaller FSR of ~100 GHz that matches with the spectral grid of wavelength-division-multiplexing (WDM) optical communication systems [92]. The large cavity length not only makes it easy to tailor the reflectance/transmittance of each element for flexible spectral engineering, but also yields higher tolerance to fabrication errors.

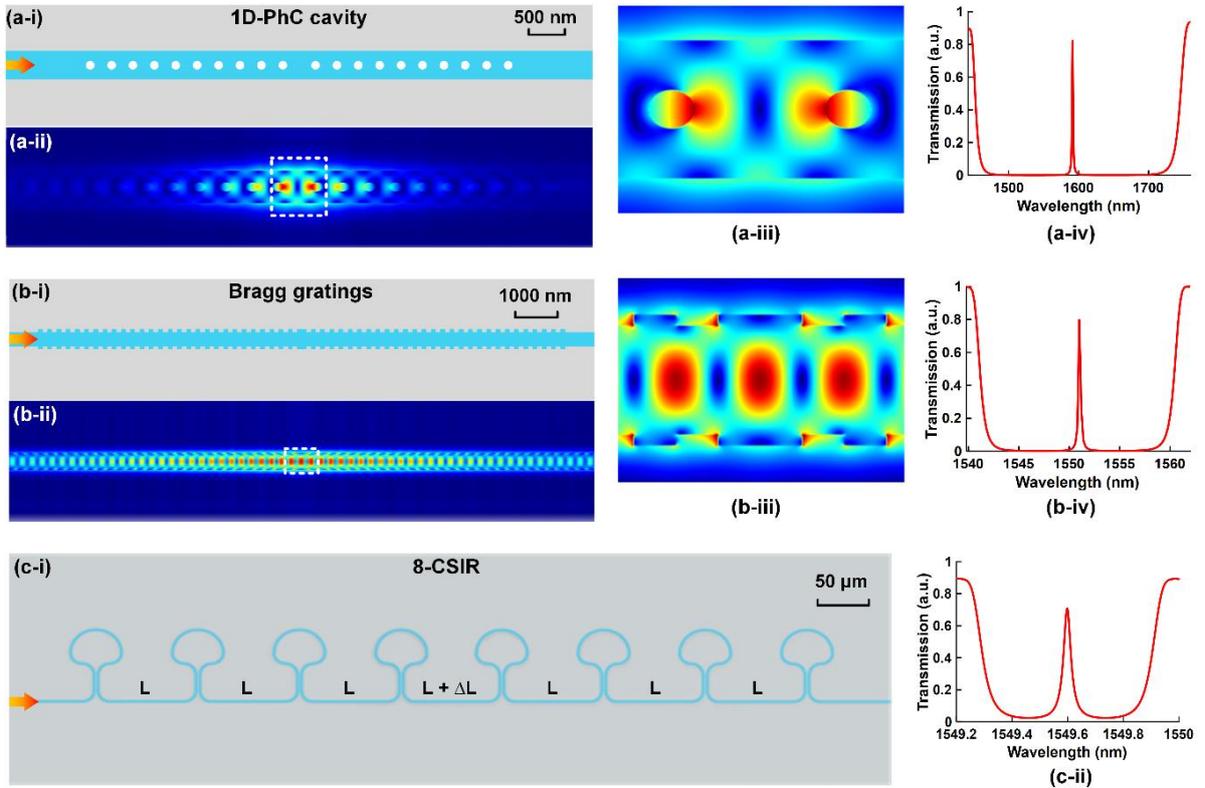

**Figure 6.** Comparison of (a) a one-dimensional photonic crystal (1D-PhC) resonant cavity, (b) Bragg gratings, and (c) a resonator formed by eight cascaded Sagnac interferometers (8-CSIR). In (a) – (b), (i) shows device schematic, (ii) shows on-resonance electric field distribution profile, (iii) shows zoon-in view of (ii), and (iv) shows power transmission spectrum. In (c), (i) shows device schematic and (ii) shows power transmission spectrum. The three devices in (a) – (c) are designed based on the SOI platform. For each device, all the elements are assumed to be identical except that an additional phase shift of π/2 is introduced to the central element.

## 2. Modeling of complex integrated Sagnac interference devices

The scattering matrix method is widely used to calculate the spectral transfer functions of RRs and complex photonic systems consisting of RRs [93, 94]. For integrated Sagnac interference devices, the calculation of their spectral transfer functions is more complicated since light waves propagate in two directions. In the following, we introduce a universal method to calculate the spectral transfer functions of complex integrated photonic systems, using the device in **Figure 7(a-i)** formed by two mutually coupled Sagnac interferometers as an example. First, the device is divided into several directional couplers and connecting waveguides, with $E_i$ ($i = 1 - 16$) denoting the optical fields at the dividing points. At each point, since there are optical fields travelling in two directions, we define the one that travels from left to right or in a clockwise direction as "+" and the opposite one as the "-" direction. Next, the scattering matrix equations showing the relation between the input and output ports of the directional couplers and the connecting waveguides can be easily obtained, as shown in **Table 1**. After that, the system input should be set, e.g., $E_1^+ = 1$ for input from Port 1, and $E_4^+ = 0$, $E_{13}^- = 0$, $E_{16}^- = 0$ for no input from Ports 2 – 4. Finally, the output spectral transfer function can be obtained by solving all the linear equations (via matrix operation or computing software such as MATLAB) to obtain the optical fields at corresponding output ports, e.g., $E_4^-$, $E_{13}^+$, and $E_{16}^+$ for Ports 2, 3, and 4, respectively. **Figure 7(a-ii)** shows the intensity spectral responses at Ports 2 – 4 for device in **Figure 7(a-i)** with input from Port 1, which are plot based on the spectral transfer functions calculated using the above method. Note that this method can be used to calculate spectral transfer fucntions for not only Sagnac interference devices with bidirectional light propagation, but also RRs with unidirectional light propagation, and even hybrid systems including both Sagnac interferometers and RRs.

**Table 1. Definitions of structural parameters of device in Figure 7 and the corresponding scattering matrix equations.**

| | | | | | | |
|---|---|---|---|---|---|---|
| **Figure 7(a-i)** | Connecting waveguides | Structural parameters | Structure | Length | Transmission factor | Phase shift |
| | | | Bus waveguides ($i = 1, 2$) | $L_{wi}$ | $a_{wi}$ | $\varphi_{wi}$ |
| | | | Sagnac loops ($i = 1, 2$) | $L_{si}$ | $a_{si}$ | $\varphi_{si}$ |

| | | Bus waveguides | Sagnac loops |
|---|---|---|---|
| | Scattering matrix equations [a] | $\begin{bmatrix} E_9^+ \\ E_5^- \end{bmatrix} = T_{w1} \begin{bmatrix} E_5^+ \\ E_9^- \end{bmatrix},$ | $\begin{bmatrix} E_3^+ \\ E_2^- \end{bmatrix} = T_{s1}^{1/2} \begin{bmatrix} E_2^+ \\ E_3^- \end{bmatrix}, \begin{bmatrix} E_6^+ \\ E_7^- \end{bmatrix} = T_{s1}^{1/2} \begin{bmatrix} E_7^+ \\ E_6^- \end{bmatrix},$ |
| | | $\begin{bmatrix} E_{12}^+ \\ E_8^- \end{bmatrix} = T_{w2} \begin{bmatrix} E_8^+ \\ E_{12}^- \end{bmatrix},$ | $\begin{bmatrix} E_{11}^+ \\ E_{10}^- \end{bmatrix} = T_{s2}^{1/2} \begin{bmatrix} E_{10}^+ \\ E_{11}^- \end{bmatrix}, \begin{bmatrix} E_{14}^+ \\ E_{15}^- \end{bmatrix} = T_{s2}^{1/2} \begin{bmatrix} E_{15}^+ \\ E_{14}^- \end{bmatrix},$ |

| | Structural parameters | Field transmission coefficient ($i = 1 - 4$) | | | $t_i$ |
|---|---|---|---|---|---|
| Directional couplers (DCs) | | Field cross-coupling coefficient ($i = 1 - 4$) | | | $\kappa_i$ |
| | | DC$_1$ | DC$_2$ | DC$_3$ | DC$_4$ |
| | Scattering matrix equations [b] | $\begin{bmatrix} E_2^+ \\ E_6^- \end{bmatrix} = S_1 \begin{bmatrix} E_1^+ \\ E_5^- \end{bmatrix},$ | $\begin{bmatrix} E_8^+ \\ E_7^- \end{bmatrix} = S_2 \begin{bmatrix} E_4^+ \\ E_3^- \end{bmatrix},$ | $\begin{bmatrix} E_{14}^- \\ E_{13}^+ \end{bmatrix} = S_3 \begin{bmatrix} E_{10}^- \\ E_9^+ \end{bmatrix}, \begin{bmatrix} E_{16}^+ \\ E_{12}^- \end{bmatrix} = S_4 \begin{bmatrix} E_{15}^- \\ E_{11}^+ \end{bmatrix},$ |
| | | $\begin{bmatrix} E_1^- \\ E_5^+ \end{bmatrix} = S_1 \begin{bmatrix} E_2^- \\ E_6^+ \end{bmatrix},$ | $\begin{bmatrix} E_4^- \\ E_3^+ \end{bmatrix} = S_2 \begin{bmatrix} E_8^- \\ E_7^+ \end{bmatrix},$ | $\begin{bmatrix} E_{10}^+ \\ E_9^- \end{bmatrix} = S_3 \begin{bmatrix} E_{14}^+ \\ E_{13}^- \end{bmatrix}, \begin{bmatrix} E_{15}^+ \\ E_{11}^- \end{bmatrix} = S_4 \begin{bmatrix} E_{16}^- \\ E_{12}^+ \end{bmatrix},$ |

| | Input | Equations | $E_1^+ = 1, \ E_4^+ = 0, \ E_{13}^- = 0, \ E_{16}^- = 0$ |
|---|---|---|---|
| **Figure 7(b-i)** | All units | Scattering matrix equations [c] | $\begin{bmatrix} E_2^+ \\ E_1^- \end{bmatrix} = \begin{bmatrix} T_{SI-1} & R_{SI-1} \\ R_{SI-1} & T_{SI-1} \end{bmatrix} \begin{bmatrix} E_1^+ \\ E_2^- \end{bmatrix}, \begin{bmatrix} E_3^+ \\ E_2^- \end{bmatrix} = T_{w1} \begin{bmatrix} E_2^+ \\ E_3^- \end{bmatrix}, \begin{bmatrix} E_4^+ \\ E_3^- \end{bmatrix} = T_{AD-RR} \begin{bmatrix} E_3^+ \\ E_4^- \end{bmatrix},$ |
| | | | $\begin{bmatrix} E_5^+ \\ E_4^- \end{bmatrix} = T_{w2} \begin{bmatrix} E_4^+ \\ E_5^- \end{bmatrix}, \begin{bmatrix} E_6^+ \\ E_5^- \end{bmatrix} = \begin{bmatrix} T_{SI-2} & R_{SI-2} \\ R_{SI-2} & T_{SI-2} \end{bmatrix} \begin{bmatrix} E_5^+ \\ E_6^- \end{bmatrix},$ |
| | Input | Equations | $E_1^+ = 1, \ E_6^- = 0$ |

[a] $T_{wi}$ ($i = 1, 2$) $= a_{wi}\exp(-j\varphi_{wi})$ and $T_{si}^{1/2}$ ($i = 1, 2$) $= a_{si}^{1/2} \exp(-j\varphi_{si}/2)$ are the field transfer functions of the bus waveguides and half-length of the Sagnac interferometers, respectively.

[b] $S_i$ ($i = 1 - 4$) $= \begin{bmatrix} t_i & j\kappa_i \\ j\kappa_i & t_i \end{bmatrix}$, are the field transfer functions of the directional couplers.

[c] $T_{AD-RR}$ is the field transfer function of the add-drop MRR. $R_{SI-i}$ and $T_{SI-i}$ ($i = 1, 2$) are the field transmission and reflection functions for the two Sagnac interferometers, respectively.

For complex photonic systems, the spectral transfer functions of independent basic units such as a single RR or a single Sagnac inteferometer can be substituted as a whole into the scattering matrix equations to simplify the calculation. Here, "independent basic units" delineates the units that have energy exchanges with other parts only via connecting waveguides. **Figure 7(b-i)** shows a hybrid resonator consisting of an AD-RR sandwiched between two

Sagnac interferometers. The corresponding scattering matrix equations and intensity spectral responses are shown in **Table I** and **Figure 7(b-i)**, respectively. By substituting the spectral transfer functions of the AD-RR and the Sagnac inteferometers, the total number of scatering matrix equations is 12, in contrast to 32 for completely dividing the device into directional couplers and connecting waveguides.

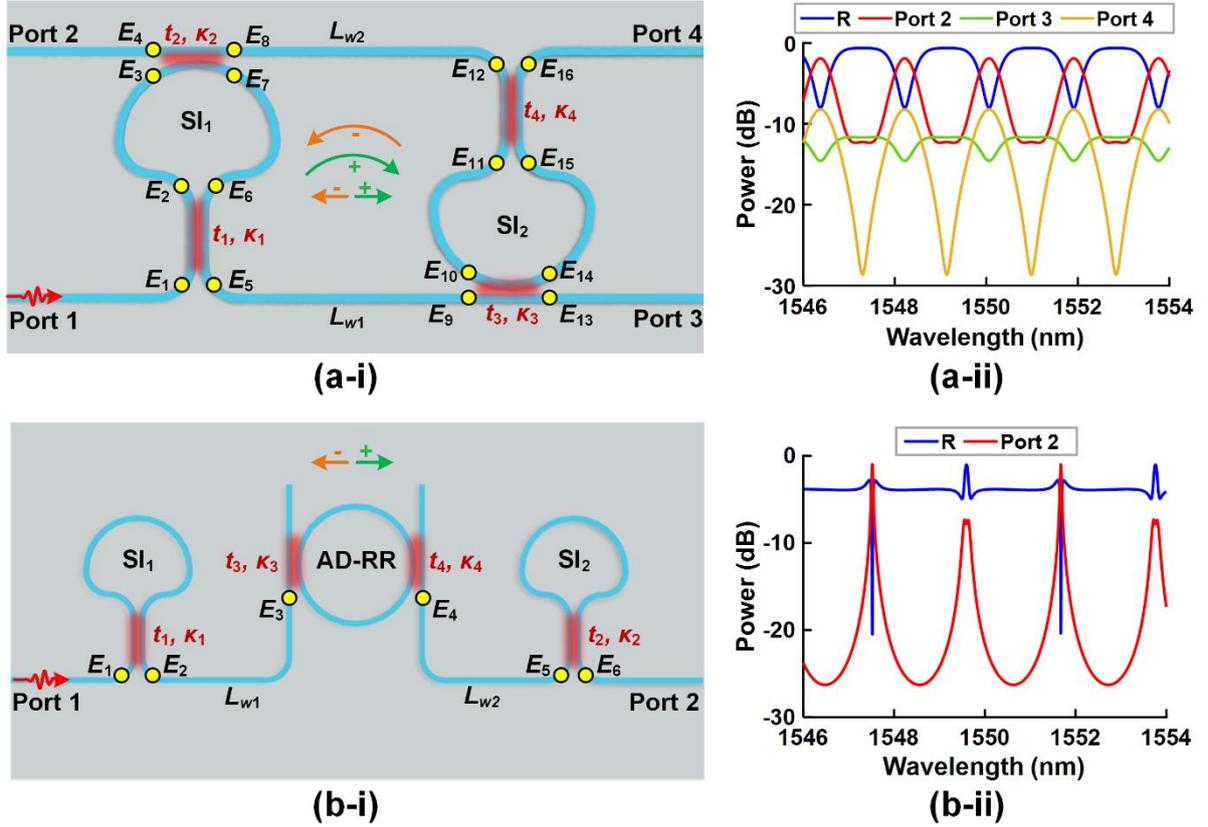

**Figure 7.** Modelling complex integrated photonic filters based on the scattering matrix method. (a) A resonator formed by two mutually coupled Sagnac interferometers (SI). (i) shows the schematic illustration, which is divided into directional couplers and connecting waveguides. (ii) shows the calculated intensity spectral responses at Port 2 – 4 with input from Port 1. (b) A resonator formed by an add-drop RR (AD-RR) sandwiched between two Sagnac interferometers. (i) shows schematic illustration, which is divided into three basic units including a AD-RR and two Sagnac interferometers. (ii) shows the calculated intensity spectral responses at the reflection port (R) and Port 2.

## 3. Device design and tuning

The accurate control of the coupling strength between optical waveguides is fundamentally needed for the design and implementation of not only Sagnac interference devices but also MZIs and RRs. As shown in **Figure 3**, the MZI, RR, and Sagnac interferometers all contain directional couplers. In a directional coupler formed by two closely placed optical waveguides with mutual energy coupling (**Figure 8(a)**), the coupling strength can be changed by varying either the interaction length or the separation gap between them. According to the coupled mode theory [95, 96], the operation principle of a directional coupler can be simplified and explained based on the phase matching condition between the two fundamental eigenmodes of the coupled waveguides, which are commonly termed the even and odd modes, or the symmetric and anti-symmetric modes. **Figure 8(b)** shows the mode profile of the even and odd modes of a directional coupler formed by two parallel silicon wire waveguides. The optical power oscillates between the two waveguides as the modes travel with different propagation constants, and after each distance termed the cross-over length, $L_x$, the optical power totally transfers from one waveguide to the other. The $L_x$ can be given by [96]:

$$L_x = \frac{\lambda}{2(n_{eff,even} - n_{eff,odd})} \tag{7}$$

where $\lambda$ is the light wavelength, $n_{eff,\,even}$ and $n_{eff,\,odd}$ are the effective indices of the two modes, respectively. For a straight coupling length of $L_c$, the field coupling coefficient, $\kappa$, can be expressed as [96]:

$$\kappa = \sin(\frac{\pi}{2}\cdot\frac{L_c}{L_x}) \tag{8}$$

**Figure 8(c-i)** shows $\kappa$ as a function of the gap width $G$ between the two silicon wire waveguides, which was calculated based on **Eqs. (7)** and **(8)**. As can be seen, for a fixed straight coupling length of $L_c$ that is smaller than $L_x$, the coupling strength of the directional coupler can be enhanced by reducing the gap width. This is because the decrease in gap width results in a smaller $L_x$ and hence a larger $\kappa$ according to **Eq. (8)**. Note that the decrease in $\kappa$ for $G < 50$ nm in **Figure 8(c-ii)** is attributed to the fact that $L_x$ in this range is smaller than the fixed $L_c = 14$ μm used in the simulation. For practical devices, the minimum achievable gap width depends on the particular fabrication techniques employed. For electron beam lithography, it is typically

between 50 nm and 150 nm. Whereas for deep ultraviolet (e.g., 193 nm or 248 nm) lithography, it is typically above 150 nm.

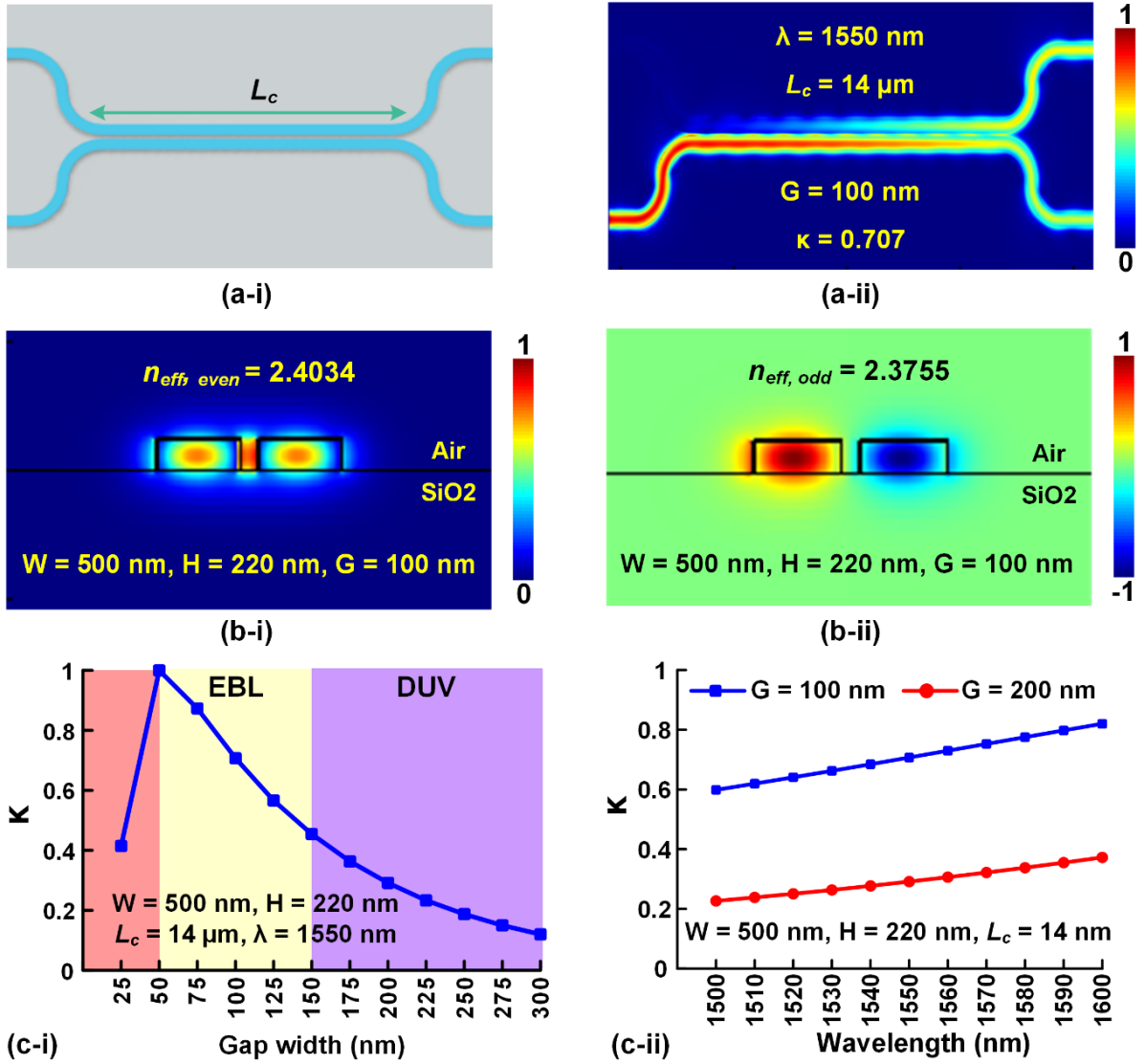

**Figure 8**. Design of directional couplers in integrated photonic devices. (a) A directional coupler formed by two silicon wire waveguides, (i) shows the device schematic and (ii) shows the simulated electric field distribution when the directional coupler works as a 3-dB coupler. (b) Simulated mode profiles of (i) even and (ii) odd modes of the directional coupler in (a). (c) Coupling coefficient $\kappa$ of the directional coupler in (a) as functions of (i) gap width G between the two waveguides and (ii) light wavelength $\lambda$. The ranges of gap width that can be achieved via state-of-the-art electron beam lithography and deep ultraviolet lithography are labeled in (c-i). In (a) – (c), the width and height of the silicon wire waveguides are assumed to be $W$ = 500 nm and $H$ = 220 nm, respectively.

**Figure 8(c-ii)** shows $\kappa$ as a function of incident wavelength, which was calculated based on **Eqs. (7)** and **(8)** after taking account of the waveguide dispersion (including both the material and structure dispersion). As can be seen, the coupling strength of the directional coupler is wavelength dependent due to the existence of dispersion. For a gap width of 100 nm, the $\kappa$

varies from ~0.599 to ~0.820 in a wavelength range of 1500 nm – 1600 nm, and from ~0.662 to ~0.741 in the telecom C-band from 1530 nm to 1565 nm. Whereas for a larger gap width of 200 nm, the change of $\kappa$ with wavelength becomes more gradual, which only varies from ~0.263 to ~0.314 in the C-band. For practical devices, the coupling strength can only be regarded as a wavelength-independent constant in a small wavelength range, as we assumed in previous **Eqs. (1) – (6)** and **Table 1**. Whereas for devices with large operation bandwidths, the wavelength dependence of the coupling strength needs to be considered.

For passive integrated photonic devices, the response spectra are fixed except for tiny variations with environmental factors such as temperature. In practical applications, active tuning of the passive devices is often needed, either to achieve the optimized device performance or to meet the requirements of different applications. The tuning can be achieved by introducing thermo-optic micro-heaters [97-99] or PN junctions [100, 101]. The former has typical response times on the order of $10^{-3}$ s or $10^{-6}$ s, whereas the latter can achieve faster tuning on the order of $10^{-9}$ s or even lower.

**Figure 9** shows the device configurations of tunable Sagnac interferometers, where a tunable MZI coupler replaces the directional coupler in the Sagnac interferometer in **Figure 3(d)**. The effective coupling strength of the MZI coupler can be externally controlled by adjusting the phase difference $\Delta\varphi$ between the two arms. In principle, by integrating a micro-heater along one arm of the MZI coupler (**Figures 9(a) and (b)**) to introduce additional $\pi/2$ phase shift, the reflectivity of the Sagnac interferometer can be tuned from 0% to 100% (**Figure 9(e)**). The tuning of the Sagnac interferometer can also be achieved by integrating PN junctions along the MZI coupler. **Figures 9(c) and (d)** show the device configurations with two PN junctions operating in the common and differential modes, respectively. In the common mode, the phase shifts along the two arms varies symmetrically, which does not introduce any changes in the effective coupling strength and hence the reflectivity. Whereas for the differential mode with the phase shifts along the two arms varying asymmetrically, the effective coupling strength is changed, thus resulting in a variation in the reflectivity. The tuning efficiency is also doubled compared with the devices in **Figures 9(a) and (b)** that have only one phase shifter (**Figure 9(f)**).

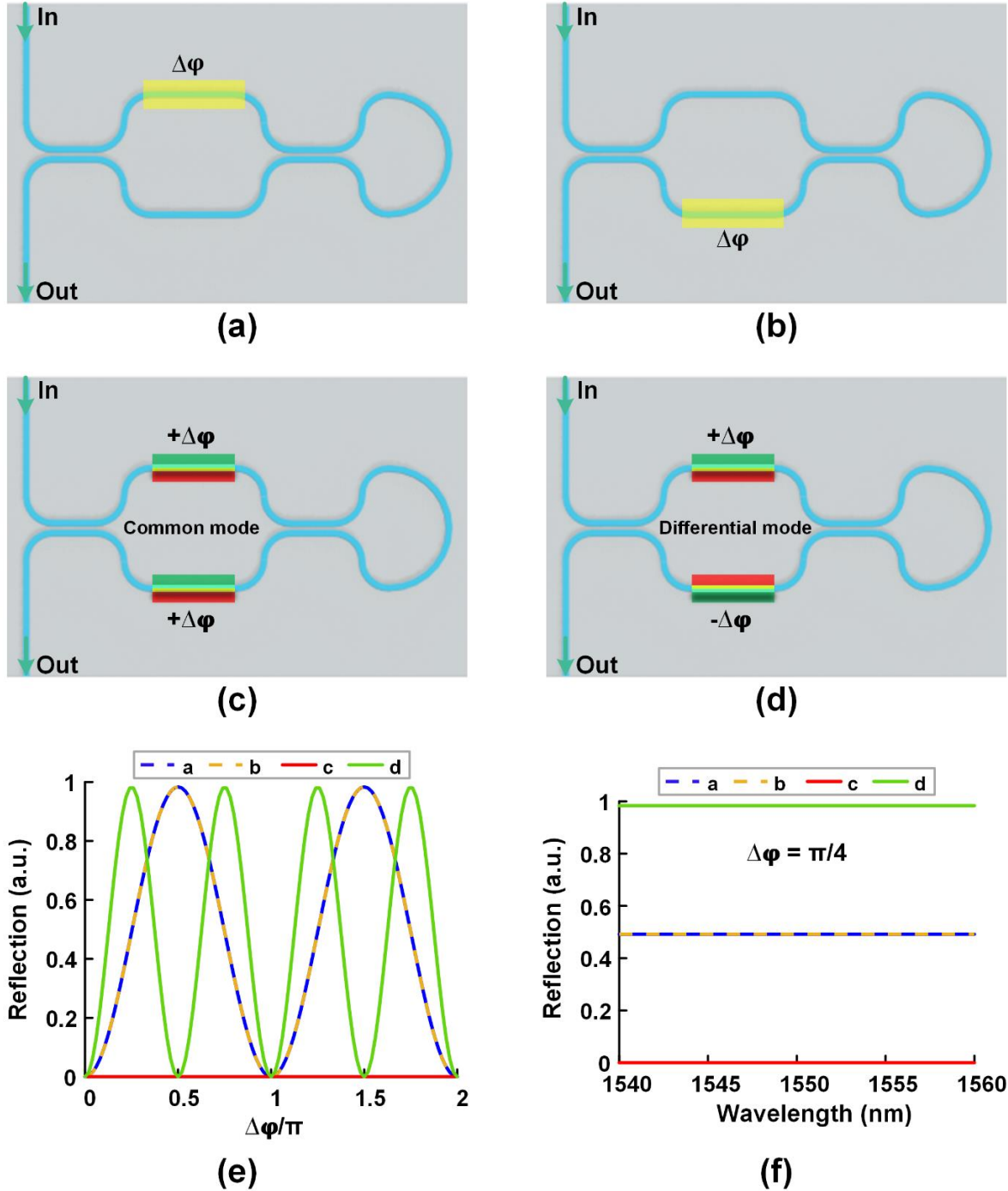

**Figure 9**. Tunable Sagnac interferometers with MZI couplers. (a) – (b) Schematics of tunable Sagnac interferometers with a micro-heater integrated along one arm of the MZI coupler. (c) – (d) Schematics of tunable Sagnac interferometers with PN junctions integrated along both arms of the MZI coupler. (e) – (f) Reflectivity of the Sagnac interferometers in (a) – (d) as functions of phase difference between the two arms Δφ and wavelength, respectively.

## III. Integrated Sagnac interference devices

Integrated Sagnac inteferometers that feature a simple configuration, low wavelength dependence, and high scalability are versatile for implementing a variety of functional photonic devices. Along with the advances in fabrication techniques, a series of applications of integrated Sagnac interference devices have been continuously demonstrated. In this section, we present our latest results for integrated Sagnac interference devices. These include reflection mirrors, optical gyroscopes, basic filters, wavelength (de)interleavers, optical analogues of quantum physics, and others.

### 1. Reflection mirrors

A direct application of integrated Sagnac interferometers is reflection mirrors, which are fundamental components in PICs [102]. Compared to reflection mirrors based on Bragg gratings [103], PhCs [104], and coupled RRs [105], Sagnac loop reflection mirrors (SLRMs) are advantageous by simultaneously providing high fabrication tolerance, broad reflection band, and high flexibility in tuning the reflectance. **Figure 10** shows a variety of photonic integrated systems including SLRMs.

**Figure 10(a)** shows an arrayed waveguide grating (AWG) system fabricated on an SOI wafer [106], where an array of SLRMs implemented based on $1 \times 2$ multi-mode interference (MMI) couplers were employed to reflect the signals back to the arrayed waveguides. **Figure 10(b)** shows a silicon optical delay line system consisting of 13 cascaded RRs and a SLRM [107], where the light reflected from the SLRM passed through the cascaded RRs for a second time, thus allowing for a doubled delay-bandwidth product. **Figure 10(c)** shows a monolithically integrated multi-wavelength laser on an indium phosphide (InP) wafer [108], where each laser cavity was formed by two SLRMs, together with an AWG and multiple semiconductor optical amplifiers (SOAs) that serve as frequency selection and signal amplification modules, respectively. **Figure 10(d)** shows a silicon-based optical driver for fiber optic gyroscopes [109]. Similar to the multi-wavelength laser in **Figure 10(c)**, there is a FP laser cavity between the two SLRMs, which were designed to have different reflectivities to direct the generated light towards the subsequent module. **Figure 10(e)** shows another type of integrated multi-wavelength laser [110], where coherent laser frequency combs were generated based on optical parametric oscillation in the high-Q SiN RR, with a SLRM serving as the laser output coupler.

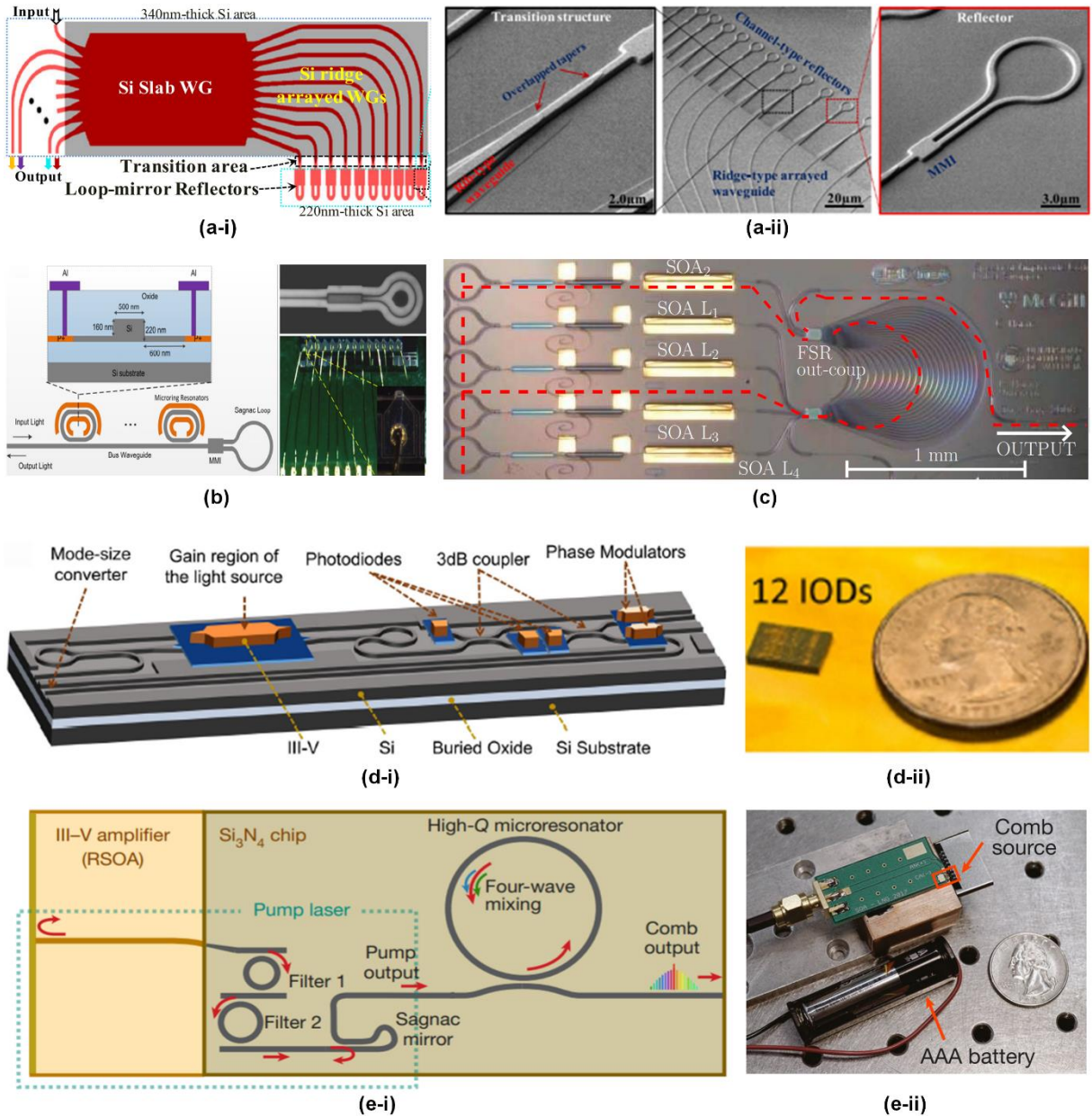

**Figure 10**. Integrated Sagnac loop reflection mirrors (SLRMs). (a) A schematic and SEM images of a silicon arrayed waveguide grating including multiple SLRMs. WG: waveguide grating. (b) A schematic and images of a silicon optical delay line terminated with a SLRM. (c) A microscope image of an InP multi-wavelength laser with an array of SLRMs. SOA: semiconductor optical amplifier. (d) A schematic and a photograph of a silicon-based optical driver (IOD) for fiber optic gyroscopes including SLRMs. (e) A schematic and a photograph of an integrated optical frequency comb generator with a SLRM. (a) Adapted with permission [106]. Copyright 2018, IEEE. (b) Adapted with permission [107]. Copyright 2014, Optica Publishing Group. (c) Adapted with permission [108]. Copyright 2010, Springer. (d) Adapted with permission [109]. Copyright 2017, Optica Publishing Group. (e) Adapted with permission [110]. Copyright 2018, Springer Nature Limited.

## 2. Optical gyroscopes

Optical gyroscopes providing a powerful approach for highly precise rotation measurement have formed the backbone of inertial navigation systems in modern ships, airplanes, satellites, submarines, and spacecraft [111, 112]. In an optical gyroscope, the Sagnac effect induces a time difference between light waves traveling in opposite directions around a circular path, which increases with the rotation rate and can be measured via optical interference. This time difference is detected as a phase shift when the circular path is open or a frequency shift when the circular path is closed.

The advances in technologies for integrated device fabrication as well as accurate measurement of Sagnac interference in PICs have enabled chip-scale optical gyroscopes, which provide significantly reduced size, weight, and power consumption (SWaP) compared to conventional ring laser gyroscopes [43] and fiber optic gyroscopes [34, 113]. In concert with the development of ultra-low-loss waveguides [114-117] and ultra-high-Q resonators [118-120], various schemes have been proposed to realize chip-scale optical gyroscopes. Here we classify them into three categories, including interferometric optical gyroscopes (IOGs), passive resonant optical gyroscopes (PROGs), and Brillouin ring laser gyroscopes (BRLGs). In the PROGs and BRLGs, ultrahigh-Q resonators are employed. Note that although some ultrahigh-Q whispering-gallery-mode (WGM) cavities are off-chip discrete devices, recent progress in fabrication techniques has enabled their full integration in PICs [121, 122], therefore they are considered as integrated devices. We also note other types of optical gyroscopes using integrated III-V semiconductor ring laser cavities as both optical sources and sensing elements [123, 124], which was first proposed in the 1980's [125] and gradually replaced by the PROGs and BRLGs after 2010, mainly due to the fact that the backscattering, mode competition, and lock-in effect inside the ring laser cavities make it quite complex to generate stable and low-noise beat signals for angular rate measurement [111, 126]. In **Table 2**, we summarize and compare the performance of the state-of-the-art integrated optical gyroscopes.

**Table 2**. Performance comparison of the state-of-the-art integrated optical gyroscopes.

| Type | Integrated components | Platform | Bias drift [a] (deg/h) | Sensitivity [b] (deg/h) | ARW [c] (deg/h$^{1/2}$) | Ref. |
|---|---|---|---|---|---|---|
| **IOG** | Coiled waveguide | SiN | 58.7 | – | 8.52 | [127] |
| **IOG** | Coiled waveguide | Si | – | 184680 | – | [128] |
| **IOG** | Coiled waveguide | Silica | 7.32 | – | 1.26 | [129] |
| **IOG** | All the passive components and a thermo-optic micro-heater | Si | – | 2304 | – | [130] |
| **IOG** | A MZI switch, two RRs with thermo-optic micro-heaters, two germanium photodiodes, and several waveguide couplers | Si | 21600 | 432000 | 650 | [131] |
| **PROG** | RR | Silica | – | – | – | [132] |
| **PROG** | RR | Silica | – | – | – | [133] |
| **PROG** | RR | Silica | 324 | – | – | [134] |
| **PROG** | RR | Polymer | – | 324 | – | [135] |
| **PROG** | RR | InP | – | 10 | – | [136] |
| **PROG** | A spiral resonator coupled to a straight bus waveguide through a MMI coupler | InP | 1 | 10 | – | [137] |
| **PROG** | RR | Silica | 14.4 | 3.74 | – | [138] |
| **PROG** | All parts are integrated on an optical micro-bench | CaF$_2$ | 3 | – | 0.02 | [139] |
| **PROG** | Microrod resonator | Silica | – | 7200 | – | [140] |
| **BRLG** | Disk resonator | Silica | – | 22 | – | [141] |
| **BRLG** | RR | SiN | – | 90000 | – | [142] |
| **BRLG** | Wedge resonator | Silica | – | – | – | [143] |
| **BRLG** | Wedge resonator | Silica | 3.6 | 5 | 0.068 | [144] |

[a] Bias drift (deg/h): deviation from the mean value of the output rate in the bias stability measurement.
[b] Sensitivity (deg/h): minimum detectable angular rate or angular velocity of the gyroscope.
[c] Angle random walk (ARW, deg/h$^{1/2}$): noise contribution to the rotation angle value, describing the average deviation or error that occurs during signal integration.

## A. Interferometric optical gyroscopes (IOGs)

In integrated IOGs, long coiled waveguides are usually employed as sensing modules to accumulate the Sagnac phase shift. **Figure 11** shows typical integrated IOGs implemented based on different material platforms such as SOI, SiN, and silica.

An IOG with an SiN coiled waveguide has been demonstrated (**Figure 11(a)**) [127], where the coiled waveguide had a length of ~3 m and a low propagation loss < 0.78 dB/m, yielding an angle random walk (ARW) of ~8.52 deg/h$^{1/2}$ and a bias drift of ~58.7 deg/h. Subsequently, another IOG with an SOI coiled waveguide has been reported (**Figure 11(b)**) [128], where the multi-mode coiled waveguide had a length of ~2.76 cm and included 15 crossings, with an average propagation loss of ~1.33 dB/cm and an average crossing loss of ~0.08 dB. An IOG with a 2.14-m-long SiO$_2$ coiled waveguide has also been investigated (**Figure 11(c)**) [129], which was connected to other optical devices of the gyroscope system via fiber tail coupling. By optimizing the bend radius and the space between adjacent loops, a low insertion loss of ~8.37 dB was achieved for the fabricated coiled waveguide with 11 crossings, yielding a low bias drift of ~7.32 deg/h and a low ARW of ~1.26 deg/h$^{1/2}$.

**Figure 11(d)** shows a silicon IOG consisting of two mode multiplexers, two 3-dB couplers, a thermo-optic micro-heater, two bent connecting waveguides, several grating couplers, and a coiled waveguide [130]. Two counter-propagating modes in the coiled waveguide induced a constant phase difference, which was engineered to achieve significantly improved detection sensitivity for the system. Compared to conventional IOGs, the IOG assisted by mode interference did not require any phase modulators or circulators, yielding both reduced system complexity and strong capability for monolithic integration.

A hybrid integrated IOG with a compact footprint of ~2 mm$^2$ has been reported [131], which consisted of an electro-optic MZI switch, two RRs with thermo-optic micro-heaters, two germanium photodiodes, and several waveguide couplers on a single SOI chip (**Figure 11(e)**). The MZI switch was used to change the direction of light injected into the RRs and hence the output toggles between the two photodiodes. The output signals from the two photodiodes were summed and mixed with the reference frequency to extract the amplitude related to the rotation rate information. Owing to the significant reduction of thermal fluctuations and mismatch enabled by the reciprocity of such system, sensitive detection of small phase shift down to 3 nrad was achieved, which is 30 times smaller than typical fibre-optic gyroscopes.

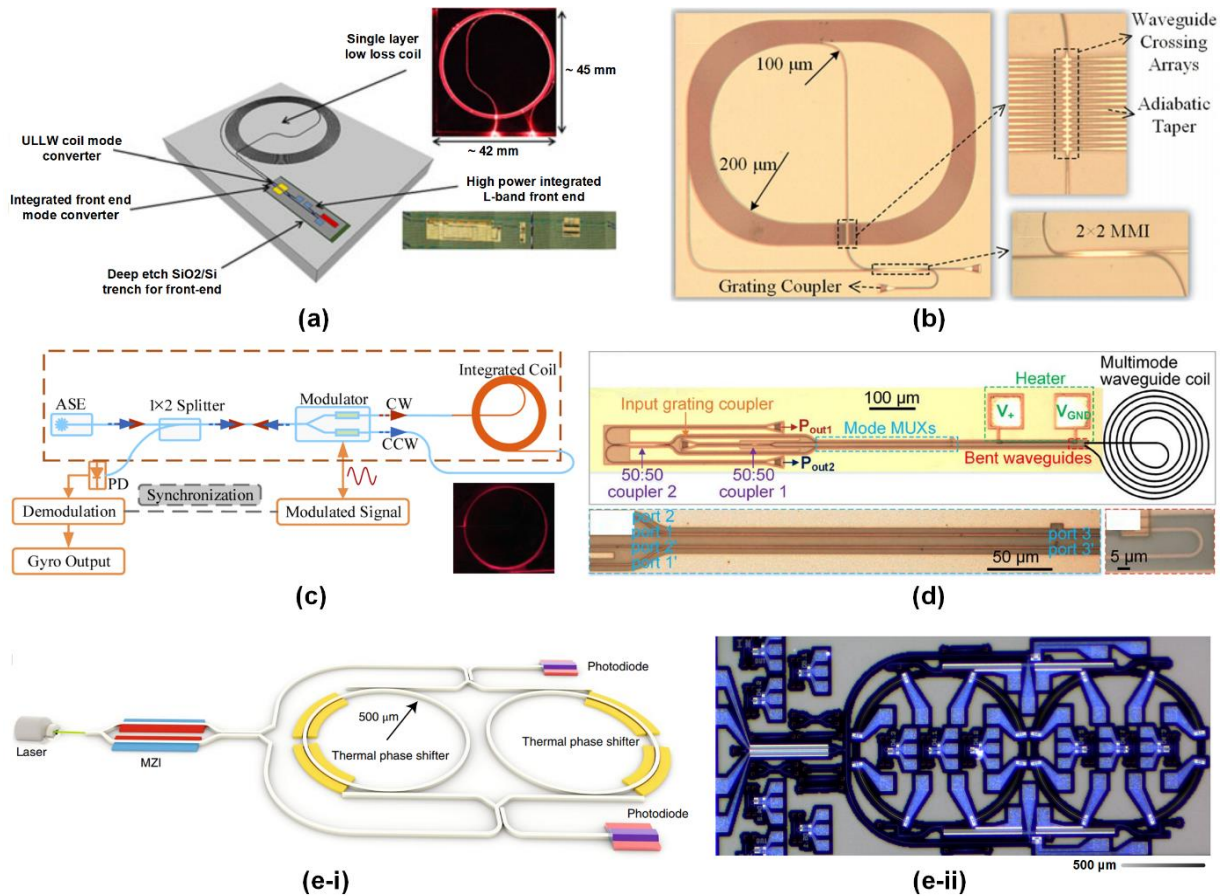

**Figure 11**. Integrated interferometric optical gyroscopes (IOGs). (a) An IOG with an SiN coil waveguide. (b) An IOG with an SOI coiled waveguide. (c) An IOG with a $SiO_2$ coiled waveguide. (d) A silicon IOG with improved sensitivity enabled by mode interference. (e) A hybrid integrated IOG with reciprocal sensitivity enhancement. (i) and (ii) shows the schematic and the practical device, respectively. (a) Adapted with permission [127]. Copyright 2017, IEEE. (b) Adapted with permission [128]. Copyright 2018, Springer Nature Limited. (c) Adapted with permission [129]. Copyright 2020, Optica Publishing Group. (d) Adapted with permission [130]. Copyright 2019, Springer Nature Limited. (e) Adapted with permission [131]. Copyright 2018, Springer Nature Limited.

## B. Passive resonant optical gyroscopes (PROGs)

In contrast to the use of long coiled waveguides as sensing modules in integrated IOGs, the sensing modules in integrated PROGs are implemented by passive resonators, thus allowing for more compact device footprint. Moreover, unlike the IOGs with their sensitivities being restricted by the transmission loss of the long coiled waveguides, the PROGs based on high-Q optical resonators show advantages in achieving high sensitivities [129, 132, 145]. **Figure 12** shows several typical integrated PROGs.

A PROG based on silica planar lightwave circuit (PLC) resonator with a circumference of ~14.8 cm and a propagation loss of ~2.4 dB/m was first demonstrated (**Figure 12(a)**) [132], where the system configuration enabled countermeasures for noise induced by both backscattering and polarization fluctuations. In the experimental demonstration, the former was effectively suppressed by using binary phase shift keying modulation to compensate the spectral response of the thermo-optic modulator and the electronic gating, whereas the latter was lowered by adjusting the spacing between the two eigenstates of polarization to control the waveguide birefringence.

A PROG based on a silica RR with a diameter of ~4 cm has been demonstrated (**Figure 12(b)**) [133], where laser frequency modulation spectroscopy technique was used for signal detection in the open-loop operation PROG system. A dynamic range between $-2.0 \times 10^3$ and $2.0 \times 10^3$ rad/s was experimentally achieved for the PROG, and the slope of the linear fit for the equivalent gyroscope rotation was about 0.330 mV/(deg/s) based on the $-900$ to 900 kHz equivalent frequency. By using a trapezoidal phase modulation technique to achieve real-time compensation for the output of the gyroscope, another PROG based on a silica RR with a diameter of ~6 cm has also been reported [134] (**Figure 12(c)**). The experimental demonstration verified that the deviation induced by noise and short-term drift was significantly reduced after the compensation, yielding a bias stability of ~0.09 deg/s with an integration time of 10 s over 3000 s. Other modulation techniques have also been investigated, either to improve the output from signal detection modules [146-149], or to reduce the backscattering [145, 150, 151] and backreflection [145, 152-154] noise.

The use of a polymer RR with a diameter of ~1 cm as the sensing element of a PROG has been investigated (**Figure 12(d)**) [135], where the fabricated polymer RR had a propagation loss of ~0.5 dB/cm and a Q factor of ~$10^5$. The theoretically estimated shot-noise-limited sensitivity of such PROG was < 0.09 deg/s, reaching the level of rate-grade gyroscope. Another PROG with a InP RR as the sensing element has also been investigated (**Figure 12(e)**) [136]. The fabricated RR with a diameter of ~26 mm had a propagation loss of ~0.45 dB/cm and a Q factor of ~$1 \times 10^6$, leading to theoretically estimated shot-noise-limited resolution of ~10 deg/h.

**Figure 12(f)** shows an InP spiral resonator coupled to a straight bus waveguide through a MMI coupler, which was designed as the sensing element of a PROG [137]. The fabricated InP spiral resonator had a total length of ~60 mm and a Q factor of ~$5.9 \times 10^5$, resulting in theoretically estimated resolution and bias drift of ~150 deg/h and ~4 deg/h, respectively.

A PROG based on a Ge-doped silica RR with a circumference of ~7.9 cm has been demonstrated (**Figure 12(g)**) [138], where the RR was pigtailed with single-polarization fibers instead of polarization maintaining fibers to reduce the polarization induced error. In the experimental demonstration, a high bias stability of ~0.004 deg/s over 1 h was achieved, with the detection resolution exceeding the earth's rotation rate. In addition, a PROG based on a WGM cavity with a diameter of ~7 mm has been reported (**Figure 12(h)**) [139]. The WGM cavity featured both high Q factors (with a finesse of $10^5$) and low Rayleigh backscattering (< 10 ppm), yielding an ARW of ~0.02 deg/$h^{1/2}$ and a bias drift of ~3 deg/h at a cavity driven power of ~80 μW. **Figure 12(i)** shows a schematic of another PROG based on a silica micro-rod resonator with a diameter of ~2.8 mm [140], where symmetry breaking induced by Kerr nonlinear interaction between the counter-propagating light waves was utilized to significantly enhance the responsivity by about 4 orders of magnitude.

In addition to the experimental reports mentioned above, there has been significant theoretical work proposing new device designs for the PROGs, such as those based on coupled resonators [155-159], PhC cavities [160-162], and plasmonic devices [163], with the aim of increasing the sensitivity or reducing the footprint.

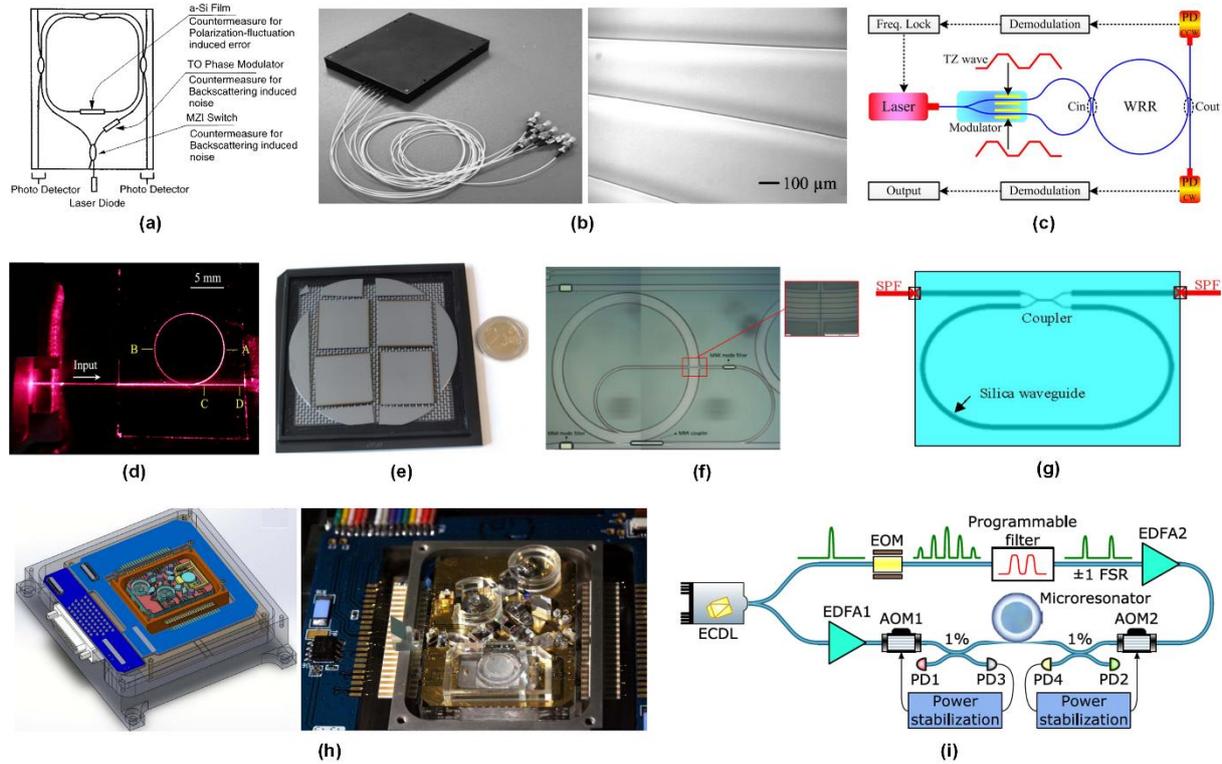

**Figure 12**. Integrated passive-resonant optical gyroscopes (PROGs). (a) A PROG based on a silica planar lightwave circuit (PLC) resonator with countermeasures for noises induced by backscattering and polarization fluctuation. (b) A packaged silica RR used for a PROG with open-loop operation. (c) A PROG based on a silica RR using trapezoidal phase modulation technique. (d) A centimeter-scale polymer RR designed as sensing element of a PROG. (e) Four RR chips on a InP wafer as sensing elements for a PROG. (f) An InP spiral resonator coupled to a straight bus waveguide through a MMI coupler designed as sensing element of a PROG. (g) A Ge-doped silica RR coupled with single-polarization fibers (SPFs) used for a PROG. (h) A PROG based on a WGM cavity. (i) A PROG based on a silica micro-rod resonator. (a) Adapted with permission [132]. Copyright 2000, Optica Publishing Group. (b) Adapted with permission [133]. Copyright 2014, IOP Publishing. (c) Adapted with permission [134]. Copyright 2015, Optica Publishing Group. (d) Adapted with permission [135]. Copyright 2016, Elsevier B.V. (e) Adapted with permission [136]. Copyright 2013, Optica Publishing Group. (f) Adapted with permission [137]. Copyright 2015, IEEE. (g) Adapted with permission [138]. Copyright 2017, Optica Publishing Group. (h) Adapted with permission [139]. Copyright 2017, Optica Publishing Group. (i) Adapted with permission [140]. Copyright 2021, Optica Publishing Group.

**C. Brillouin ring laser gyroscopes (BRLGs)**

With the unique property of narrowing the pump laser linewidth, highly coherent and ultra-low phase noise emission, and operation across wide wavelength ranges (e.g., from visible to infrared [164, 165]), ring lasers based on stimulated Brillouin scattering (SBS) are promising candidates for high-spectral-purity optical sources [166-169]. BRLGs were first studied in optical fibers [170] and more recently in integrated platforms [142-144]. Compared to PROGs, BRLGs based on narrow-linewidth Brillouin lasing can yield better sensitivity. **Figure 13** shows typical integrated BRLGs demonstrated experimentally.

A BRLG based on a high-Q silica micro-disk resonator with a diameter of ~18 mm has been demonstrated (**Figure 13(a)**) [141], where the frequency shift induced by Sagnac interference was measured by using a single pump to trigger Brillouin lasing in a cascaded fashion. In the rotation-rate measurement, a sensitivity of ~15 deg/h/Hz$^{1/2}$ and a rotation rate down to ~22 deg/h were achieved.

**Figure 13(b)** shows another BRLG demonstrated using a Brillouin laser chip and discrete front-end gyroscope components on a rotation table [142]. The Brillouin laser was realized based on a high-Q SiN RR with a circumference of ~74 mm, achieving a fundamental linewidth down to ~0.7 Hz. By applying a rotation rate varying from 0 to 75 deg/s, the BRLGs achieved linear operation with a scale factor of ~152 Hz/deg/s.

**Figure 13(c)** shows a BRLG with enhanced response to rotations when working near an exceptional point where multiple eigenstates coalesce [143]. In the experimental demonstration, precise control of the counter-propagating laser modes with a high stability was achieved via phase matching of the Brillouin gain and the dispersion of the silica wedge resonator. Four-times improvement of the Sagnac scale factor was observed by measuring rotations with an amplitude of about 1 revolution per hour.

**Figure 13(d)** shows a BRLG integrated on a silicon chip [144], where counter-propagating Brillouin lasing was generated by counter-pumping a high-Q silica wedge resonator with a diameter of ~36 mm. The fabricated device was used for measuring the Earth's rotation rate with both high sensitivity and stability, achieving an ARW noise of ~0.068 deg/h$^{1/2}$ and a bias drift of ~3.6 deg/h.

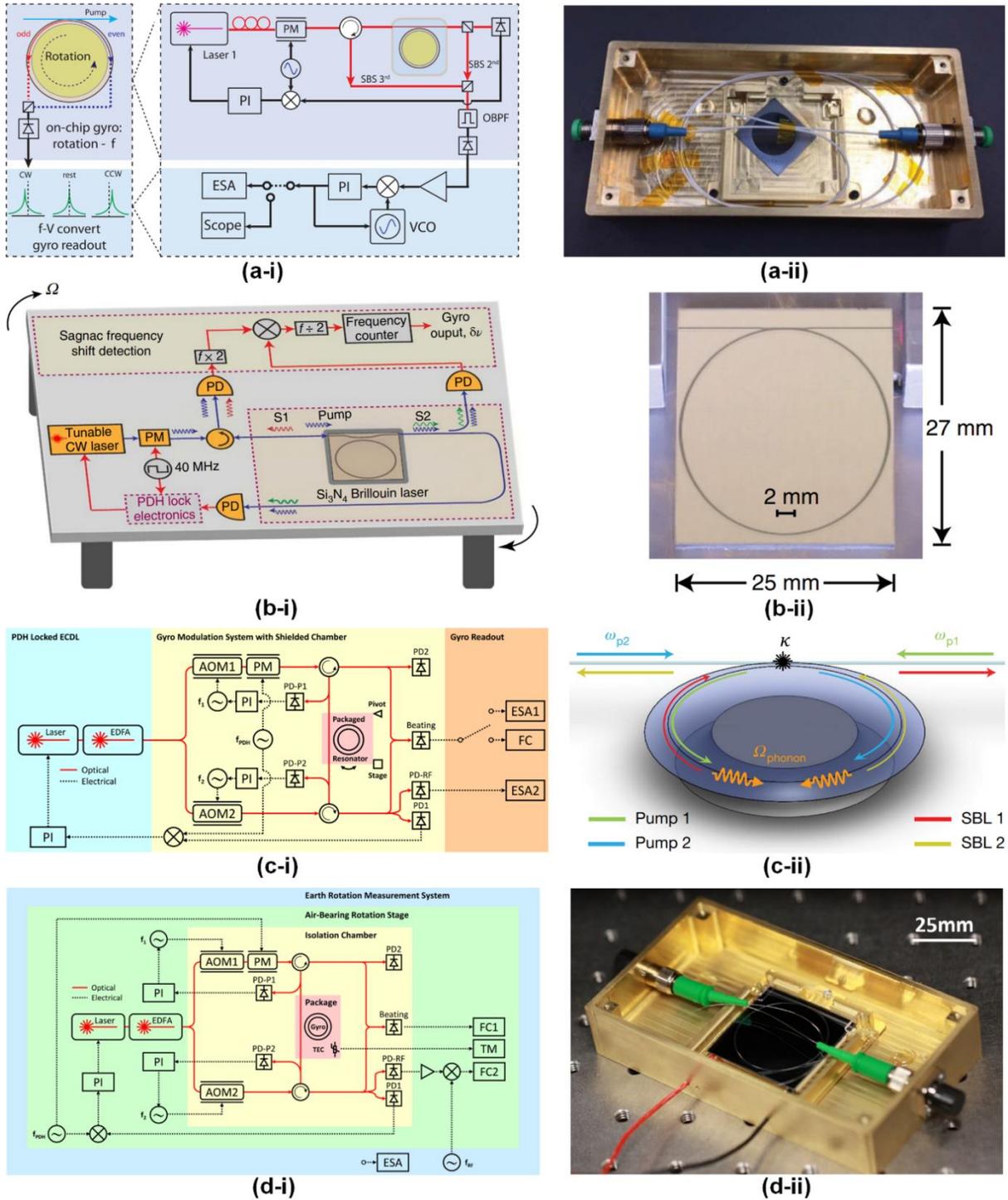

**Figure 13**. Integrated Brillouin ring laser gyroscopes (BRLGs). (a) A BRLG based on a high-Q silica micro-disk resonator. (i) shows the system schematic and (ii) shows the packaged gyroscope. (b) A BRLG based on a high-Q SiN RR. (i) shows the system schematic and (ii) shows the fabricated SiN Brillouin laser chip. (c) A BRLG based on a high-Q silica wedge resonator. (i) shows the system schematic and (ii) shows illustration of the dual stimulated

Brillouin scattering laser process in the wedge resonator. (d) A BRLG system based on high-Q silica wedge resonator used for earth rotation measurement. (i) shows the system schematic and (ii) shows the packaged gyroscope. (a) Adapted with permission [141]. Copyright 2017, Optica Publishing Group. (b) Adapted with permission [142]. Copyright 2018, Springer Nature Limited. (c) Adapted with permission [143]. Copyright 2019, Springer Nature Limited. (d) Adapted with permission [144]. Copyright 2020, Springer Nature Limited.

## 3. Basic filters

Basic filters such as comb, Butterworth, Bessel, Chebyshev, and elliptic filters are of fundamental importance for signal filtering and processing in communications and computing systems [100, 171-174]. Integrated photonic resonators provide an attractive solution to realize these filters in the optical domain and with compact device footprint. Compared to devices with unidirectional light propagation such as RRs, Sagnac interference devices with bidirectional light propagation provide more versatile mode interference that can be tailored for realizing a range of basic optical filters. In **Table 3**, we compare different basic optical filters, including both Sagnac and non-Sagnac interference devices.

**Table 3. Comparison of integrated basic optical filters.**

| Filter name | Key characteristics of spectral response | Non-Sagnac interference device | Sagnac interference device |
|---|---|---|---|
| **Comb filter** | With an amplitude response consisting of a series of regularly spaced notches or peaks that resemble a comb. | [175-177] | [100, 174, 178-180] |
| **Butterworth filter** | With a flat bandpass amplitude response | [181-183] | [56, 180, 184-188] |
| **Bessel filter** | With a linear phase response (i.e., constant group delay) over the amplitude passband | [181, 189] | [180, 184, 185, 187, 188] |
| **Chebyshev Type I filter** | With passband ripples and flat stopband response | [181, 190, 191] | [187, 188] |
| **Chebyshev Type II filter** | With stopband ripples and flat passband response | [190-192] | [187] |
| **Elliptic filter** | With ripples in both passband and stopband, thus providing the steepest roll-off than other types of filters. | [191, 192] | [187] |

A tunable silicon photonic comb filter formed by two cascaded Sagnac interferometers has been demonstrated (**Figure 14(a)**) [100], which had 54 filtering channels with a spacing of ~115 GHz in the wavelength range of 1510 − 1560 nm. Interleaved PN junctions were implemented to electrically tune the comb filter, with both blue and red shift of the comb channels being achieved. By replacing the directional couplers in the Sagnac interferometers with tunable MZI couplers, this comb filter was upgraded to achieve both wavelength and bandwidth tuning (**Figure 14(b)**) [174], where 93 filtering channels with a spacing of ~40 GHz were obtained in the wavelength range of 1535 − 1565 nm. Micro-heaters were integrated to thermally tune the MZI couplers, achieving wavelength tuning with an efficiency of ~0.019 nm/mW as well as continuous bandwidth tuning from ~6 GHz to ~25 GHz.

**Figure 14(c)** shows a silicon photonic Butterworth filter formed by four cascaded Sagnac interferometers [180]. Both bandwidth tuning from ~8.50 GHz to ~20.25 GHz and central wavelength tuning exceeding one FSR were demonstrated by tuning the thermo-optic micro-heaters integrated along the coupling regions and the connecting waveguides between adjacent Sagnac interferometers, respectively.

An integrated optical filter formed by a self-coupled Sagnac interferometer has been proposed (**Figure 14(d)**) [184]. By properly designing the device structural parameters, both Butterworth and Bessel filters can be realized. Since the filter shape of such device resulted from mutual coupling between two counter-propagated modes in the same resonant cavity, it showed higher power-efficiency in phase tuning as well as higher tolerance to fabrication errors compared to RRs.

Another integrated optical filter formed by three cascaded Sagnac interferometers coupled to a top bus waveguide has also been investigated (**Figure 14(e)**) [187]. By tailoring coherent mode interference in such a device consisting of both FIR and IIR filter elements, its spectral response was engineered to achieve Butterworth, Bessel, Chebyshev, and elliptic filters with broad filtering bandwidths and high extinction ratios.

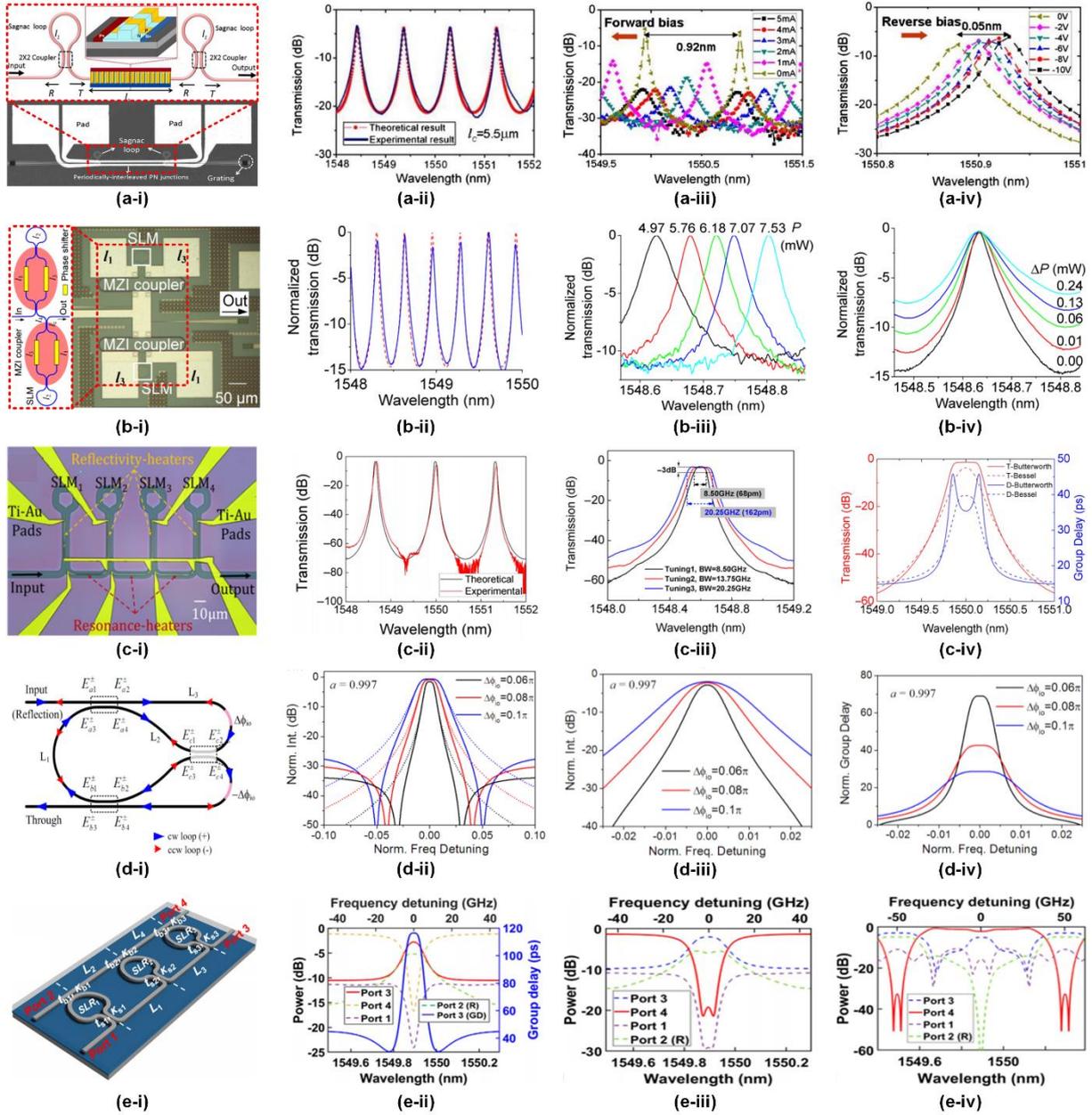

**Figure 14.** Basic optical filters formed by intergraded Sagnac interferometers. (a) An electrically tunable silicon photonic comb filter formed by two cascaded Sagnac interferometers. (b) A thermally tunable silicon photonic comb filter formed by two cascaded Sagnac interferometers with MZI couplers. (c) A silicon photonic reconfigurable Butterworth filter formed by four cascaded Sagnac interferometers. (d) Butterworth and Bessel filters based on a resonator formed by a self-coupled Sagnac interferometer. (e) Butterworth, Bessel, Chebyshev, and elliptic filters based on a resonator formed by three cascaded Sagnac interferometers coupled to a top bus waveguide. In (a) – (e), (i) shows the microscope image or schematic of the device, and (ii) – (iv) show the spectral response. (a) Adapted with permission [100]. Copyright 2013, Optica Publishing Group. (b) Adapted with permission [174]. Copyright 2016, Optica Publishing Group. (c) Adapted with permission [180]. Copyright 2021, Optica Publishing Group. (d) Adapted with permission [184]. Copyright 2011, Optica Publishing Group. (e) Adapted with permission [187]. Copyright 2021, Optica Publishing Group.

## 4. Wavelength (de)interleavers

Wavelength (de)interleavers are key components for signal multiplexing and demultiplexing in WDM optical communication systems [193-195]. To date, various schemes have been proposed to realise compact chip-scale interleavers based on RRs [196, 197], MZIs [195, 198, 199], and Sagnac interferometers [70, 179, 194, 200]. To achieve high filtering roll-off, these devices usually include multiple cascaded subunits. Compared to (de)interleavers composed of RRs or MZIs, (de)interleavers formed by Sagnac interferometers can achieve the same level of filtering flatness and roll-off with fewer subunits, due to the stronger coherent mode interference within more compact device footprint enabled by the bidirectional light propagation as well as the SW resonator nature. In **Table 4**, we compare the performance of the state-of-the-art integrated wavelength (de)interleavers based on Sagnac interference.

**Table 4. Performance comparison of integrated wavelength (de)interleavers based on Sagnac interference. ER: extinction ratio. CS: channel spacing. IL: insertion loss. SI: Sagnac interferometer.**

| Device structure | Integrated platform | Device footprint (µm$^2$) | ER (dB) | CS (GHz) | IL (dB) | Ref. |
|---|---|---|---|---|---|---|
| 7 coupled SIs | SOI | ~320 × 150 | ~20 | ~100 | ~8.0 | [194] |
| 4 cascaded SIs | SOI | ~125 × 376 | ~20 | – | ~6.0 | [201] |
| 2 cascaded SIs in a Saganc interfering loop | SOI | ~120 × 60 | ~25 | – | ~7.3 | [200] |
| 2 cascaded SIs with MZI couplers in a Saganc interfering loop | SOI | ~736 × 523 | ~20 | – | ~6.0 | [179] |
| 1D-PhC FP cavity in a Saganc interfering loop | SOI | ~64 × 70 | ~20 | ~2370 | ~0.5 | [202] |
| A MZI structure with cascaded SIs in the two arms | N/A [a] | N/A [a] | ~29 | ~50 | ~1.0 | [186] |
| 2 parallel SIs coupled to a bus waveguide | N/A [a] | N/A [a] | ~32 | ~50 | ~0.8 | [70] |
| 2 coupled SIs with a feedback loop | N/A [a] | N/A [a] | ~13 | – | ~0.4 | [203] |

[a] This is simulation work.

**Figure 15(a)** shows a passive silicon photonic interleaver based on coupled Sagnac interferometers formed by a self-coupled optical waveguide [194]. Compared to ring-assisted MZI interleavers [204-206], the high-order filtering capability of the multi-stage Sagnac interferometers enabled both a reduced footprint and an increased extinction ratio. The fabricated device exhibited a flat-top spectral response with a steep roll-off, achieving an extinction ratio of ~20 dB and an insertion loss of ~8 dB in the C-band.

A tunable silicon photonic interleaver based on a Michelson-Gires-Tournois interferometer formed by cascaded Sagnac interferometers has also been demonstrated (**Figure 15(b)**) [201], where thermo-optic micro-heaters were integrated to tune the phase shifts along the waveguides and hence the filtering center wavelength. The SW resonator nature of cascaded Sagnac interferometers yielded both a small device footprint of ~125 × 376 $\mu m^2$ and a high tuning efficiency of ~0.04 nm/mW for the fabricated device. The interleaver had a channel spacing of ~2.5 nm (i.e., ~312 GHz) and achieved a high 20-to-3 dB bandwidth ratio of ~1.37.

**Figure 15(c)** shows another tunable silicon photonic interleaver formed by incorporating two Sagnac interferometers in an interfering loop [200]. Similar to the device in **Figure 15(b),** a micro-heater was employed to thermally tune the phase shift along the connecting waveguide between the two Sagnac interferometers, enabling a tunable center wavelength across the entire FSR with an efficiency of ~0.08 nm/mW. To achieve flat-top filtering, the coupling strengths of the directional couplers were optimized by setting the second-order derivative of the intensity transfer function to zero. An operation bandwidth of ~60 nm and a 20-to-3 dB bandwidth ratio of ~1.42 were achieved for the fabricated device with a footprint of ~120 × 60 $\mu m^2$.

By replacing the directional couplers of the Sagnac interferometers with MMI-assisted tunable MZI couplers, the interleaver in **Figure 15(c)** was modified to provide an additional degree of freedom in tuning the extinction ratio (**Figure 15(d)**) [179]. By tuning the micro-heaters along one of the MZI couplers and the connecting waveguide between the two Sagnac interferometers, tunable extinction ratio from 11.8 dB to 24.0 dB and center wavelength with an efficiency of ~0.0193 nm/mW were demonstrated, respectively.

**Figure 15(e)** shows another silicon photonic interleaver modified on the basis of the device in **Figure 15(c)**, where the two Sagnac interferometers in the interfering loop were replaced by etched holes to form a 1D-PhC cavity [202], yielding a reduced footprint and an increased FSR for coarse WDM applications. The fabricated device consisted of two identical interleavers that could separate the reflection light from the input, thus avoiding additional off-chip circulators. Other attractive features included a compact footprint of ~64 × 70 μm$^2$, a low insertion loss of ~0.5 dB, and a large channel spacing of ~19 nm.

In addition to the experimental work mentioned above, there have been theoretical investigations. An optical interleaver based on a MZI structure with cascaded Sagnac interferometers in its two arms has been proposed (**Figure 15(f)**) [186]. The designs for such interleavers with channel spacings of 200 GHz, 50 GHz, and 25 GHz were provided, together with a detail analysis for the influence of the reflectivity and the number of Sagnac interferometers on the insertion loss, channel spacing, and extinction ratio. Another optical interleaver formed by two parallel Sagnac interferometers coupled to a top bus waveguide has also been investigated (**Figure 15(g)**) [70]. The hybrid nature of such device, which includes both TW and SW filter elements as well as FIR and IIR filter elements, enables strong mode interference in a compact device footprint and hence good filtering flatness for wavelength interleaving. **Figure 15(h)** shows an optical interleaver featuring a simple design and high fabrication tolerance [203], which consists two coupled Sagnac interferometers formed by a self-coupled optical waveguide. The high fabrication tolerance is enabled by using a single self-coupled waveguide, so the length fabrication errors in different segments would not induce any asymmetry of the filter shape.

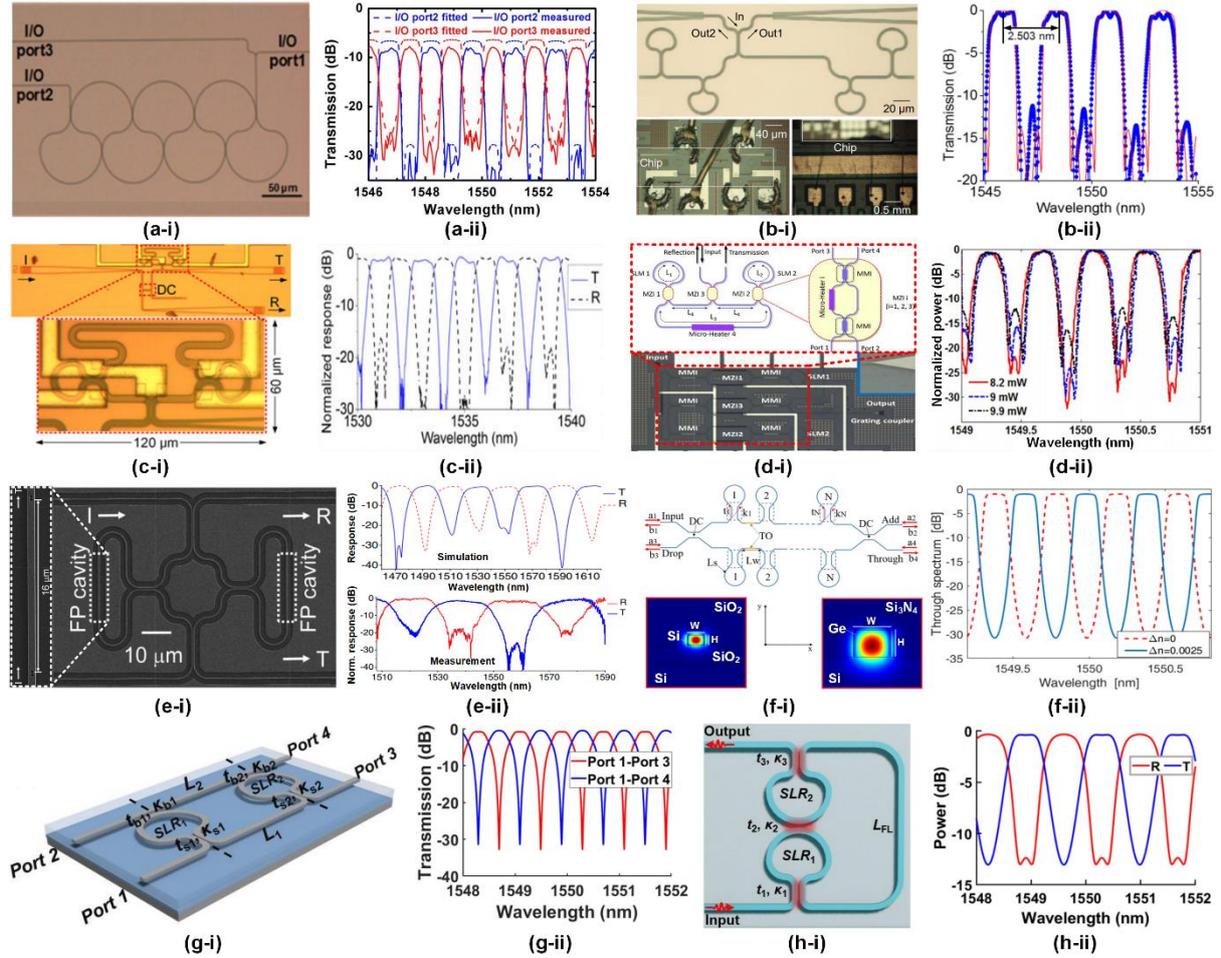

**Figure 15.** Wavelength (de)interleavers formed by intergraded Sagnac interferometers. (a) A passive silicon photonic interleaver based on coupled Sagnac interferometers formed by a self-coupled optical waveguide. (b) A tunable silicon photonic interleaver based on Michelson-Gires-Tournois interferometer formed by cascaded Sagnac interferometers. (c) A tunable silicon photonic interleaver based on a FP cavity formed by two Sagnac interferometers in an interfering loop. (d) A tunable silicon photonic interleaver formed by Sagnac interferometers with tunable MZI couplers. (e) A passive silicon photonic interleaver based on a 1D-PhC FP cavity in an interfering loop. (f) An integrated optical interleaver based on a MZI structure with cascaded Sagnac interferometers in the two arms. (g) An integrated optical interleaver formed by two parallel Sagnac interferometers coupled to a top bus waveguide. (h) An integrated optical interleaver based on two coupled Sagnac interferometers with a feedback loop formed by a self-coupled optical waveguide. In (a) – (h), (i) shows the microscope image and/or schematic of the device, and (ii) shows the spectral response. (a) Adapted with permission [194]. Copyright 2016, Optica Publishing Group. (b) Adapted with permission [201]. Copyright 2016, Optica Publishing Group. (c) Adapted with permission [200]. Copyright 2017, IEEE. (d) Adapted with permission [179]. Copyright 2018, Optica Publishing Group. (e) Adapted with permission [202]. Copyright 2018, Optica Publishing Group. (f) Adapted with permission [186]. Copyright 2018, IEEE. (g) Adapted with permission [70]. Copyright 2021, Optica Publishing Group. (h) Adapted with permission [203]. Copyright 2022, SPIE.

## 5. Optical analogues of quantum physics

Similar to the interaction between quantum states in multi-level atoms, coherent mode interference in coupled resonators can yield optical analogues of many quantum phenomena in atomic or condense matter physics, such as electromagnetically induced transparency (EIT), electromagnetically induced absorption (EIA), Autler–Towns splitting (ATS), and Fano resonances. These optical analogues have been utilized in a variety of applications such as light storage [207-209], sensing [210-212], dispersion engineering [213, 214], photonic computing [97], and signal multicasting [215, 216]. A variety of integrated photonic devices, including those based on Sagnac interference or others, have been used to realize optical analogues of EIT, EIA, ATS, and Fano resonances. Compared to the devices formed by TW resonators such as RRs, the devices formed by Sagnac interferometers show advantages in terms of device footprint due to their SW resonator nature. The strong mode interference within compact resonant cavities can also yield increased Q factors and reduced FSRs [217, 218]. In **Table 5**, we summarize optical analogues of quantum physics generated by integrated photonic devices based on Sagnac interference. In the following, we discuss them in detail.

Optical analogues of ATS in a close-loop resonator formed by two Sagnac interferometers were utilized for selective millimetre-wave (MMW) signal generation (**Figure 16(a)**) [219]. By varying the coupling strength of the central directional coupler, the spectral range between the split resonances was changed, thus enabling the extraction of frequencies with different intervals and hence the generation of MMW signals at different frequencies. In the experimental demonstration, ~39-GHz and ~29-GHz MMW signals were generated by using two passive devices with different coupling strengths of the central directional couplers.

By replacing the central directional coupler of the device in **Figure 16(a)** with a MZI coupler and integrating a micro-heater along one arm to tune the phase shift, a similar device was employed as a tunable photonic analog computer to solve differential-equations (**Figure 16(b)**) [97]. The split resonances arising from the optical analogues of ATS were self-aligned, therefore there were neither unequal thermal wavelength drifts nor the need for accurate wavelength alignment as in the case of cascaded RRs. An experimental demonstration was performed using 10-Gb/s optical Gaussian and super-Gaussian signals as the input, and the results showed good agreement with theory. In Ref. [220], a tunable spectral range between the split resonances from zero to the entire FSR was demonstrated for a device with the same structure.

**Table 5. Comparison of optical analogues of quantum physics in integrated photonic devices based on Sagnac interference. SI: Sagnac interferometer.**

| Device structure | Integrated platform | Corresponding atomic physics | Demonstration of dynamic tuning | Ref. |
|---|---|---|---|---|
| A close-loop resonator formed by 2 SIs connected via a directional coupler | SOI | ATS | No | [219] |
| A close-loop resonator formed by 2 SIs connected via a MZI coupler | SOI | ATS | Yes | [97, 220] |
| 2 coupled SIs formed by a self-coupled optical waveguide | SOI | EIT | Yes | [221] |
| 2 coupled SIs formed by a bottom RR and a top S-bend waveguide | SiN | EIT | Yes | [222] |
| 2 cascaded self-coupled optical waveguides including 4 SIs | SOI | EIT | Yes | [223] |
| 4 coupled SIs | SOI | EIT | Yes | [224] |
| 2 cascaded SIs with MZI couplers embedded in a RR | SOI | EIT | Yes | [225] |
| 3, 4, and 8 cascaded SIs | SOI | Multiple energy level splitting | No | [56] |
| An add-drop RR coupled to an FP cavity formed by cascaded SIs | SOI | Fano resonance | Yes | [226] |
| Two coupled FP cavities formed by cascaded SIs | SOI | Fano resonance | No | [227] |
| A zig-zag-like resonator formed by 3 coupled SIs | N/A [a] | Fano resonance | N/A [a] | [187] |
| 3 SIs formed by a self-coupled waveguide | N/A [a] | Fano resonance | N/A [a] | [188] |

[a] This is simulation work.

Optical analogues of EIT in coupled Sagnac interferometers formed by a self-coupled optical waveguide have also been investigated, first via theoretical simulation [228], and followed by an experimental demonstration based on the SOI platform (**Figure 16(c)**) [221]. The Sagnac interference in such resonator allowed for the co-excitation of the CW and CCW resonance modes in the same cavity, which was engineered to realize different filtering functions. Single-channel, dual-channel, and broad stopband spectral responses were realized for the passive devices with different coupling strengths of the directional couplers. Dynamic tuning was also demonstrated by replacing the directional couplers with MZI couplers and integrating p-i-n diodes to electrically tune the phase shift via the free-carrier effect of silicon.

**Figure 16(d**) shows a tunable optical filter formed by a bottom racetrack RR and a top S-bend waveguide that vertically coupled with each other [222]. The coupling between the racetrack RR and the S-bend waveguide induced Sagnac interference in the device. By engineering their coupling strength, optical analogues of EIT were generated by the fabricated devices based on the SiN platform. A tunable resonance wavelength was demonstrated by integrating a micro-heater along the bottom racetrack RR to tune the phase shift. In contrast, the micro-heaters along the top S-bend waveguide had little influence on the spectral response, reflecting the thermal stability of such device.

**Figure 16(e**) shows a tunable silicon photonic filter formed by cascading two self-coupled optical waveguides in **Figure 16(c)** [223]. Optical analogues of EIT and high-order bandstop filtering were observed for the measured spectral responses of the fabricated devices. Dynamic tuning of the spectral response was also demonstrated by applying different electrical powers to either a micro-heater or a p-i-n diode along the connecting waveguide. Another silicon photonic resonator modified on the basis of the device in **Figure 16(c)** is shown in **Figure 16(f)** [224], which consists four coupled Sagnac interferometers. Optical analogues of EIT were generated in the resonator when there was weak coupling for the two outer directional couplers and strong coupling for the two inner directional couplers. Dynamic tuning of the spectral response of the device was demonstrated by integrating a micro-heater to tune the phase shift along the feedback waveguide.

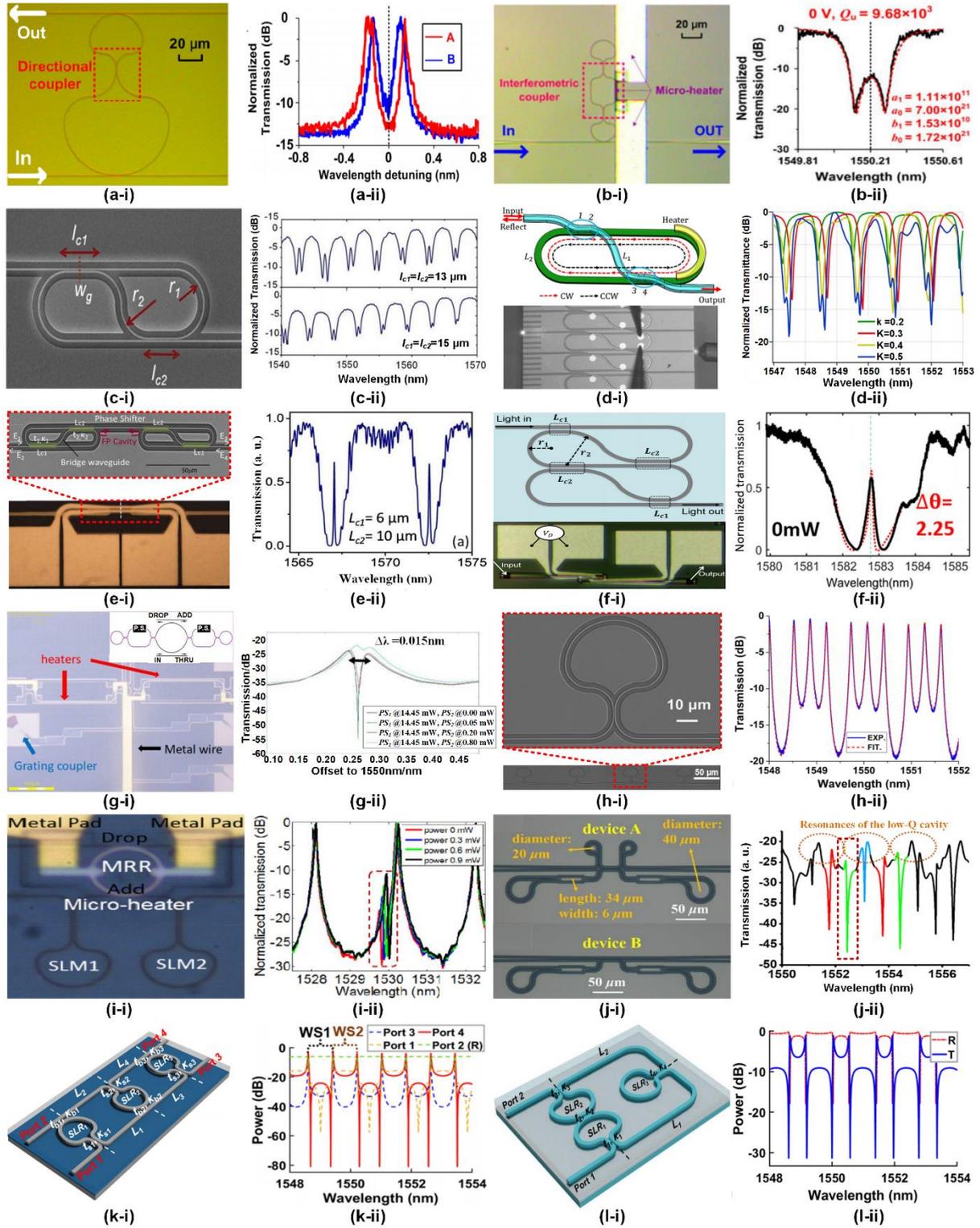

**Figure 16**. Optical analogues of quatum physics in integrated photonic devices based on Sagnac interference. (a) Millimeter-wave signal generation based on optical analogues of ATS in a silicon photonic resonator consisting of two Sagnac interferometers connected via a directional coupler. (b) A tunable differential-equation solver based on engineering optical analogues of ATS in a silicon photonic resonator consisting of two Sagnac interferometers connected via a MZI coupler. (c) Optical analogues of EIT generated by a silicon photonic resonator consisting of coupled Sagnac interferometers formed by a self-coupled optical waveguide. (d) Optical analogues of EIT generated by a SiN photonic resonator formed by a bottom racetrack RR and a top S-bend waveguide that vertically coupled with each other. (e) A tunable optical filter based on a silicon photonic resonator formed by cascading two self-coupled optical waveguides in (c). (f) Optical analogues of EIT generated by a silicon photonic resonator consisting of four coupled Sagnac interferometers formed by a self-coupled optical waveguide. (g) A tunable

optical filter based on engineering optical analogues of EIT in a silicon photonic resonator consisting of two cascaded Sagnac interferometers embedded in a RR. (h) Optical analogues of multiple energy level splitting in a silicon photonic resonator formed by multiple cascaded Sagnac interferometers. (i) Optical analogues of Fano resonances generated by a silicon photonic resonator consisting of an add-drop RR and an FP cavity formed by two cascaded Sagnac interferometers. (j) Optical analogues of Fano resonances generated by a silicon photonic resonator consisting of two FP cavities formed by cascaded Sagnac interferometers. (k) Optical analogues of Fano resonances generated by a zig-zag-like structure formed by three inversely coupled Sagnac interferometers. (l) Optical analogues of Fano resonances in a resonator consisting of three Sagnac interferometers formed by a self-coupled optical waveguide. In (a) – (l), (i) shows the microscope image and/or schematic of the device, and (ii) shows the spectral response. (a) Adapted with permission [219]. Copyright 2016, Elsevier B.V. (b) Adapted with permission [97]. Copyright 2015, IEEE. (c) Adapted with permission [221]. Copyright 2013, IEEE. (d) Adapted with permission [222]. Copyright 2018, Optica Publishing Group. (e) Adapted with permission [223]. Copyright 2013, Optica Publishing Group. (f) Adapted with permission [224]. Copyright 2018, Optica Publishing Group. (g) Adapted with permission [225]. Copyright 2017, Optica Publishing Group. (h) Adapted with permission [56]. Copyright 2018, AIP Publishing LLC. (i) Adapted with permission [226]. Copyright 2017, Optica Publishing Group. (j) Adapted with permission [227]. Copyright 2019, Optica Publishing Group. (k) Adapted with permission [187]. Copyright 2021, Optica Publishing Group. (l) Adapted with permission [188]. Copyright 2021. IEEE.

The generation of optical analogues of EIT based on a resonant cavity consisting of two cascaded Sagnac interferometers embedded in a RR has also been demonstrated (**Figure 16(g)**) [225, 229]. The two Sagnac interferometers formed an FP cavity inside the RR, and the coherent mode interference between them enabled the generation of EIT-like resonances. Tunable extinction ratio and bandwidth of the EIT-like spectrum were demonstrated by tuning the micro-heaters along the MZI couplers of the Sagnac interferometers.

Optical analogues of multiple energy level splitting in resonators formed by multiple cascaded Sagnac interferometers have also been investigated, first via theoretical simulation [230], followed by an experimental demonstration using silicon photonic devices (**Figure 16(h)**) [56]. Coherent mode interference in these devices was tailored by engineering the reflectivity of the Sagnac interferometers and the phase shifts along the connecting waveguides, which enabled the generation of multiple split resonances with potential applications for enhanced light trapping [185], wavelength multicasting [215, 216], and RF spectral shaping [231, 232].

Optical analogues of Fano resonances, which feature an asymmetric resonant lineshape, have formed the basis for many sensors and switches in photonics and plasmonics [228, 233-235] [233, 236, 237]. Here, we present our latest results on integrated photonic devices based on Sagnac interference.

A silicon photonic resonator consisting of an add-drop RR and an FP cavity formed by two cascaded Sagnac interferometers was employed for generating optical analogues of Fano resonances (**Figure 16(i)**) [226]. Fano-like resonances arising from the coherent mode interference between the RR and the FP cavity were generated when they were weakly coupled with each other. The fabricated device achieved a maximum extinction ratio of ~23.2 dB and a maximum slope rate (defined as the ratio of the extinction ratio to the wavelength difference between the Fano-like resonance peak and notch) of ~252 dB/nm. Wavelength tuning via the co-integrated micro-heater along the RR was also demonstrated, achieving an efficiency of ~0.23 nm/mW.

**Figure 16(j)** shows another silicon photonic device that was used for generating optical analogues of Fano resonances [227], where two FP cavities with quite different Q factors formed by cascaded Sagnac interferometers were weakly coupled with each other. The high-Q and low-Q cavities served as discrete-like and continuum-like states, respectively, and the coherent interference between them enabled the generation of Fano-like resonances. A maximum extinction ratio of ~22.3 dB and corresponding slope rate of ~413 dB/nm were achieved for the fabricated device. In Ref. [101], a silicon electro-optic modulator was demonstrated based on a device with a similar structure, where the resonance wavelengths of the optical analogues of Fano resonance were tuned by integrating a PN junction phase shifter for carrier-depletion refractive-index modulation, achieving extinction ratios of ~2.8 dB and ~3.4 dB for 20-Gb/s and 10-Gb/s on-off keying (OOK) signals, respectively.

Optical analogues of Fano resonances generated by a resonator with a zig-zag-like structure formed by three inversely coupled Sagnac interferometers have also been investigated (**Figure 16(k)**) [187]. By engineering coherent mode interference in such devices consisting of both FIR and IIR filter elements, periodical Fano-like resonances can be generated, with a high extinction ratio of ~76.3 dB and a high slope rate of ~998 dB/nm being achieved in theoretical simulations. **Figure 16(l)** shows another resonator structure capable of generating optical analogues of Fano resonances [188], which consists of three Sagnac interferometers formed by a self-coupled optical waveguide. Similar to the device in **Figure 16(h)**, this device has a high tolerance to the length fabrication errors. In theoretical simulations, an extinction ratio of ~30.2 dB and a slope rate of ~748 dB/nm were achieved.

## 6. Other applications

In addition to the applications discussed above, there are other applications for integrated photonic devices based on Sagnac interference, such as Q factor enhancement, FSR broadening, sensing, and quantum optics.

The enhancement of the Q factor of a silicon add-drop RR has been realized by embedding it in an integrated FP cavity formed by two cascaded Sagnac interferometers (**Figure 17(a)**) [238], where the FP cavity reshaped the transmission spectrum of the RR, yielding both an increased Q factor and extinction ratio. Up to 11-times enhancement of the Q factor and 8-dB improvement in the extinction ratio were achieved for the fabricated device. The enhancement of the Q factor of a silicon 1D-PhC cavity has also been realized by connecting an integrated SLRM (**Figure 17(b)**) [239], where the light reflected back from the SLRM was recycled by the 1D-PhC cavity to enable the Q factor enhancement. The theoretically estimated increase in the Q factor compared to the device without the SLRM was up to ~79.5%.

A silicon photonic resonator with an ultra-wide FSR has been realized by embedding a Sagnac interferometer in a RR (**Figure 17(c)**) [240]. The Sagnac interferometer was designed to have strong reflection for all the resonances except for one, which resulted in only a single resonance with a high extinction ratio in a wide spectral range. Dynamic tuning of the resonance wavelength was demonstrated by using a Sagnac interferometer with a MZI coupler and tuning the phase shift along its one arm via a co-integrated micro-heater. A tuning range of over 55 nm was achieved for a low power of ~16 mW.

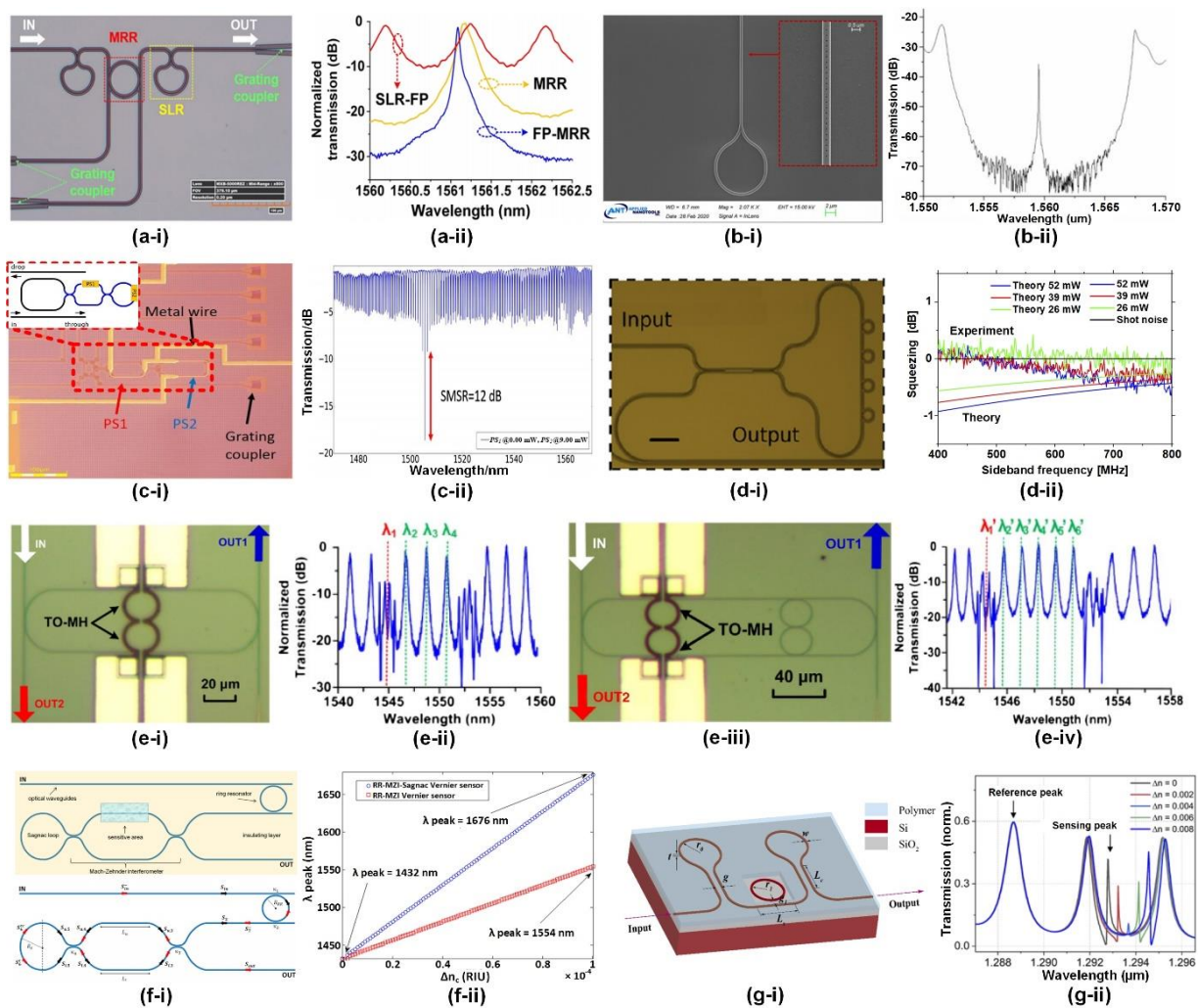

**Figure 17**. Other applications of integrated photonic devices based on Sagnac interference. (a) Q factor enhancement of a silicon add-drop RR via spectral reshaping by an FP cavity formed by two cascaded Sagnac interferometers. (b) Q factor enhancement of a silicon 1D-PhC cavity by connecting an integrated SLRM. (c) A silicon photonic resonator with an ultra-wide FSR enabled by embedding a Sagnac interferometer with a MZI coupler in a RR. (d) Broadband quadrature squeezing generation based on an SiN resonator consisting of four RRs coupled to a Sagnac interferometer. (e) Wavelength selective switches based on Sagnac interference in silicon RRs with nested pairs of subrings. (f) An integrated photonic sensor formed by a RR and a Sagnac interferometer with a MZI coupler. (g) An integrated photonic sensor consisting of a RR coupled to an FP cavity formed by two cascaded Sagnac interferometers. (a) Adapted with permission [238]. Copyright 2017, AIP Publishing LLC. (b) Adapted with permission [239]. Copyright 2021, MDPI (Basel, Switzerland). (c) Adapted with permission [240]. Copyright 2017, Optica Publishing Group. (d) Adapted with permission [241]. Copyright 2020, AIP Publishing LLC. (e) Adapted with permission [242]. Copyright 2015, Optica Publishing Group. (f) Adapted with permission [243]. Copyright 2017, Elsevier B.V. (g) Adapted with permission [244]. Copyright 2021, Optica Publishing Group.

Broadband quadrature squeezing has been realized based on an SiN resonator consisting of four RRs coupled to a Sagnac interferometer (**Figure 17(d)**) [241]. Two counter-propagating bright squeezed states were generated and re-interfered in the Sagnac interferometer to create a single quadrature squeezed state, which yielded both reduced spurious noise and optical power. The measured quadrature squeezing level was ~0.45 dB in the telecom band.

On-chip 1 × 2 wavelength selective switches have also been implemented based on Sagnac interference in silicon RRs with nested pairs of subrings (**Figure 17(e)**) [242], which induced coherent interference between the CW and CCW modes in the outer RRs and hence resonance splitting at certain resonance wavelengths. The fabricated devices achieved extinction ratios >16 dB and processing bandwidths >25 GHz. By using thermal micro-heaters to tune the phase shifts along the nested subrings, experimental demonstrations of dynamic channel routing using fabricated devices with one and two pairs of subrings were performed for 10 Gb/s non-return-to-zero signal.

An integrated photonic sensor formed by a RR and a Sagnac interferometer with a MZI coupler has been proposed (**Figure 17(f)**) [243]. The incorporation of the Sagnac interferometer supporting counter-propagating light transmission enabled a 2-fold enhancement in the refractive index sensing performance. The theoretically estimated wavelength sensitivity was more than $2.5 \times 10^3$ µm/RIU (refractive index unit). **Figure 17(g)** shows another type of integrated photonic sensor consisting of a RR coupled to a FP cavity formed by two cascaded Sagnac interferometers [244]. The transmission spectrum of this device was comprised of Fano-like resonances and FP oscillations, which were used as sensing and reference peaks, respectively. The theoretically estimated sensitivity was 220 nm/RIU. The multiplexing capability of this sensor concept was also investigated by introducing multiple RRs with different radii.

## IV. Challenges and perspectives

As evidenced by the substantial body of work presented and referenced here, the past decade has witnessed a rapid growth in research on integrated photonic devices based on Sagnac interference for a wide range of applications. These devices not only have reduced footprint and improved scalability compared to their conventional counterparts implemented by spatial light or optical fiber devices, but also show many new features and capabilities compared to integrated photonic devices based on MZIs, RRs, PhC cavities, and Bragg gratings. Despite the current success, there is still much room for future development. In this section, we discuss the open challenges and exciting opportunities of this field.

As mentioned in **Section II**, due to the existence of dispersion induced by both the waveguide material and structure, the coupling strengths of directional couplers (**Figure 18(a)**) can no longer be regarded as a wavelength-independent constant for devices with broad operation bandwidths such as reflection mirrors and wavelength (de)interleavers. To reduce the wavelength dependence for integrated optical couplers, many novel coupler designs have been proposed, as shown in **Figures 18 (b) – (j)**. By introducing an intermediate phase delay in a MZI coupler consisting of two directional couplers with different coupling strengths (**Figure 18(b)**), the effective coupling strength no longer monotonically increases with wavelength as in the case for directional couplers, which has been used for reducing the wavelength dependence of MZI couplers [245]. Similarly, curved directional couplers (**Figure 18(c)**) can also achieve wavelength-flattened coupling strengths by introducing a phase mismatch between the modes in the two bent waveguides [246]. On the basis of the curved directional couplers, combined straight and curved directional couplers (**Figure 18(d)**) have been proposed [247], where the straight coupled waveguide sections provide an additional degree of freedom to engineer the transmission characteristic. Asymmetric-waveguide-assisted directional couplers (**Figure 18e**) can mitigate the wavelength dependence by using asymmetric waveguides to generate a slight phase shift between the two symmetric couplers [248]. By employing tapered waveguides to adiabatically convert the mode of a single waveguide into either even or odd mode of two coupled waveguides, adiabatic couplers (**Figure 18(f)**) with no power coupling between different modes also enable wavelength-flattened coupling strengths [249]. MMI couplers (**Figure 18(g)**) that show advantages in achieving compact footprint and high fabrication tolerance can be properly designed to achieve broadband wavelength insensitivity [250-252]. Wavelength-insensitive sub-wavelength grating (SWG) couplers (**Figure 18(h)**) have also been reported [253, 254], where SWGs were embedded in a directional coupler to engineer the dispersion properties of the optical modes within the coupling section. Similarly, MMI-SWG couplers (**Figure 18(i)**) with ultrabroadband operation bandwidth have been realized by using SWGs to engineer the dispersion properties of the MMI section [255, 256]. Broadband SWG-adiabatic couplers (**Figure 18(j)**) that combines the advantages of SWGs and adiabatic couplers have also been demonstrated [257, 258], where tapered SWG waveguides were used to achieve adiabatic mode evolution in a more compact volume than conventional adiabatic couplers.

In directional couplers, apart from the coupling in the central straight regions as expressed by **Eq. (8)** in **Section 2**, the coupling between the input/output bending waveguides also affects the coupling strengths of practical devices. Therefore, the coupling contributions of these bending waveguides should be considered in a more accurate modelling, where **Eq. (8)** can be modified as [96]:

$$\kappa = \sin\left(\frac{\pi}{2} \cdot \frac{(L_c + L_b)}{L_x}\right) \tag{9}$$

where $L_b$ is the effective additional coupling length introduced by the bending waveguides. To minimize the difference induced by the coupling between bending waveguides, small waveguide bending radii that do not induce significant bending loss are preferable. Increasing the gap width could be another option, although this could also result in a longer straight coupling region to achieve comparable coupling strength. An approximate value of $L_b$ in **Eq. (9)** can be obtained via 3D-FDTD simulations. For fabricated devices, more accurate values of $L_b$ can be derived from the measured power split ratios, which, in turn, can be used as empirical values for the design of similar devices.

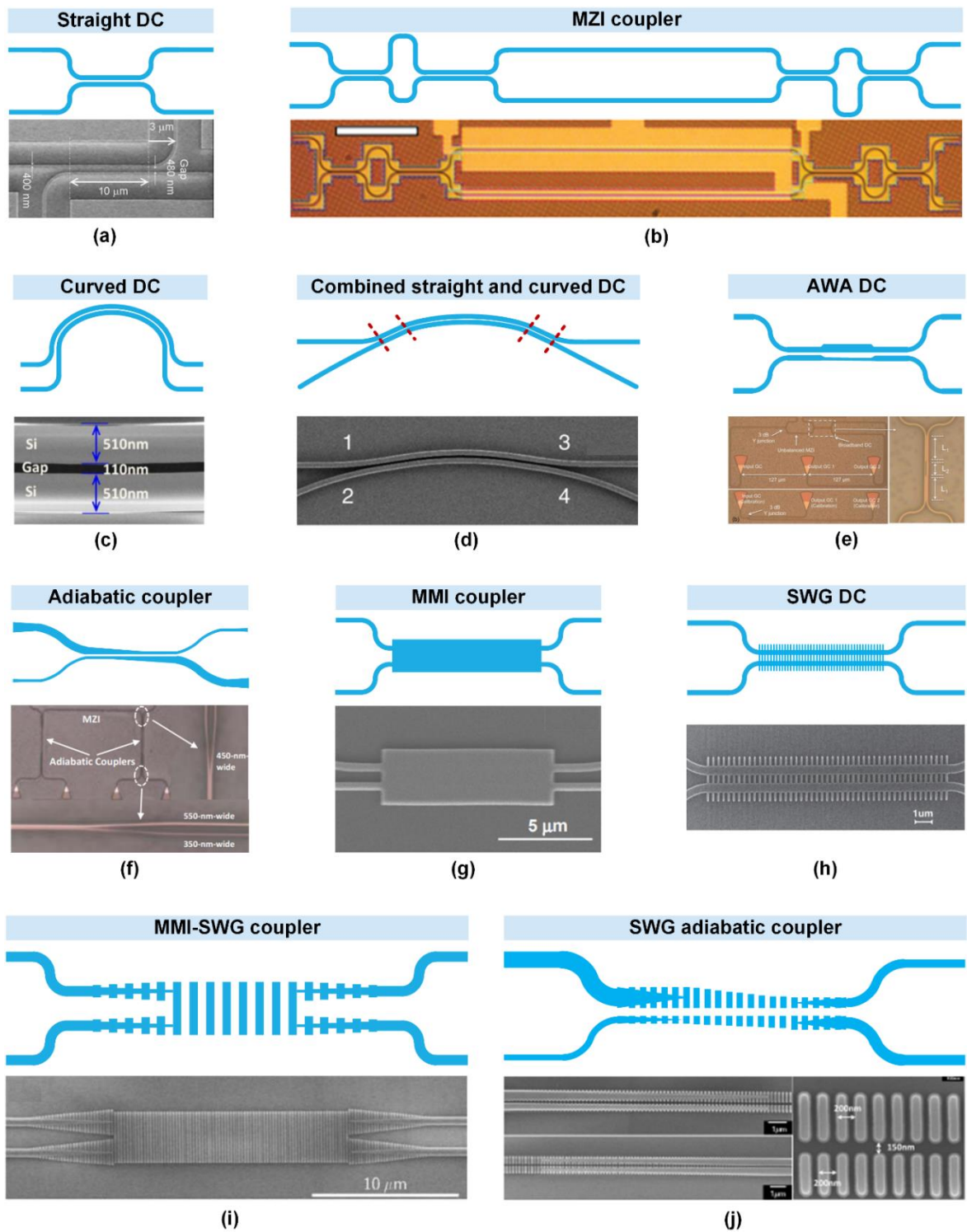

**Figure 18**. Integrated optical couplers with different structures. (a) Straight directional coupler (DC). (b) MZI coupler. (c) Curved DC. (d) Combined straight and curved DC. (e) Asymmetric-waveguide-assisted (AWA) DC. (f) Adiabatic coupler. (g) MMI coupler. (h) Sub-wavelength grating (SWG) DC. (i) MMI-SWG coupler. (j) SWG adiabatic coupler. In (a) − (f), the figures above show the device schematics, and the figures below show the fabricated devices. (a) Adapted with permission [259]. Copyright 2006, Optica Publishing Group. (b) Adapted with permission [245]. Copyright 2009, Optica Publishing Group. (c) Adapted with permission [246]. Copyright 2016, Optica Publishing Group. (d) Adapted with permission [247]. Copyright 2017, Springer Nature LimId. (e) Adapted with permission [248]. Copyright 2015, Optica Publishing Group. (f) Adapted with permission [249]. Copyright 2015, Optica Publishing Group. (g) Adapted with permission [251]. Copyright 2007, Optica Publishing Group. (h) Adapted with permission [254]. Copyright 2016, IEEE. (i) Adapted with permission [255]. Copyright 2016, John Wiley & Sons, Inc. (j) Adapted with permission [257]. Copyright 2016, Optica Publishing Group.

In tandem with the development of PICs, many tunable integrated optical couplers have been demonstrated by using thermo-optic [97, 260] or electro-optic [261, 262] effects to tune the refractive index and hence the waveguide phase shift. Except for the widely employed tunable MZI couplers [97, 179, 225], compact tunable directional couplers have been demonstrated by integrating thermo-optic micro-heaters above the coupling region [180, 260], where phase velocity mismatch between the coupled modes of the waveguides induced by thermal gradient allows for dynamic tuning of the coupling strength. Although integrated thermo-optic and electro-optic phase shifters have been widely used for state-of-the-art PICs, they suffer from limitations with respect to relatively small refractive index changes on the order of $10^{-3}$ or $10^{-4}$, which result in long tunable regions as well as high power consumptions. In addition, their volatile nature necessitates a continuous power supply to maintain their working states. Recently, phase-change materials have shown great potential to implement high-performance tunable directional couplers due to the strong nonvolatile modulation of their refractive indices upon the phase transition between amorphous and crystalline states over broad bands [263, 264].

Similar to integrated photonic devices based on SWGs and PhC cavities, there is bidirectional light propagation in integrated Sagnac interference devices. The backward light transmission at the input ports could induce damage to the laser sources, which needs to be properly managed for practical systems. For laser output injected into integrated photonic devices via fiber-to-chip coupling, commercial fiber-optic isolators can be employed for managing the back reflected light coupled into the input optical fiber. Whereas for light input from integrated laser sources, integrated optical isolators that enable nonreciprocal light transmission are needed. According to the Lorentz reciprocity theorem, nonreciprocal light transmission cannot be achieved in linear, nonmagnetic, and time-invariant systems [265], as the case for most linear integrated photonic devices. In the past decade, there has been a rapid surge in a variety of nonreciprocal optical devices in chip scale, either by employing magneto-optic materials [266-269] or introducing different asymmetric nonlinear effects such as thermo-optic nonlinearity [265, 270, 271], SBS [272-274], optomechanically induced transparency [275-277], and nonreciprocal Kerr effect [278]. These devices have achieved notable performance, although still face challenges in terms of large-scale on-chip integration for commercial products as well as simultaneously achieving efficient, fast, and stable time modulation, hinting more exciting new breakthroughs in the future. It should also be noted that the bidirectional light transmission in integrated Sagnac interference devices could lead to undesired signals at unused output ports, and the light reflected from these ports could induce distortions on the transmission spectra. Therefore, these ports should be properly designed (e.g., terminated with MMI structures or grating couplers [81, 219, 279]) to dissipate the undesired signals.

In practical applications, reducing the thermal drift induced by temperature variation is widely required for many integrated photonic devices, including those based on Sagnac interference. This is particularly true for silicon photonic devices given the large thermo-optic coefficient (TOC) of silicon (~$1.86 \times 10^{-4}$ K$^{-1}$ [280]). To address such limitation, several approaches have been proposed to reduce the temperature sensitivity of integrated photonic devices. These can be classified into four main categories, each of which has pros and cons, and the best option should be well tailored to a particular application scenario. The first one is to achieve active stabilization of the device temperature by using local temperature controllers, which normally comes at the expense of complex feedback systems, increased power consumption, and added cost. The second exploits other integrated platforms that have small TOCs, such as SiN (with a TOC of ~$2.5 \times 10^{-5}$ K$^{-1}$ [281]), silicon oxynitride (with a TOC of ~$1.8 \times 10^{-5}$ K$^{-1}$ [282]), silicon carbide (with a TOC of ~$2.8 \times 10^{-5}$ K$^{-1}$ [283]), and high-index doped silica glass (with a TOC close to that of silica, i.e., ~$1.1 \times 10^{-5}$ K$^{-1}$ [61, 65]). The third introduces cladding materials (e.g., polymers [284, 285] and titanium oxide [286, 287]) that have negative TOCs to compensate the positive TOC. This can be applied to both FIR and IIR filers, but usually requires accurate control of the cladding thickness and waveguide geometry. The last one implements devices having waveguide sections with different TOCs [288, 289], which does not require active control but only works for the FIR filters (e.g., MZIs and AWGs).

As presented in **Section III**, SLRMs with a high flexibility in tuning their reflectivity as well as a high fabrication tolerance have already been employed as functional building blocks in many integrated photonic systems. To implement SLRMs with broad operation bandwidths, the optical couplers in the SLRMs need to be specially designed to reduce the wavelength dependence, as those mentioned in **Figure 18**. In addition, to reduce the footprint of SLRMs for compact integration, MMI couplers can be employed to replace the directional couplers. Since in theory the reflectivities of SLRMs are only affected by the coupling regions, the circumferences of the Sagnac loops can also be reduced unless there is significant bending loss induced by the small bending radii.

Integrated optical gyroscopes show significantly reduced device footprint and power consumption compared to conventional well-established bulk ring laser gyroscopes [43] and fiber optic gyroscopes [34, 113], for which their size, cost, complexity of assembling, and operability in harsh environments limit their applicability despite their excellent performance in terms of precision and stability. Moreover, to implement gyroscopes in integrated form also yields high scalability for implementation of sophisticated gyroscope arrays that can perform more complicated functions. The continuous improvement in technologies of micro/nano device fabrication as well as advances in accurately measuring Sagnac interference in small volumes is beginning to open the door for manufacturable integrated optical gyroscopes with high performance. On the other hand, state-of-the-art integrated optical gyroscopes still face several limitations that hinder their practical deployment for wide applications. First, implementing integrated optical gyroscopes with high sensitivity and precision poses a challenge for device fabrication, where fabrication errors as well as mismatch between different components could induce significant performance degradation by introducing extra loss and noise. Second, there are also demanding requirements for accurately measuring the very weak and slowly varying response in integrated optical gyroscopes, for which highly efficient and stable light coupling between the integrated components and the other functional modules are critically needed. Third, although many schemes of integrated optical gyroscopes have been proposed, the lack of simplified and universal schemes hampers the development of relevant commercial products. Finally, current work on integrated optical gyroscopes only demonstrates some integrated submodules, there are still challenges to achieve monotonically integrated optical gyroscope systems. Although just using integrated coiled waveguides or resonators to replace their bulky counterparts already yields significant benefits in terms of size, cost, and complexity, there is much more to be gained by increasing the level of integration for the overall system. In principle, all the components can be integrated on the same chip. For example, the optical components such as lasers [290, 291], electro-optic modulators [292, 293] and photodetectors [294-296] have already been heterogeneously integrated on silicon chips. The electrical components in the read-out modules such as amplifiers, adders, and mixers also have their integrated forms [142, 144, 297] that can potentially be co-integrated. All of these pave the way for implementing the entire gyroscope system on a single chip in the future.

In addition to the common issues mentioned above for all types of integrated optical gyroscopes, there are still specific issues to be addressed for integrated IOGs, PROGs, and BRLGs. For integrated IOGs, long coiled waveguides have already been demonstrated based on silicon, SiN, and silica platforms. Compared to silicon coiled waveguides, SiN and silica coiled waveguides have much lower propagation loss that is desirable for achieving high sensitivities, while the SiN and silica platforms suffer from limitations with respect to co-integration with other components such as lasers, modulators, and photodetectors. To address this, heterogeneously integrated SiN coiled waveguides and silicon devices [298] could be a possible solution, where specially designed couplers that enable efficient and stable light coupling between the SiN and silicon modules [299] are needed for minimizing the insertion loss and increasing the sensitivity. For integrated PROGs and BRLGs based on high-Q microresonators, apart from the current demonstrations using microresonators made from silica, SiN, CaF$_2$, InP, and polymer, many other material platforms can be exploited. This is particularly true given the fact that a range of material platforms have been developed for fabricating high-Q microresonators used for generating optical microcombs [300-302], which mainly include doped silica [303, 304], magnesium fluorides (MgF$_2$) [305, 306], aluminum nitride (AlN) [307, 308], diamond [309], lithium niobate (LiNbO$_3$) [292, 310, 311], aluminum gallium arsenide (AlGaAs) [312, 313], silicon carbide (SiC) [314], tantalum pentoxide (Ta$_2$O$_5$) [315], and gallium phosphide (GaP) [316]. For WGM cavities implemented based on bulk optics, recently there have been exciting advances in developing fabrication methods for their on-chip integration [121, 122]. For waveguide-based microresonators, spiral ring resonators [317, 318] can be employed to increase the lengths of interference paths while maintaining the device footprint. To further improve the sensitivity of integrated PROGs and BRLGs, the shot noise can be reduced by increasing the Q factors of microresonators via modified device structure and fabrication [238, 319, 320]. Other noise sources such as polarization fluctuations, backscattering, and Kerr effect should also be well suppressed by optimizing the hardware implementation of practical systems. Some detailed methods have been proposed in Refs. [132, 138, 153, 321-329].

For classical filters based on integrated Sagnac interference devices, there have already been investigations of a range of basic network synthesis filters [330] such as Butterworth, Bessel, Chebyshev, and elliptic filters. Compared to integrated photonic resonators formed by RRs, these devices realized by engineering the coherent mode interference in resonators with bidirectional light propagation show advantages with respect to device footprint and diversity of mode interference. To achieve the desired filter shapes, both the coupling strengths of the optical couplers and the lengths of the connecting waveguides in the integrated Sagnac interference devices need to be accurately designed and controlled. On the other hand, as the field grows, more types of classical filters are expected to be investigated. These include not only other network synthesis filters such as Linkwitz–Riley filters [331, 332], Legendre–Papoulis filters [333, 334], and Gaussian filters [335, 336], but also image impedance filters such as Zobel network filters, lattice (all-pass) filters, and general image filters [337, 338].

Similar to the classical filters, wavelength (de)interleavers based on integrated Sagnac interference devices can also reap the great dividend of strong mode interference within small volumes enabled by the bidirectional light propagation and self-aligned resonances. For wavelength (de)interleavers, large extinction ratios, high filtering roll-off, low insertion loss, and broad operation bandwidths are highly desirable in practical applications. State-of-the-art wavelength (de)interleavers based on integrated Sagnac interference devices have already achieved notable performance in each of these parameters [70, 179, 186, 194, 200-203], and yet still face challenges in balancing the trade-offs among them. In theory, the extinction ratios and filtering roll-off can be improved by cascading more subunits. Whereas for practical devices, this would also impose more stringent requirements for device fabrication and normally occurs in accompany with deteriorated filter shapes and operation bandwidths. To overcome these limitations, active tuning mechanisms can be introduced to compensate for the fabrication errors of the passive devices. Another attractive solution is to implement the wavelength (de)interleavers based on single self-coupled optical waveguides [188, 194], where the random fabrication errors in the lengths of different parts do not cause any asymmetry in the filter shape, thus yielding a high fabrication tolerance.

In the past decade, integrated Sagnac interference devices have been engineered to realize a range of optical analogues of quantum phenomena in atomic or condense matter physics, such as EIT, EIA, ATS, and Fano resonances [97, 187, 188, 219-227]. These optical analogues, although originating from different underlying physics, are related to each other and sometimes can transit from one to another. For example, although both the EIT and the ATS are featured by a transparency window in their transmission spectra, the former results from Fano interferences enabled by the coupling of a discrete transition to a continuum [339, 340], whereas the latter is not related to interference effects and stems from the splitting of energy levels driven by strong fields [341, 342]. By changing the coupling strength between two coupled resonant cavities, successful transition between the EIT and the ATS has been demonstrated [343]. Understanding the differences and connections between these optical analogues is critical for the design and implementation of devices. Detailed discussions on these have been given in Refs. [236, 237, 339, 343]. In recent years, an interesting phenomenon called bound states in the continuum (BICs) has attracted significant attention by providing a promising solution for engineering light trapping and high-Q resonances in photonic resonators [344, 345]. The transitions from Fano resonances to BICs can be achieved by changing the structural parameters or the excitation conditions of Fano resonance devices. This offers new opportunities for realizing BICs based on integrated Sagnac interference devices. In addition, Fano resonances in integrated photonic devices have been engineered to achieve nonreciprocal light transmission [346-348], indicating the possible use of integrated Sagnac interference devices for implementing optical isolators. Apart from the optical analogues mentioned above, optical analogues of many other coherent quantum effects such as Rabi splitting [349-351] and parity–time symmetry [352-354] remain to be explored. Although the majority of the current work on the optical analogues of quantum physics remains to be proof-of-concept demonstrations, it is expected that their practical applications in sensing, optical switching, nonreciprocal transmission, and data storage will increase along with this fast growing field.

Engineering Sagnac interference in integrated photonic devices to achieve new functionalities holds promise for many potential applications. For example, by introducing SLRMs at the output ports of nonlinear integrated photonic devices, the reflected light will pass through the devices twice, yielding doubled interaction lengths for enhanced nonlinear performance without sacrificing device footprint. This can also be used for enhancing light-matter interaction in sensors as well as hybrid integrated photonic devices incorporating polymers [355-359], liquid crystals [360-362], or 2D materials [363-369]. Sagnac interference could also be introduced to optical microcomb devices [301, 370, 371] to facilitate mode locking and soliton control. Recently, by engineering backscattered light from the microresonator to the pump laser cavity, turnkey soliton microcomb generation without the need of complex startup protocols and feedback control circuitry has been realized [372]. The Q factor enhancement of resonant cavities achieved by exploiting Sagnac interference [56, 187, 238, 239] can be employed for implementing low-linewidth lasers, high-sensitivity sensors, and high-efficiency nonlinear optical devices. Given the bulky size and complex structure of state-of-the-art microscopy systems based on Sagnac interference [373-375], integrated Sagnac interference devices hint at the implementation of miniatured microscopy systems with reduced SWaP. By connecting multiple basic modules of Sagnac interference devices, more complex filtering arrays or banks can be implemented, which could find possible applications for optical routers [376-378], phase array antennas [379-381], microwave photonic beamformers [382-384], and optical neural networks [385-387]. Apart from spectral filtering, temporal signal processing functions of integrated Sagnac interference devices can also be investigated, which have potential applications for high-speed image processing [388-390] and neuromorphic computing [304, 391, 392]. Although the computing accuracy of state-of-the-art photonic hardware is still not as high as their electronic counterpart, a recent demonstration of fast self-calibrating PICs [393] provides a promising way towards overcoming such drawback. In addition, increasing applications of integrated Sagnac interference devices in quantum optics are expected, which would bring new capabilities and improved performance for quantum optical sources [394-396], sensing [45, 397-399] and nondemolition measurements [400] (e.g., gravitational-wave detection [401, 402]). Although state-of-the-art integrated Sagnac interference devices are mainly implemented based on SOI, SiN, and Group III/IV platforms, their implementation in other integrated device platforms such as $LiNbO_3$ [292, 310, 311], chalcogenide glass [64, 403], AlN [404, 405], and $Ta_2O_5$ [315] is of fundamental importance and could be interesting topics of future work.

Sagnac based integrated interference devices realized in nanophotonic chips will have a significant impact not just on linear photonic filters but potentially on nonlinear photonic chips including microcombs [406-427] and their subsequent applications to RF and microwave photonics. [428-457] The advanced filtering functions achievable with Sagnac devices offer a key and fundamental capability for this wide range of linear and nonlinear photonic devices, as well as potentially for quantum optics [458-470] as well as nonlinear devices based on graphene oxide. [471-488]

The general nature as well as the clear exciting conditions of Sagnac interference make it capable of engineering interference processes across many branches of physics, providing a powerful source of a number of innovative concepts and possible applications. Although many real applications of engineering Sagnac interference in integrated photonic devices are still in their infancy and yet to be implemented, the realization of relevant devices based on integrated platforms is already taking a big step towards compact cost-effective commercial products that can be easily used by the broad community. Along with the continuous improvement in integrated device fabrication technologies as well as the broadening of the application scope, it is anticipated that research on integrated Sagnac interference devices will continue to thrive, in parallel with the development of commercial products that will enable eventually bridge the gap between laboratory-based research and practical industrial applications.

## V. Conclusion

Sagnac interference, as a classical optical interferometry configuration named after its creator – French physicist Georges Sagnac, has laid the foundation for a variety of modern optical systems with wide applications to reflection manipulation, precision measurements, and spectral engineering. In the past decade, the implementation of Sagnac interference components and systems in integrated form has brought new vitality to this field, yielding a vast array of functional devices with superior performance and new features. In this paper, we present our latest results for functional integrated photonic devices based on Sagnac interference. We present the theory of integrated Sagnac interference devices and highlight their comparison with other integrated building blocks such as MZIs, RRs, PhC cavities, and Bragg gratings. We then present our latest results for integrated Sagnac interference devices for a range of applications including reflection mirrors, optical gyroscopes, basic filters, wavelength (de)interleavers, optical analogues of atomic systems. We discuss the current challenges and future perspectives. Along with future growth of this field, ever more researchers and engineers will carry the torch of Georges Sagnac who had a lifelong passion for optics [489]. The synergy between advances in device fabrication technologies and the expanded applications with demanding requirements will be a strong driving force for the performance improvement and wide deployment of integrated Sagnac interference devices as well as the extensive translation of them into commercial products.

**Conflict of interest**

The authors declare no conflicts of interest.

**References**

[1] R. Hanbury Brown, *The intensity interferometer. Its applications to astronomy*. London, Taylor & Francis; New York, Halsted Press, 1974.
[2] J. D. Monnier, "Optical interferometry in astronomy," *Reports on Progress in Physics,* vol. 66, no. 5, p. 789, 2003.
[3] A. Labeyrie, S. G. Lipson, and P. Nisenson, *An Introduction to Optical Stellar Interferometry*. Cambridge: Cambridge University Press, 2010.
[4] M. Islam, M. M. Ali, M.-H. Lai, K.-S. Lim, and H. Ahmad, "Chronology of Fabry-Perot interferometer fiber-optic sensors and their applications: a review," *Sensors,* vol. 14, no. 4, pp. 7451-7488, 2014.
[5] B. Mitra, A. Shelamoff, and D. Booth, "An optical fibre interferometer for remote detection of laser generated ultrasonics," *Measurement Science and Technology,* vol. 9, no. 9, p. 1432, 1998.
[6] P. J. Caber, "Interferometric profiler for rough surfaces," *Appl Opt,* vol. 32, no. 19, pp. 3438-41, Jul 1 1993.
[7] P. De Groot and L. Deck, "Surface profiling by analysis of white-light interferograms in the spatial frequency domain," *Journal of modern optics,* vol. 42, no. 2, pp. 389-401, 1995.
[8] R. P. Dionisio, "Interferometry applications in all-optical communications networks," *Optical Interferometry,* p. 167, 2017.
[9] D. W. Berry and H. M. Wiseman, "Quantum optics on a chip," *Nature Photonics,* vol. 3, no. 6, pp. 317-319, 2009.
[10] M. Chekhova and Z. Ou, "Nonlinear interferometers in quantum optics," *Advances in Optics and Photonics,* vol. 8, no. 1, pp. 104-155, 2016.
[11] Y. Gao, Q. Gan, Z. Xin, X. Cheng, and F. J. Bartoli, "Plasmonic Mach–Zehnder interferometer for ultrasensitive on-chip biosensing," *ACS nano,* vol. 5, no. 12, pp. 9836-9844, 2011.

[12]   A. Santos *et al.*, "Tunable Fabry-Pérot interferometer based on nanoporous anodic alumina for optical biosensing purposes," *Nanoscale research letters,* vol. 7, no. 1, pp. 1-4, 2012.

[13]   W. Bachalo and M. Houser, "Optical interferometry in fluid dynamics research," *Optical Engineering,* vol. 24, no. 3, p. 243455, 1985.

[14]   K. M. van Delft *et al.*, "Micromachined Fabry− Pérot interferometer with embedded nanochannels for nanoscale fluid dynamics," *Nano letters,* vol. 7, no. 2, pp. 345-350, 2007.

[15]   M. Kaschke, K.-H. Donnerhacke, and M. S. Rill, *Optical devices in ophthalmology and optometry: technology, design principles and clinical applications*. Wiley-VCH Verlag GmbH & Co. KGaA, 2014.

[16]   M. Françon, "Optical interferometry," in *Neutron interferometry*, 1979.

[17]   T. Kreis, *Handbook of holographic interferometry: optical and digital methods*. John Wiley & Sons, 2006.

[18]   T. Young, "II. The Bakerian Lecture. On the theory of light and colours," *Philosophical transactions of the Royal Society of London,* no. 92, pp. 12-48, 1802.

[19]   T. Young, "I. The Bakerian Lecture. Experiments and calculations relative to physical optics," *Philosophical transactions of the Royal Society of London,* no. 94, pp. 1-16, 1804.

[20]   M. A. X. Born and E. Wolf, "Chapter VII - elements of the theory of interference and interferometers," in *Principles of Optics (Sixth Edition)*, M. A. X. Born and E. Wolf, Eds.: Pergamon, 1980, pp. 256-369.

[21]   A. Banishev, J. Wang, and M. Bhowmick, *Optical Interferometry*. IntechOpen, 2017.

[22]   H. Lloyd, "On a New Case of Interference of the Rays of Light," *The Transactions of the Royal Irish Academy,* vol. 17, pp. 171-177, 1831.

[23]   P. Hariharan, *Basics of Interferometry* Second ed. Burlington: Academic Press, 2007.

[24]   G. Sagnac, "L'éther lumineux démontré par l'effet du vent relatif d'éther dans un interféromètre en rotation uniforme," *CR Acad. Sci.,* vol. 157, pp. 708-710, 1913.

[25]   E. J. Post, "Sagnac effect," *Reviews of Modern Physics,* vol. 39, no. 2, p. 475, 1967.

[26]   A. H. Rosenthal, "Regenerative circulatory multiple-beam interferometry for the study of light-propagation effects," *JOSA,* vol. 52, no. 10, pp. 1143-1148, 1962.

[27]   W. M. Macek and D. Davis Jr, "Rotation Rate Sensing with Traveling-Wave Ring Lasers," *Applied Physics Letters,* vol. 2, no. 3, pp. 67-68, 1963.

[28]   G. Stedman, K. Schreiber, and H. Bilger, "On the detectability of the Lense–Thirring field from rotating laboratory masses using ring laser gyroscope interferometers," *Classical and Quantum Gravity,* vol. 20, no. 13, p. 2527, 2003.

[29]   G. Stedman, R. Hurst, and K. Schreiber, "On the potential of large ring lasers," *Optics Communications,* vol. 279, no. 1, pp. 124-129, 2007.

[30]   K. U. Schreiber, T. Klügel, J.-P. Wells, R. Hurst, and A. Gebauer, "How to detect the chandler and the annual wobble of the earth with a large ring laser gyroscope," *Physical Review Letters,* vol. 107, no. 17, p. 173904, 2011.

[31]   F. Bosi *et al.*, "Measuring gravitomagnetic effects by a multi-ring-laser gyroscope," *Physical Review D,* vol. 84, no. 12, p. 122002, 2011.

[32]   E. Udd, *Applications Of The Fiber Optic Sagnac Interferometer* (O-E/Fiber LASE '88). SPIE, 1989.

[33]   B. Culshaw, "The optical fibre Sagnac interferometer: an overview of its principles and applications," *Measurement Science and Technology,* vol. 17, no. 1, p. R1, 2005.

[34]   H. C. Lefevre, *The Fiber-Optic Gyroscope, Second Edition*. Artech House Publishers, 2014.

[35]   R. B. Brown, "NRL Memorandum Report 1871," Naval Research Laboratory, Washington, D.C1968.

[36]   T. A. Birks and P. Morkel, "Jones calculus analysis of single-mode fiber Sagnac reflector," *Appl Opt,* vol. 27, no. 15, pp. 3107-13, Aug 1 1988.

[37]   D. B. Mortimore, "Fiber loop reflectors," *Journal of lightwave technology,* vol. 6, no. 7, pp. 1217-1224, 1988.

[38]   L. D. Miller *et al.*, "A Nd(3+)-doped cw fiber laser using all-fiber reflectors," *Appl Opt,* vol. 26, no. 11, pp. 2197-201, Jun 1 1987.

[39]   R. Dyott, V. Handerek, and J. Bello, "Polarization holding directional couplers using D fiber," in *Fiber Optic Couplers, Connectors, and Splice Technology*, 1984, vol. 479, pp. 23-29: International Society for Optics and Photonics.

[40]   K. C. Kao and G. A. Hockham, "Dielectric-fibre surface waveguides for optical frequencies," in *Proceedings of the Institution of Electrical Engineers*, 1966, vol. 113, no. 7, pp. 1151-1158: IET.

[41]   V. Vali and R. W. Shorthill, "Fiber ring interferometer," *Appl Opt,* vol. 15, no. 5, pp. 1099-100, May 1 1976.

[42]   S. a. K. Ezekiel, G.E., *Laser Inertial Rotation Sensors*. Bellingham, Wash. : Society of Photo-optical Instrumentation Engineers, 1978.

[43]   W. Chow, J. Gea-Banacloche, L. Pedrotti, V. Sanders, W. Schleich, and M. Scully, "The ring laser gyro," *Reviews of Modern Physics,* vol. 57, no. 1, p. 61, 1985.

[44]   H. C. Lefevre, *The fiber-optic gyroscope*. Artech house, 2014.

[45] E. Moan *et al.*, "Quantum rotation sensing with dual Sagnac interferometers in an atom-optical waveguide," *Physical review letters,* vol. 124, no. 12, p. 120403, 2020.

[46] J. Du and C. Shu, "Cascaded and multisection Sagnac interferometers for scalable and tunable all-optical OFDM DEMUX," *Journal of lightwave technology,* vol. 31, no. 14, pp. 2307-2313, 2013.

[47] E. A. Kuzin, B. Ibarra Escamilla, D. E. Garcia-Gomez, and J. W. Haus, "Fiber laser mode locked by a Sagnac interferometer with nonlinear polarization rotation," *Opt Lett,* vol. 26, no. 20, pp. 1559-61, Oct 15 2001.

[48] B. Ibarra-Escamilla, E. Kuzin, D. Gomez-Garcia, F. Gutierrez-Zainos, S. Mendoza-Vazquez, and J. Haus, "A mode-locked fibre laser using a Sagnac interferometer and nonlinear polarization rotation," *Journal of Optics A: Pure and Applied Optics,* vol. 5, no. 5, p. S225, 2003.

[49] A. Starodumov, L. Zenteno, D. Monzon, and E. De La Rosa, "Fiber Sagnac interferometer temperature sensor," *Applied Physics Letters,* vol. 70, no. 1, pp. 19-21, 1997.

[50] J. Ma, Y. Yu, and W. Jin, "Demodulation of diaphragm based acoustic sensor using Sagnac interferometer with stable phase bias," *Optics express,* vol. 23, no. 22, pp. 29268-29278, 2015.

[51] P. Bouyer, "The centenary of Sagnac effect and its applications: From electromagnetic to matter waves," *Gyroscopy and Navigation,* vol. 5, no. 1, pp. 20-26, 2014.

[52] "2010 Pioneer Award," *IEEE Transactions on Aerospace and Electronic Systems,* vol. 47, no. 3, pp. 2289-2299, 2011.

[53] J. Napoli, *20 years of KVH fiber optic gyro technology: the evolution from large, low performance FOGs to compact, precise FOGs and FOG-based inertial systems* (SPIE Commercial + Scientific Sensing and Imaging). SPIE, 2016.

[54] E. Udd and I. Udd Scheel, *Mars or bust! 40 years of fiber optic sensor development* (SPIE Commercial + Scientific Sensing and Imaging). SPIE, 2017.

[55] X. Wu and L. Tong, "Optical microfibers and nanofibers," *Nanophotonics,* vol. 2, no. 5-6, pp. 407-428, 2013.

[56] J. Wu, T. Moein, X. Xu, and D. J. Moss, "Advanced photonic filters based on cascaded Sagnac loop reflector resonators in silicon-on-insulator nanowires," *APL Photonics,* vol. 3, no. 4, p. 046102, 2018.

[57] S. Stopinski, L. Augustin, and R. Piramidowicz, "Single-Frequency Integrated Ring Laser for Application in Optical Gyroscope Systems," *IEEE Photonics Technology Letters,* vol. 30, no. 9, pp. 781-784, 2018.

[58] S. Latkowski *et al.*, "Monolithically integrated widely tunable laser source operating at 2 μm," *Optica,* vol. 3, no. 12, pp. 1412-1417, 2016/12/20 2016.

[59] W. Bogaerts *et al.*, "Silicon microring resonators," *Laser & Photonics Reviews,* vol. 6, no. 1, pp. 47-73, 2012.

[60] S. Feng, T. Lei, H. Chen, H. Cai, X. Luo, and A. W. Poon, "Silicon photonics: from a microresonator perspective," *Laser & photonics reviews,* vol. 6, no. 2, pp. 145-177, 2012.

[61] D. J. Moss, R. Morandotti, A. L. Gaeta, and M. Lipson, "New CMOS-compatible platforms based on silicon nitride and Hydex for nonlinear optics," *Nature Photonics,* vol. 7, no. 8, pp. 597-607, 2013.

[62] W. Bogaerts and L. Chrostowski, "Silicon Photonics Circuit Design: Methods, Tools and Challenges," *Laser & Photonics Reviews,* vol. 12, no. 4, p. 1700237, 2018.

[63] Z. Yan *et al.*, "A monolithic InP/SOI platform for integrated photonics," *Light: Science & Applications,* vol. 10, no. 1, p. 200, 2021/09/26 2021.

[64] B. J. Eggleton, B. Luther-Davies, and K. Richardson, "Chalcogenide photonics," *Nature Photonics,* vol. 5, no. 3, pp. 141-148, 2011.

[65] M. Ferrera *et al.*, "Low-power continuous-wave nonlinear optics in doped silica glass integrated waveguide structures," *Nature Photonics,* vol. 2, no. 12, pp. 737-740, 2008.

[66] A. Yariv, "Universal relations for coupling of optical power between microresonators and dielectric waveguides," *Electronics letters,* vol. 36, no. 4, pp. 321-322, 2000.

[67] T. Ye, Y. Zhou, C. Yan, Y. Li, and Y. Su, "Chirp-free optical modulation using a silicon push-pull coupling microring," *Optics letters,* vol. 34, no. 6, pp. 785-787, 2009.

[68] B. Nakarmi, T. Q. Hoai, Y.-H. Won, and X. Zhang, "Short-pulse controlled optical switch using external cavity based single mode Fabry-Pérot laser diode," *Optics express,* vol. 22, no. 13, pp. 15424-15436, 2014.

[69] B. Nakarmi, H. Chen, Y. H. Won, and S. Pan, "Microwave frequency generation, switching, and controlling using single-mode FP-LDs," *Journal of Lightwave Technology,* vol. 36, no. 19, pp. 4273-4281, 2018.

[70] H. Arianfard, J. Wu, S. Juodkazis, and D. J. Moss, "Advanced Multi-Functional Integrated Photonic Filters Based on Coupled Sagnac Loop Reflectors," *Journal of Lightwave Technology,* vol. 39, no. 5, pp. 1400-1408, 2021/03/01 2021.

[71] M. Ferrera *et al.*, "On-chip CMOS-compatible all-optical integrator," *Nat Commun,* vol. 1, no. 1, p. 29, Jun 15 2010.

[72]   L. He, Ş. K. Özdemir, J. Zhu, W. Kim, and L. Yang, "Detecting single viruses and nanoparticles using whispering gallery microlasers," *Nature nanotechnology,* vol. 6, no. 7, pp. 428-432, 2011.
[73]   F. Vollmer and L. Yang, "Review Label-free detection with high-Q microcavities: a review of biosensing mechanisms for integrated devices," *Nanophotonics,* vol. 1, no. 3-4, pp. 267-291, 2012.
[74]   A. Shitikov, I. Bilenko, N. Kondratiev, V. Lobanov, A. Markosyan, and M. Gorodetsky, "Billion Q-factor in silicon WGM resonators," *Optica,* vol. 5, no. 12, pp. 1525-1528, 2018.
[75]   J. E. Heebner, V. Wong, A. Schweinsberg, R. W. Boyd, and D. J. Jackson, "Optical transmission characteristics of fiber ring resonators," *IEEE journal of quantum electronics,* vol. 40, no. 6, pp. 726-730, 2004.
[76]   F. Xia, L. Sekaric, and Y. Vlasov, "Ultracompact optical buffers on a silicon chip," *Nature Photon,* vol. 1, no. 1, pp. 65-71, 2007.
[77]   R. W. Boyd, D. J. Gauthier, and A. L. Gaeta, "Applications of slow light in telecommunications," *Optics and Photonics News,* vol. 17, no. 4, pp. 18-23, 2006.
[78]   F. Liu, Q. Li, Z. Zhang, M. Qiu, and Y. Su, "Optically tunable delay line in silicon microring resonator based on thermal nonlinear effect," *IEEE Journal of Selected Topics in Quantum Electronics,* vol. 14, no. 3, pp. 706-712, 2008.
[79]   F. Liu *et al.*, "Compact optical temporal differentiator based on silicon microring resonator," *Opt Express,* vol. 16, no. 20, pp. 15880-6, Sep 29 2008.
[80]   T. Yang *et al.*, "All-optical differential equation solver with constant-coefficient tunable based on a single microring resonator," *Scientific reports,* vol. 4, no. 1, pp. 1-6, 2014.
[81]   J. Wu *et al.*, "Compact tunable silicon photonic differential-equation solver for general linear time-invariant systems," *Optics express,* vol. 22, no. 21, pp. 26254-26264, 2014.
[82]   Y. Lu, F. Liu, M. Qiu, and Y. Su, "All-optical format conversions from NRZ to BPSK and QPSK based on nonlinear responses in silicon microring resonators," *Opt Express,* vol. 15, no. 21, pp. 14275-82, Oct 17 2007.
[83]   L. Zhang, J. Y. Yang, Y. Li, M. Song, R. G. Beausoleil, and A. E. Willner, "Monolithic modulator and demodulator of differential quadrature phase-shift keying signals based on silicon microrings," *Opt Lett,* vol. 33, no. 13, pp. 1428-30, Jul 1 2008.
[84]   L. Zhang *et al.*, "Microring-based modulation and demodulation of DPSK signal," *Opt Express,* vol. 15, no. 18, pp. 11564-9, Sep 3 2007.
[85]   J. Foresi *et al.*, "Photonic-bandgap microcavities in optical waveguides," *Nature,* vol. 390, no. 6656, pp. 143-145, 1997.
[86]   P. B. Deotare, M. W. McCutcheon, I. W. Frank, M. Khan, and M. Lončar, "High quality factor photonic crystal nanobeam cavities," *Applied Physics Letters,* vol. 94, no. 12, p. 121106, 2009.
[87]   M. H. Haron, D. D. Berhanuddin, B. Y. Majlis, and A. R. Md Zain, "Double-peak one-dimensional photonic crystal cavity in parallel configuration for temperature self-compensation in sensing," *Appl Opt,* vol. 60, no. 6, pp. 1667-1673, Feb 20 2021.
[88]   P. Prabhathan, V. Murukeshan, Z. Jing, and P. V. Ramana, "Compact SOI nanowire refractive index sensor using phase shifted Bragg grating," *Optics express,* vol. 17, no. 17, pp. 15330-15341, 2009.
[89]   M. Burla, L. R. Cortés, M. Li, X. Wang, L. Chrostowski, and J. Azaña, "Integrated waveguide Bragg gratings for microwave photonics signal processing," *Optics express,* vol. 21, no. 21, pp. 25120-25147, 2013.
[90]   X. Wang *et al.*, "Precise control of the coupling coefficient through destructive interference in silicon waveguide Bragg gratings," *Optics letters,* vol. 39, no. 19, pp. 5519-5522, 2014.
[91]   X. Wang, W. Shi, R. Vafaei, N. A. Jaeger, and L. Chrostowski, "Uniform and sampled Bragg gratings in SOI strip waveguides with sidewall corrugations," *IEEE Photonics Technology Letters,* vol. 23, no. 5, pp. 290-292, 2010.
[92]   *G.wdm.1*, 2012.
[93]   A. Yariv, "Critical coupling and its control in optical waveguide-ring resonator systems," *IEEE Photonics Technology Letters,* vol. 14, no. 4, pp. 483-485, 2002.
[94]   J. Wu *et al.*, "Nested Configuration of Silicon Microring Resonator With Multiple Coupling Regimes," *IEEE Photonics Technology Letters,* vol. 25, no. 6, pp. 580-583, 2013.
[95]   A. Yariv and P. Yeh, *Photonics: Optical Electronics in Modern Communications*. Oxford University Press, 2007.
[96]   L. Chrostowski and M. Hochberg, "Fundamental Building Blocks," in *Silicon Photonics Design: From Devices to Systems* Cambridge: Cambridge University Press, 2015, pp. 92-161.
[97]   J. Wu *et al.*, "On-Chip Tunable Second-Order Differential-Equation Solver Based on a Silicon Photonic Mode-Split Microresonator," *Journal of Lightwave Technology,* vol. 33, no. 17, pp. 3542-3549, 2015.
[98]   T. Zhao *et al.*, "Independently tunable double Fano resonances based on waveguide-coupled cavities," *Optics Letters,* vol. 44, no. 12, pp. 3154-3157, 2019/06/15 2019.

[99]   T. Hu et al., "Tunable Fano resonances based on two-beam interference in microring resonator," *Applied Physics Letters,* vol. 102, no. 1, p. 011112, 2013.

[100]  X. Sun et al., "Tunable silicon Fabry–Perot comb filters formed by Sagnac loop mirrors," *Optics Letters,* vol. 38, no. 4, pp. 567-569, 2013.

[101]  H. Du et al., "A Si Optical Modulator Based on Fano-Like Resonance," *IEEE Photonics Technology Letters,* vol. 33, no. 21, pp. 1209-1212, 2021.

[102]  W. B. Elmer, *The Optical Design of Reflectors*, 2nd ed. John Wiley & Sons, 1980, p. 290.

[103]  A. Cutolo, M. Iodice, A. Irace, P. Spirito, and L. Zeni, "An electrically controlled Bragg reflector integrated in a rib silicon on insulator waveguide," *Applied physics letters,* vol. 71, no. 2, pp. 199-201, 1997.

[104]  P. G. O'Brien, N. P. Kherani, A. Chutinan, G. A. Ozin, S. John, and S. Zukotynski, "Silicon photovoltaics using conducting photonic crystal back-reflectors," *Advanced Materials,* vol. 20, no. 8, pp. 1577-1582, 2008.

[105]  I. Chremmos and N. Uzunoglu, "Reflective properties of double-ring resonator system coupled to a waveguide," *IEEE photonics technology letters,* vol. 17, no. 10, pp. 2110-2112, 2005.

[106]  Q. Fang et al., "Folded Silicon-Photonics Arrayed Waveguide Grating Integrated With Loop-Mirror Reflectors," *IEEE Photonics Journal,* vol. 10, no. 4, pp. 1-8, 2018.

[107]  J. Xie, L. Zhou, Z. Zou, J. Wang, X. Li, and J. Chen, "Continuously tunable reflective-type optical delay lines using microring resonators," *Optics express,* vol. 22, no. 1, pp. 817-823, 2014.

[108]  P. Munoz et al., "Multi-wavelength laser based on an Arrayed Waveguide Grating and Sagnac loop reflectors monolithically integrated on InP," in *Proceedings of the 15th European Conference on Integrated Optics*, 2010.

[109]  M. A. Tran, T. Komljenovic, J. C. Hulme, M. Kennedy, D. J. Blumenthal, and J. E. Bowers, "Integrated optical driver for interferometric optical gyroscopes," *Optics express,* vol. 25, no. 4, pp. 3826-3840, 2017.

[110]  B. Stern, X. Ji, Y. Okawachi, A. L. Gaeta, and M. Lipson, "Battery-operated integrated frequency comb generator," *Nature,* vol. 562, no. 7727, pp. 401-405, 2018.

[111]  V. Passaro, A. Cuccovillo, L. Vaiani, M. De Carlo, and C. E. Campanella, "Gyroscope technology and applications: A review in the industrial perspective," *Sensors,* vol. 17, no. 10, p. 2284, 2017.

[112]  A. Lawrence, *Modern inertial technology: navigation, guidance, and control*. Springer Science & Business Media, 2001.

[113]  H. C. Lefèvre, "The fiber-optic gyroscope, a century after Sagnac's experiment: The ultimate rotation-sensing technology?," *Comptes Rendus Physique,* vol. 15, no. 10, pp. 851-858, 2014/12/01/ 2014.

[114]  D. Urbonas, R. F. Mahrt, and T. Stoferle, "Low-loss optical waveguides made with a high-loss material," *Light Sci Appl,* vol. 10, no. 1, p. 15, Jan 12 2021.

[115]  T. Horikawa, D. Shimura, and T. Mogami, "Low-loss silicon wire waveguides for optical integrated circuits," *MRS Communications,* vol. 6, no. 1, pp. 9-15, 2016.

[116]  H. Lee, T. Chen, J. Li, O. Painter, and K. J. Vahala, "Ultra-low-loss optical delay line on a silicon chip," *Nature Communications,* vol. 3, no. 1, p. 867, 2012/05/29 2012.

[117]  J. F. Bauters et al., "Planar waveguides with less than 0.1 dB/m propagation loss fabricated with wafer bonding," *Optics express,* vol. 19, no. 24, pp. 24090-24101, 2011.

[118]  S. Honari, S. Haque, and T. Lu, "Fabrication of ultra-high Q silica microdisk using chemo-mechanical polishing," *Applied Physics Letters,* vol. 119, no. 3, p. 031107, 2021.

[119]  H. Lee et al., "Chemically etched ultrahigh-Q wedge-resonator on a silicon chip," *Nature Photonics,* vol. 6, no. 6, pp. 369-373, 2012.

[120]  D. T. Spencer, J. F. Bauters, M. J. Heck, and J. E. Bowers, "Integrated waveguide coupled Si 3 N 4 resonators in the ultrahigh-Q regime," *Optica,* vol. 1, no. 3, pp. 153-157, 2014.

[121]  K. Y. Yang et al., "Bridging ultrahigh-Q devices and photonic circuits," *Nature Photonics,* vol. 12, no. 5, pp. 297-302, 2018.

[122]  D. Farnesi et al., "Metamaterial engineered silicon photonic coupler for whispering gallery mode microsphere and disk resonators," *Optica,* vol. 8, no. 12, pp. 1511-1514, 2021.

[123]  M. Sorel et al., "Alternate oscillations in semiconductor ring lasers," *Opt Lett,* vol. 27, no. 22, pp. 1992-4, Nov 15 2002.

[124]  H. Cao et al., "Large S-section-ring-cavity diode lasers: directional switching, electrical diagnostics, and mode beating spectra," *IEEE photonics technology letters,* vol. 17, no. 2, pp. 282-284, 2005.

[125]  O. Kenji, "Semiconductor ring laser gyro," Patent 60,148,185, 1985, 1985.

[126]  F. Dell'Olio, T. Tatoli, C. Ciminelli, and M. N. Armenise, "Recent advances in miniaturized optical gyroscopes," *Journal of the European optical society-Rapid publications,* vol. 9, 2014.

[127]  S. Gundavarapu et al., "Interferometric Optical Gyroscope Based on an Integrated Si3N4 Low-Loss Waveguide Coil," *Journal of Lightwave Technology,* vol. 36, no. 4, pp. 1185-1191, 2018.

[128]  B. Wu, Y. Yu, J. Xiong, and X. Zhang, "Silicon Integrated Interferometric Optical Gyroscope," *Scientific Reports,* vol. 8, no. 1, p. 8766, 2018/06/08 2018.

[129] D. Liu *et al.*, "Interferometric optical gyroscope based on an integrated silica waveguide coil with low loss," *Optics Express,* vol. 28, no. 10, pp. 15718-15730, 2020/05/11 2020.

[130] B. Wu, Y. Yu, and X. Zhang, "Mode-assisted Silicon Integrated Interferometric Optical Gyroscope," *Scientific Reports,* vol. 9, no. 1, p. 12946, 2019/09/10 2019.

[131] P. P. Khial, A. D. White, and A. Hajimiri, "Nanophotonic optical gyroscope with reciprocal sensitivity enhancement," *Nature Photonics,* vol. 12, no. 11, pp. 671-675, 2018.

[132] K. Suzuki, K. Takiguchi, and K. Hotate, "Monolithically Integrated Resonator Microoptic Gyro on Silica Planar Lightwave Circuit," *Journal of Lightwave Technology,* vol. 18, no. 1, p. 66, 2000/01/01 2000.

[133] N. Liang, G. Lijun, K. Mei, and C. Tuoyuan, "Waveguide-type optical passive ring resonator gyro using frequency modulation spectroscopy technique," *Journal of Semiconductors,* vol. 35, p. 124008, December 01, 2014 2014.

[134] J. Wang, L. Feng, Y. Tang, and Y. Zhi, "Resonator integrated optic gyro employing trapezoidal phase modulation technique," *Optics Letters,* vol. 40, no. 2, pp. 155-158, 2015/01/15 2015.

[135] G. Qian *et al.*, "Demonstrations of centimeter-scale polymer resonator for resonant integrated optical gyroscope," *Sensors and Actuators A: Physical,* vol. 237, pp. 29-34, 2016.

[136] C. Ciminelli, F. Dell'Olio, M. N. Armenise, F. M. Soares, and W. Passenberg, "High performance InP ring resonator for new generation monolithically integrated optical gyroscopes," *Optics Express,* vol. 21, no. 1, pp. 556-564, 2013/01/14 2013.

[137] C. Ciminelli *et al.*, "A high-Q InP resonant angular velocity sensor for a monolithically integrated optical gyroscope," *IEEE Photonics Journal,* vol. 8, no. 1, pp. 1-19, 2015.

[138] J. Zhang, H. Ma, H. Li, and Z. Jin, "Single-polarization fiber-pigtailed high-finesse silica waveguide ring resonator for a resonant micro-optic gyroscope," *Optics Letters,* vol. 42, no. 18, pp. 3658-3661, 2017/09/15 2017.

[139] W. Liang *et al.*, "Resonant microphotonic gyroscope," *Optica,* vol. 4, no. 1, p. 114, 2017.

[140] J. M. Silver *et al.*, "Nonlinear enhanced microresonator gyroscope," *Optica,* vol. 8, no. 9, p. 1219, 2021.

[141] J. Li, M.-G. Suh, and K. Vahala, "Microresonator Brillouin gyroscope," *Optica,* vol. 4, no. 3, p. 346, 2017.

[142] S. Gundavarapu *et al.*, "Sub-hertz fundamental linewidth photonic integrated Brillouin laser," *Nature Photonics,* vol. 13, no. 1, pp. 60-67, 2019/01/01 2019.

[143] Y.-H. Lai, Y.-K. Lu, M.-G. Suh, Z. Yuan, and K. Vahala, "Observation of the exceptional-point-enhanced Sagnac effect," *Nature,* vol. 576, no. 7785, pp. 65-69, 2019/12/01 2019.

[144] Y.-H. Lai *et al.*, "Earth rotation measured by a chip-scale ring laser gyroscope," *Nature Photonics,* vol. 14, no. 6, pp. 345-349, 2020.

[145] Y. M. He, F. H. Yang, W. Yan, W. H. Han, and Z. F. Li, "Asymmetry Analysis of the Resonance Curve in Resonant Integrated Optical Gyroscopes," (in eng), *Sensors (Basel, Switzerland),* vol. 19, no. 15, p. 3305, 2019.

[146] M. Lei, L. Feng, and Y. Zhi, "Sensitivity improvement of resonator integrated optic gyroscope by double-electrode phase modulation," *Appl Opt,* vol. 52, no. 30, pp. 7214-9, Oct 20 2013.

[147] H. Ma, J. Zhang, L. Wang, and Z. Jin, "Double closed-loop resonant micro optic gyro using hybrid digital phase modulation," *Optics Express,* vol. 23, no. 12, pp. 15088-15097, 2015/06/15 2015.

[148] Y. Zhi, L. Feng, M. Lei, and K. Wang, "Low-delay, high-bandwidth frequency-locking loop of resonator integrated optic gyro with triangular phase modulation," *Applied Optics,* vol. 52, no. 33, pp. 8024-8031, 2013/11/20 2013.

[149] H. Ma, X. Zhang, Z. Jin, and C. Ding, "Waveguide-type optical passive ring resonator gyro using phase modulation spectroscopy technique," *Optical Engineering,* vol. 45, no. 8, p. 080506, 2006.

[150] H. Ma, Z. He, and K. Hotate, "Reduction of Backscattering Induced Noise by Carrier Suppression in Waveguide-Type Optical Ring Resonator Gyro," *Journal of Lightwave Technology,* vol. 29, no. 1, pp. 85-90, 2011.

[151] Y. Zhi, L. Feng, J. Wang, and Y. Tang, "Reduction of backscattering noise in a resonator integrated optic gyro by double triangular phase modulation," *Applied Optics,* vol. 54, no. 1, pp. 114-122, 2015/01/01 2015.

[152] L. Feng, M. Lei, H. Liu, Y. Zhi, and J. Wang, "Suppression of backreflection noise in a resonator integrated optic gyro by hybrid phase-modulation technology," *Applied Optics,* vol. 52, no. 8, pp. 1668-1675, 2013/03/10 2013.

[153] J. Wang, L. Feng, Q. Wang, H. Jiao, and X. Wang, "Suppression of backreflection error in resonator integrated optic gyro by the phase difference traversal method," *Optics Letters,* vol. 41, no. 7, pp. 1586-1589, 2016/04/01 2016.

[154] L. Feng *et al.*, "Transmissive resonator optic gyro based on silica waveguide ring resonator," *Optics Express,* vol. 22, no. 22, pp. 27565-27575, 2014/11/03 2014.

[155] J. Scheuer and A. Yariv, "Sagnac effect in coupled-resonator slow-light waveguide structures," *Phys Rev Lett,* vol. 96, no. 5, p. 053901, Feb 10 2006.

[156]    C. Sorrentino, J. R. E. Toland, and C. P. Search, "Ultra-sensitive chip scale Sagnac gyroscope based on periodically modulated coupling of a coupled resonator optical waveguide," *Optics Express,* vol. 20, no. 1, pp. 354-363, 2012/01/02 2012.

[157]    Y. Zhang *et al.*, "A high sensitivity optical gyroscope based on slow light in coupled-resonator-induced transparency," *Physics Letters A,* vol. 372, no. 36, pp. 5848-5852, 2008.

[158]    C. Peng, Z. Li, and A. Xu, "Optical gyroscope based on a coupled resonator with the all-optical analogous property of electromagnetically induced transparency," *Opt Express,* vol. 15, no. 7, pp. 3864-75, Apr 2 2007.

[159]    B. Z. Steinberg, J. Scheuer, and A. Boag, "Rotation-induced superstructure in slow-light waveguides with mode-degeneracy: optical gyroscopes with exponential sensitivity," *Journal of the Optical Society of America B,* vol. 24, no. 5, pp. 1216-1224, 2007/05/01 2007.

[160]    B. Z. Steinberg and A. Boag, "Splitting of microcavity degenerate modes in rotating photonic crystals—the miniature optical gyroscopes," *Journal of the Optical Society of America B,* vol. 24, no. 1, pp. 142-151, 2007/01/01 2007.

[161]    A. Shamir and B. Z. Steinberg, "On the Electrodynamics of Rotating Crystals, Micro-Cavities, and Slow-Light Structures: From Asymptotic Theories to Exact Green's Function Based Solutions," in *2007 International Conference on Electromagnetics in Advanced Applications*, 2007, pp. 45-48.

[162]    B. Z. Steinberg, "Rotating photonic crystals: A medium for compact optical gyroscopes," *Physical Review E,* vol. 71, no. 5, p. 056621, 05/31/ 2005.

[163]    T. Zhang *et al.*, "Integrated optical gyroscope using active Long-range surface plasmon-polariton waveguide resonator," *Scientific Reports,* vol. 4, no. 1, p. 3855, 2014/01/24 2014.

[164]    R. I. Woodward, E. J. R. Kelleher, S. V. Popov, and J. R. Taylor, "Stimulated Brillouin scattering of visible light in small-core photonic crystal fibers," *Optics Letters,* vol. 39, no. 8, pp. 2330-2333, 2014/04/15 2014.

[165]    S. P. Smith, F. Zarinetchi, and S. Ezekiel, "Narrow-linewidth stimulated Brillouin fiber laser and applications," *Opt Lett,* vol. 16, no. 6, pp. 393-5, Mar 15 1991.

[166]    K. O. Hill, B. S. Kawasaki, and D. C. Johnson, "cw Brillouin laser," *Applied Physics Letters,* vol. 28, no. 10, pp. 608-609, 1976.

[167]    A. Debut, S. Randoux, and J. Zemmouri, "Linewidth narrowing in Brillouin lasers: Theoretical analysis," *Physical Review A,* vol. 62, no. 2, p. 023803, 2000.

[168]    W. Loh *et al.*, "A microrod-resonator Brillouin laser with 240 Hz absolute linewidth," *New Journal of Physics,* vol. 18, no. 4, p. 045001, 2016.

[169]    R. O. Behunin, N. T. Otterstrom, P. T. Rakich, S. Gundavarapu, and D. J. Blumenthal, "Fundamental noise dynamics in cascaded-order Brillouin lasers," *Physical Review A,* vol. 98, no. 2, p. 023832, 2018.

[170]    F. Zarinetchi, S. P. Smith, and S. Ezekiel, "Stimulated Brillouin fiber-optic laser gyroscope," *Opt Lett,* vol. 16, no. 4, pp. 229-31, Feb 15 1991.

[171]    A. V. Oppenheim, A. S. Willsky, and S. H. Nawab, *Signals and Systems: Pearson New International Edition*. Pearson Education Limited, 2013.

[172]    C. K. Madsen and J. H. Zhao, "Digital Filter Concepts for Optical Filters," in *Optical Filter Design and Analysis*, 1999, pp. 95-164.

[173]    H. Zumbahlen, "Analog Filters," in *Linear Circuit Design Handbook*, H. Zumbahlen, Ed. Burlington: Newnes, 2008, pp. 581-679.

[174]    X. Jiang *et al.*, "Wavelength and bandwidth-tunable silicon comb filter based on Sagnac loop mirrors with Mach-Zehnder interferometer couplers," *Optics express,* vol. 24, no. 3, pp. 2183-2188, 2016.

[175]    D. X. Xu *et al.*, "Archimedean spiral cavity ring resonators in silicon as ultra-compact optical comb filters," *Opt Express,* vol. 18, no. 3, pp. 1937-45, Feb 1 2010.

[176]    P. Dong, S. F. Preble, and M. Lipson, "All-optical compact silicon comb switch," *Opt Express,* vol. 15, no. 15, pp. 9600-5, Jul 23 2007.

[177]    B. G. Lee, A. Biberman, P. Dong, M. Lipson, and K. Bergman, "All-Optical Comb Switch for Multiwavelength Message Routing in Silicon Photonic Networks," *IEEE Photonics Technology Letters,* vol. 20, no. 10, pp. 767-769, 2008.

[178]    S. Zheng *et al.*, "Compact tunable photonic comb filter on a silicon platform," *Optics letters,* vol. 42, no. 14, pp. 2762-2765, 2017.

[179]    N. Zhou, S. Zheng, Y. Long, Z. Ruan, L. Shen, and J. Wang, "Reconfigurable and tunable compact comb filter and (de) interleaver on silicon platform," *Optics express,* vol. 26, no. 4, pp. 4358-4369, 2018.

[180]    R. Ge, Y. Luo, S. Gao, Y. Han, L. Chen, and X. Cai, "Reconfigurable silicon bandpass filters based on cascaded Sagnac loop mirrors," *Optics Letters,* vol. 46, no. 3, pp. 580-583, 2021.

[181]    H. C. Liu and A. Yariv, "Synthesis of high-order bandpass filters based on coupled-resonator optical waveguides (CROWs)," *Opt Express,* vol. 19, no. 18, pp. 17653-68, Aug 29 2011.

[182]    B. E. Little *et al.*, "Very High-Order Microring Resonator Filters for WDM Applications," *IEEE Photonics Technology Letters,* vol. 16, no. 10, pp. 2263-2265, 2004.

[183] F. Xia, M. Rooks, L. Sekaric, and Y. Vlasov, "Ultra-compact high order ring resonator filters using submicron silicon photonic wires for on-chip optical interconnects," *Opt Express,* vol. 15, no. 19, pp. 11934-41, Sep 17 2007.

[184] L. Zhou, T. Ye, and J. Chen, "Waveguide self-coupling based reconfigurable resonance structure for optical filtering and delay," *Opt Express,* vol. 19, no. 9, pp. 8032-44, Apr 25 2011.

[185] J. Song *et al.*, "Loop coupled resonator optical waveguides," *Optics express,* vol. 22, no. 20, pp. 24202-24216, 2014.

[186] R. A. Soref, F. De Leonardis, and V. M. Passaro, "Multiple-Sagnac-Loop Mach–Zehnder Interferometer for Wavelength Interleaving, Thermo-Optical Switching and Matched Filteri," *Journal of Lightwave Technology,* vol. 36, no. 22, pp. 5254-5262, 2018.

[187] H. Arianfard, J. Wu, S. Juodkazis, and D. J. Moss, "Three Waveguide Coupled Sagnac Loop Reflectors for Advanced Spectral Engineering," *Journal of Lightwave Technology,* vol. 39, no. 11, pp. 3478-3487, 2021/06/01 2021.

[188] H. Arianfard, J. Wu, S. Juodkazis, and D. J. Moss, "Spectral Shaping Based on Coupled Sagnac Loop Reflectors Formed by a Self-Coupled Wire Waveguide," *IEEE Photonics Technology Letters,* vol. 33, no. 13, pp. 680-683, 2021.

[189] J. K. S. Poon, J. Scheuer, Y. Xu, and A. Yariv, "Designing coupled-resonator optical waveguide delay lines," *J. Opt. Soc. Am. B,* vol. 21, no. 9, pp. 1665-1673, 2004/09/01 2004.

[190] A. Melloni and M. Martinelli, "Synthesis of direct-coupled-resonators bandpass filters for WDM systems," *Journal of Lightwave Technology,* vol. 20, no. 2, pp. 296-303, 2002.

[191] A. M. Prabhu and V. Van, "Realization of asymmetric optical filters using asynchronous coupled-microring resonators," *Opt Express,* vol. 15, no. 15, pp. 9645-58, Jul 23 2007.

[192] V. Van, "Synthesis of Elliptic Optical Filters Using Mutually Coupled Microring Resonators," *Journal of Lightwave Technology,* vol. 25, no. 2, pp. 584-590, 2007.

[193] S. Cao *et al.*, "Interleaver Technology: Comparisons and Applications Requirements," *Journal of Lightwave Technology,* vol. 22, no. 1, pp. 281-289, 2004.

[194] S. Lai, Z. Xu, B. Liu, and J. Wu, "Compact silicon photonic interleaver based on a self-coupled optical waveguide," *Applied Optics,* vol. 55, no. 27, pp. 7550-7555, 2016/09/20 2016.

[195] M. Oguma *et al.*, "Compact and Low-Loss Interleave Filter Employing Lattice-Form Structure and Silica-Based Waveguide," *Journal of Lightwave Technology,* vol. 22, no. 3, pp. 895-902, 2004.

[196] M. Gad, J. Ackert, D. Yevick, L. Chrostowski, and P. Jessop, "Ring Resonator Wavelength Division Multiplexing Interleaver," *Journal of Lightwave Technology,* vol. 29, no. 14, pp. 2102-2109, 2011.

[197] W. Jiang, Y. Zhang, Y. Guo, F. Zhu, and G. Yi, "Switchable and Flexible Comb Filter/Interleaver Based on Ring-Resonators With Tunable Couplers," *IEEE Photonics Technology Letters,* vol. 33, no. 12, pp. 607-610, 2021.

[198] X. P. Li, K. X. Chen, and L. F. Wang, "Compact and electro-optic tunable interleaver in lithium niobate thin film," *Optics Letters,* vol. 43, no. 15, pp. 3610-3613, 2018/08/01 2018.

[199] T. Mizuno, M. Oguma, T. Kitoh, Y. Inoue, and H. Takahashi, "Lattice-form CWDM interleave filter using silica-based planar lightwave circuit," *IEEE Photonics Technology Letters,* vol. 18, no. 15, pp. 1570-1572, 2006.

[200] X. Jiang *et al.*, "Design and Experimental Demonstration of a Compact Silicon Photonic Interleaver Based on an Interfering Loop With Wide Spectral Range," *Journal of Lightwave Technology,* vol. 35, no. 17, pp. 3765-3771, 2017.

[201] X. Jiang *et al.*, "Compact Silicon Photonic Interleaver Using Loop-Mirror-Based Michelson-Gires-Tournois Interferometer," in *Optical Fiber Communication Conference*, Anaheim, California, 2016, p. Tu2F.5: Optical Society of America.

[202] X. Jiang, H. Zhang, Y. Zhang, C. Qiu, and Y. Su, "Compact CWDM interleaver based on an interfering loop containing a one-dimensional Fabry-Perot cavity," *Optics Letters,* vol. 43, no. 5, pp. 1071-1074, 2018/03/01 2018.

[203] H. Arianfard, J. Wu, S. Juodkazis, and D. Moss, *Spectral shaping based on optical waveguides with advanced Sagnac loop reflectors* (SPIE OPTO). SPIE, 2022.

[204] J. Song, Q. Fang, S. H. Tao, M. B. Yu, G. Q. Lo, and D. L. Kwong, "Passive ring-assisted Mach-Zehnder interleaver on silicon-on-insulator," *Opt Express,* vol. 16, no. 12, pp. 8359-65, Jun 9 2008.

[205] H. Guan, Y. Liu, Z. Li, Y. Kuang, X. Huang, and Z. Li, "Passive silicon ring-assisted Mach–Zehnder interleavers operating in the broadband spectral range," *Applied Optics,* vol. 59, no. 27, pp. 8349-8354, 2020/09/20 2020.

[206] L. W. Luo *et al.*, "High bandwidth on-chip silicon photonic interleaver," *Opt Express,* vol. 18, no. 22, pp. 23079-87, Oct 25 2010.

[207] Y.-C. Liu, B.-B. Li, and Y.-F. Xiao, "Electromagnetically induced transparency in optical microcavities," *Nanophotonics,* vol. 6, no. 5, pp. 789-811, 2017.

[208] M. F. Yanik, W. Suh, Z. Wang, and S. Fan, "Stopping light in a waveguide with an all-optical analog of electromagnetically induced transparency," *Phys Rev Lett,* vol. 93, no. 23, p. 233903, Dec 3 2004.
[209] J. B. Khurgin, "Optical buffers based on slow light in electromagnetically induced transparent media and coupled resonator structures: comparative analysis," *Journal of the Optical Society of America B,* vol. 22, no. 5, pp. 1062-1074, 2005/05/01 2005.
[210] S. Zhang, D. A. Genov, Y. Wang, M. Liu, and X. Zhang, "Plasmon-induced transparency in metamaterials," *Phys Rev Lett,* vol. 101, no. 4, p. 047401, Jul 25 2008.
[211] N. Liu *et al.*, "Planar metamaterial analogue of electromagnetically induced transparency for plasmonic sensing," *Nano Lett,* vol. 10, no. 4, pp. 1103-7, Apr 14 2010.
[212] J. Zhu *et al.*, "On-chip single nanoparticle detection and sizing by mode splitting in an ultrahigh-Q microresonator," *Nature Photonics,* vol. 4, no. 1, pp. 46-49, 2010/01/01 2009.
[213] S. Kim *et al.*, "Dispersion engineering and frequency comb generation in thin silicon nitride concentric microresonators," *Nature Communications,* vol. 8, no. 1, p. 372, 2017/08/29 2017.
[214] X. Xue *et al.*, "Normal-dispersion microcombs enabled by controllable mode interactions," *Laser & Photonics Reviews,* vol. 9, no. 4, pp. L23-L28, 2015.
[215] Q. Li, Z. Zhang, F. Liu, M. Qiu, and Y. Su, "Dense wavelength conversion and multicasting in a resonance-split silicon microring," *Applied Physics Letters,* vol. 93, no. 8, p. 081113, 2008.
[216] M. C. M. M. Souza, L. A. M. Barea, F. Vallini, G. F. M. Rezende, G. S. Wiederhecker, and N. C. Frateschi, "Embedded coupled microrings with high-finesse and close-spaced resonances for optical signal processing," *Optics Express,* vol. 22, no. 9, pp. 10430-10438, 2014/05/05 2014.
[217] L. A. M. Barea, F. Vallini, P. F. Jarschel, and N. C. Frateschi, "Silicon technology compatible photonic molecules for compact optical signal processing," *Applied Physics Letters,* vol. 103, no. 20, p. 201102, 2013.
[218] L. A. M. Barea, F. Vallini, G. F. M. d. Rezende, and N. C. Frateschi, "Spectral Engineering With CMOS Compatible SOI Photonic Molecules," *IEEE Photonics Journal,* vol. 5, no. 6, pp. 2202717-2202717, 2013.
[219] J. Wu *et al.*, "Passive silicon photonic devices for microwave photonic signal processing," *Optics Communications,* vol. 373, pp. 44-52, 2016.
[220] G. Gao *et al.*, "Tuning of resonance spacing over whole free spectral range based on Autler-Townes splitting in a single microring resonator," *Optics Express,* vol. 23, no. 21, pp. 26895-26904, 2015/10/19 2015.
[221] X. Sun *et al.*, "Investigation of Coupling Tuning in Self-Coupled Optical Waveguide Resonators," *IEEE Photonics Technology Letters,* vol. 25, no. 10, pp. 936-939, 2013.
[222] S. Zhai, J. Feng, X. Sun, R. Akimoto, and H. Zeng, "Vertically integrated waveguide self-coupled resonator based tunable optical filter," *Optics Letters,* vol. 43, no. 15, pp. 3766-3769, 2018/08/01 2018.
[223] Z. Zou *et al.*, "Tunable two-stage self-coupled optical waveguide resonators," *Optics Letters,* vol. 38, no. 8, pp. 1215-1217, 2013/04/15 2013.
[224] H. Tang, L. Zhou, J. Xie, L. Lu, and J. Chen, "Electromagnetically Induced Transparency in a Silicon Self-Coupled Optical Waveguide," *Journal of Lightwave Technology,* vol. 36, no. 11, pp. 2188-2195, 2018.
[225] A. Li and W. Bogaerts, "Tunable electromagnetically induced transparency in integrated silicon photonics circuit," *Optics Express,* vol. 25, no. 25, pp. 31688-31695, 2017/12/11 2017.
[226] S. Zheng *et al.*, "Compact tunable electromagnetically induced transparency and Fano resonance on silicon platform," *Optics Express,* vol. 25, no. 21, pp. 25655-25662, 2017/10/16 2017.
[227] H. Du *et al.*, "Ultra-sharp asymmetric Fano-like resonance spectrum on Si photonic platform," *Optics Express,* vol. 27, no. 5, pp. 7365-7372, 2019/03/04 2019.
[228] L. Zhou, T. Ye, and J. Chen, "Coherent interference induced transparency in self-coupled optical waveguide-based resonators," *Opt Lett,* vol. 36, no. 1, pp. 13-5, Jan 1 2011.
[229] A. Li and W. Bogaerts, "Using Backscattering and Backcoupling in Silicon Ring Resonators as a New Degree of Design Freedom," *Laser & Photonics Reviews,* p. 1800244, 2019.
[230] Y. Zhang *et al.*, "Sagnac loop mirror and micro-ring based laser cavity for silicon-on-insulator," *Optics express,* vol. 22, no. 15, pp. 17872-17879, 2014.
[231] J. Wang *et al.*, "Reconfigurable radio-frequency arbitrary waveforms synthesized in a silicon photonic chip," *Nature Communications,* vol. 6, no. 1, p. 5957, 2015/01/12 2015.
[232] M. H. Khan *et al.*, "Ultrabroad-bandwidth arbitrary radiofrequency waveform generation with a silicon photonic chip-based spectral shaper," *Nature Photonics,* vol. 4, no. 2, pp. 117-122, 2010.
[233] A. E. Miroshnichenko, S. Flach, and Y. S. Kivshar, "Fano resonances in nanoscale structures," *Reviews of Modern Physics,* vol. 82, no. 3, pp. 2257-2298, 08/11/ 2010.
[234] L. Stern, M. Grajower, and U. Levy, "Fano resonances and all-optical switching in a resonantly coupled plasmonic–atomic system," *Nature Communications,* vol. 5, no. 1, p. 4865, 2014/09/08 2014.

[235]  Y. Deng, G. Cao, H. Yang, G. Li, X. Chen, and W. Lu, "Tunable and high-sensitivity sensing based on Fano resonance with coupled plasmonic cavities," *Scientific Reports,* vol. 7, no. 1, p. 10639, 2017/09/06 2017.

[236]  M. F. Limonov, M. V. Rybin, A. N. Poddubny, and Y. S. Kivshar, "Fano resonances in photonics," *Nature Photonics,* vol. 11, no. 9, pp. 543-554, 2017.

[237]  M. F. Limonov, "Fano resonance for applications," *Advances in Optics and Photonics,* vol. 13, no. 3, p. 703, 2021.

[238]  J. Wu, T. Moein, X. Xu, G. Ren, A. Mitchell, and D. J. Moss, "Micro-ring resonator quality factor enhancement via an integrated Fabry-Perot cavity," *APL Photonics,* vol. 2, no. 5, p. 056103, 2017.

[239]  M. H. Haron, B. Yeop Majlis, and A. R. M. Zain, "Increasing the Quality Factor (Q) of 1D Photonic Crystal Cavity with an End Loop-Mirror," in *Photonics*, 2021, vol. 8, no. 4, p. 99: Multidisciplinary Digital Publishing Institute.

[240]  A. Li and W. Bogaerts, "Experimental demonstration of a single silicon ring resonator with an ultra-wide FSR and tuning range," *Optics Letters,* vol. 42, no. 23, pp. 4986-4989, 2017/12/01 2017.

[241]  R. Cernansky and A. Politi, "Nanophotonic source of quadrature squeezing via self-phase modulation," *APL Photonics,* vol. 5, no. 10, p. 101303, 2020.

[242]  J. Wu *et al.*, "Compact on-chip 1 × 2 wavelength selective switch based on silicon microring resonator with nested pairs of subrings," *Photonics Research,* vol. 3, no. 1, p. 9, 2014.

[243]  B. Troia, F. De Leonardis, and V. M. N. Passaro, "Cascaded ring resonator and Mach-Zehnder interferometer with a Sagnac loop for Vernier-effect refractive index sensing," *Sensors and Actuators B: Chemical,* vol. 240, pp. 76-89, 2017.

[244]  C. Lu, H. Nikbakht, M. Karabiyik, M. Alaydrus, and B. I. Akca, "A compound optical microresonator design for self-referencing and multiplexed refractive index sensing," *Optics Express,* vol. 29, no. 25, p. 42215, 2021.

[245]  J. Van Campenhout, W. M. Green, S. Assefa, and Y. A. Vlasov, "Low-power, 2× 2 silicon electro-optic switch with 110-nm bandwidth for broadband reconfigurable optical networks," *Optics Express,* vol. 17, no. 26, pp. 24020-24029, 2009.

[246]  S. Chen, Y. Shi, S. He, and D. Dai, "Low-loss and broadband 2× 2 silicon thermo-optic Mach–Zehnder switch with bent directional couplers," *Optics letters,* vol. 41, no. 4, pp. 836-839, 2016.

[247]  G. F. Chen, J. R. Ong, T. Y. Ang, S. T. Lim, C. E. Png, and D. T. Tan, "Broadband silicon-on-insulator directional couplers using a combination of straight and curved waveguide sections," *Scientific reports,* vol. 7, no. 1, pp. 1-8, 2017.

[248]  Z. Lu *et al.*, "Broadband silicon photonic directional coupler using asymmetric-waveguide based phase control," *Optics express,* vol. 23, no. 3, pp. 3795-3808, 2015.

[249]  H. Yun, Z. Lu, Y. Wang, W. Shi, L. Christowski, and N. A. Jaeger, "2× 2 broadband adiabatic 3-dB couplers on SOI strip waveguides for TE and TM modes," in *CLEO: Science and Innovations*, 2015, p. STh1F. 8: Optical Society of America.

[250]  M. T. Hill, X. J. M. Leijtens, G. D. Khoe, and M. K. Smit, "Optimizing imbalance and loss in 2 × 2 3-dB multimode interference couplers via access waveguide width," *Journal of Lightwave Technology,* vol. 21, no. 10, pp. 2305-2313, 2003.

[251]  D.-X. Xu *et al.*, "High bandwidth SOI photonic wire ring resonators using MMI couplers," *Optics Express,* vol. 15, no. 6, pp. 3149-3155, 2007.

[252]  L. B. Soldano and E. C. M. Pennings, "Optical multi-mode interference devices based on self-imaging: principles and applications," *Journal of Lightwave Technology,* vol. 13, no. 4, pp. 615-627, 1995.

[253]  R. Halir *et al.*, "Colorless directional coupler with dispersion engineered sub-wavelength structure," *Optics Express,* vol. 20, no. 12, pp. 13470-13477, 2012/06/04 2012.

[254]  Y. Wang *et al.*, "Compact broadband directional couplers using subwavelength gratings," *IEEE Photonics Journal,* vol. 8, no. 3, pp. 1-8, 2016.

[255]  R. Halir *et al.*, "Ultra-broadband nanophotonic beamsplitter using an anisotropic sub-wavelength metamaterial," *Laser & Photonics Reviews,* vol. 10, no. 6, pp. 1039-1046, 2016.

[256]  A. Maese-Novo *et al.*, "Wavelength independent multimode interference coupler," *Optics Express,* vol. 21, no. 6, pp. 7033-7040, 2013/03/25 2013.

[257]  H. Yun *et al.*, "Broadband 2× 2 adiabatic 3 dB coupler using silicon-on-insulator sub-wavelength grating waveguides," *Optics letters,* vol. 41, no. 13, pp. 3041-3044, 2016.

[258]  H. Yun, L. Chrostowski, and N. A. F. Jaeger, "Ultra-broadband 2 × 2 adiabatic 3 dB coupler using subwavelength-grating-assisted silicon-on-insulator strip waveguides," *Optics Letters,* vol. 43, no. 8, pp. 1935-1938, 2018/04/15 2018.

[259]  H. Fukuda, K. Yamada, T. Tsuchizawa, T. Watanabe, H. Shinojima, and S. Itabashi, "Ultrasmall polarization splitter based on silicon wire waveguides," *Opt Express,* vol. 14, no. 25, pp. 12401-8, Dec 11 2006.

[260] X. Zi, L. Wang, K. Chen, and K. S. Chiang, "Mode-Selective Switch Based on Thermo-Optic Asymmetric Directional Coupler," *IEEE Photonics Technology Letters,* vol. 30, no. 7, pp. 618-621, 2018.

[261] J.-S. Kim and J. T. Kim, "Silicon electro-optic modulator based on an ITO-integrated tunable directional coupler," *Journal of Physics D: Applied Physics,* vol. 49, no. 7, p. 075101, 2016/01/15 2016.

[262] M. Thomaschewski, V. A. Zenin, C. Wolff, and S. I. Bozhevolnyi, "Plasmonic monolithic lithium niobate directional coupler switches," *Nature Communications,* vol. 11, no. 1, p. 748, 2020/02/06 2020.

[263] P. Xu, J. Zheng, J. K. Doylend, and A. Majumdar, "Low-Loss and Broadband Nonvolatile Phase-Change Directional Coupler Switches," *ACS Photonics,* vol. 6, no. 2, pp. 553-557, 2019/02/20 2019.

[264] Q. Zhang, Y. Zhang, J. Li, R. Soref, T. Gu, and J. Hu, "Broadband nonvolatile photonic switching based on optical phase change materials: beyond the classical figure-of-merit," *Optics Letters,* vol. 43, no. 1, pp. 94-97, 2018/01/01 2018.

[265] A. Li and W. Bogaerts, "Reconfigurable nonlinear nonreciprocal transmission in a silicon photonic integrated circuit," *Optica,* vol. 7, no. 1, p. 7, 2020.

[266] L. Bi *et al.*, "On-chip optical isolation in monolithically integrated non-reciprocal optical resonators," *Nature Photonics,* vol. 5, no. 12, pp. 758-762, 2011/12/01 2011.

[267] W. Yan *et al.*, "Waveguide-integrated high-performance magneto-optical isolators and circulators on silicon nitride platforms," *Optica,* vol. 7, no. 11, pp. 1555-1562, 2020/11/20 2020.

[268] D. Huang, P. Pintus, C. Zhang, Y. Shoji, T. Mizumoto, and J. E. Bowers, "Electrically Driven and Thermally Tunable Integrated Optical Isolators for Silicon Photonics," *IEEE Journal of Selected Topics in Quantum Electronics,* vol. 22, no. 6, pp. 271-278, 2016.

[269] Y. Zhang *et al.*, "Monolithic integration of broadband optical isolators for polarization-diverse silicon photonics," *Optica,* vol. 6, no. 4, p. 473, 2019.

[270] L. Fan *et al.*, "An All-Silicon Passive Optical Diode," (in English), *Science,* vol. 335, no. 6067, pp. 447-450, Jan 27 2012.

[271] M. Xu, J. Wu, T. Wang, X. Hu, X. Jiang, and Y. Su, "Push–Pull Optical Nonreciprocal Transmission in Cascaded Silicon Microring Resonators," *IEEE Photonics Journal,* vol. 5, no. 1, pp. 2200307-2200307, 2013.

[272] C.-H. Dong, Z. Shen, C.-L. Zou, Y.-L. Zhang, W. Fu, and G.-C. Guo, "Brillouin-scattering-induced transparency and non-reciprocal light storage," *Nature Communications,* vol. 6, no. 1, p. 6193, 2015/02/04 2015.

[273] J. Kim, M. C. Kuzyk, K. Han, H. Wang, and G. Bahl, "Non-reciprocal Brillouin scattering induced transparency," *Nature Physics,* vol. 11, no. 3, pp. 275-280, 2015/03/01 2015.

[274] M. Merklein, B. Stiller, K. Vu, P. Ma, S. J. Madden, and B. J. Eggleton, "On-chip broadband nonreciprocal light storage," *Nanophotonics,* vol. 10, no. 1, pp. 75-82, 2020.

[275] Z. Shen *et al.*, "Experimental realization of optomechanically induced non-reciprocity," *Nature Photonics,* vol. 10, no. 10, pp. 657-661, 2016/10/01 2016.

[276] F. Ruesink, M.-A. Miri, A. Alù, and E. Verhagen, "Nonreciprocity and magnetic-free isolation based on optomechanical interactions," *Nature Communications,* vol. 7, no. 1, p. 13662, 2016/11/29 2016.

[277] F. Ruesink, J. P. Mathew, M.-A. Miri, A. Alù, and E. Verhagen, "Optical circulation in a multimode optomechanical resonator," *Nature Communications,* vol. 9, no. 1, p. 1798, 2018/05/04 2018.

[278] L. Del Bino, J. M. Silver, M. T. M. Woodley, S. L. Stebbings, X. Zhao, and P. Del'Haye, "Microresonator isolators and circulators based on the intrinsic nonreciprocity of the Kerr effect," *Optica,* vol. 5, no. 3, pp. 279-282, 2018/03/20 2018.

[279] X. Dai, Q. Lu, and W. Guo, "Fabrication-Tolerant Polarization Rotator-Splitter based on Silicon Nitride Platform," in *2021 Optical Fiber Communications Conference and Exhibition (OFC)*, 2021, pp. 1-3.

[280] B. Frey, D. Leviton, and T. Madison, *Temperature-dependent refractive index of silicon and germanium* (SPIE Astronomical Telescopes + Instrumentation). SPIE, 2006.

[281] A. Rahim *et al.*, "Expanding the Silicon Photonics Portfolio With Silicon Nitride Photonic Integrated Circuits," *Journal of Lightwave Technology,* vol. 35, no. 4, pp. 639-649, 2017.

[282] A. Trenti *et al.*, "Thermo-optic coefficient and nonlinear refractive index of silicon oxynitride waveguides," *AIP Advances,* vol. 8, no. 2, p. 025311, 2018.

[283] P. Xing, D. Ma, K. J. A. Ooi, J. W. Choi, A. M. Agarwal, and D. Tan, "CMOS-Compatible PECVD Silicon Carbide Platform for Linear and Nonlinear Optics," *ACS Photonics,* vol. 6, no. 5, pp. 1162-1167, 2019/05/15 2019.

[284] J. Teng *et al.*, "Athermal Silicon-on-insulator ring resonators by overlaying a polymer cladding on narrowed waveguides," *Opt Express,* vol. 17, no. 17, pp. 14627-33, Aug 17 2009.

[285] P. Alipour, E. S. Hosseini, A. A. Eftekhar, B. Momeni, and A. Adibi, "Athermal performance in high-Q polymer-clad silicon microdisk resonators," *Opt Lett,* vol. 35, no. 20, pp. 3462-4, Oct 15 2010.

[286] B. Guha, J. Cardenas, and M. Lipson, "Athermal silicon microring resonators with titanium oxide cladding," *Opt Express,* vol. 21, no. 22, pp. 26557-63, Nov 4 2013.

[287] S. Feng et al., "Athermal silicon ring resonators clad with titanium dioxide for 1.3μm wavelength operation," *Optics Express,* vol. 23, no. 20, pp. 25653-25660, 2015/10/05 2015.
[288] L. Lu et al., "CMOS-compatible temperature-independent tunable silicon optical lattice filters," *Opt Express,* vol. 21, no. 8, pp. 9447-56, Apr 22 2013.
[289] A. Trita, A. Thomas, and A. Rickman, "CMOS compatible athermal silicon photonic filters based on hydrogenated amorphous silicon," *Optics Express,* vol. 30, no. 11, pp. 19311-19319, 2022/05/23 2022.
[290] Z. Wang et al., "Room-temperature InP distributed feedback laser array directly grown on silicon," *Nature Photonics,* vol. 9, no. 12, pp. 837-842, 2015/12/01 2015.
[291] D. Liang and J. E. Bowers, "Recent progress in lasers on silicon," *Nature Photonics,* vol. 4, no. 8, pp. 511-517, 2010/08/01 2010.
[292] C. Wang et al., "Integrated lithium niobate electro-optic modulators operating at CMOS-compatible voltages," *Nature,* vol. 562, no. 7725, pp. 101-104, 2018/10/01 2018.
[293] M. He et al., "High-performance hybrid silicon and lithium niobate Mach–Zehnder modulators for 100 Gbit s−1 and beyond," *Nature Photonics,* vol. 13, no. 5, pp. 359-364, 2019/05/01 2019.
[294] J. Michel, J. Liu, and L. C. Kimerling, "High-performance Ge-on-Si photodetectors," *Nature Photonics,* vol. 4, no. 8, pp. 527-534, 2010.
[295] L. Li et al., "High-performance flexible waveguide-integrated photodetectors," *Optica,* vol. 5, no. 1, pp. 44-51, 2018/01/20 2018.
[296] I. Goykhman et al., "On-Chip Integrated, Silicon–Graphene Plasmonic Schottky Photodetector with High Responsivity and Avalanche Photogain," *Nano Letters,* vol. 16, no. 5, pp. 3005-3013, 2016/05/11 2016.
[297] A. H. Atabaki et al., "Integrating photonics with silicon nanoelectronics for the next generation of systems on a chip," *Nature,* vol. 556, no. 7701, pp. 349-354, 2018/04/01 2018.
[298] C. Xiang et al., "Laser soliton microcombs heterogeneously integrated on silicon," *Science,* vol. 373, no. 6550, pp. 99-103, Jul 2 2021.
[299] C. Xiang et al., "High-performance lasers for fully integrated silicon nitride photonics," *Nat Commun,* vol. 12, no. 1, p. 6650, Nov 17 2021.
[300] T. J. Kippenberg, A. L. Gaeta, M. Lipson, and M. L. Gorodetsky, "Dissipative Kerr solitons in optical microresonators," *Science,* vol. 361, no. 6402, p. eaan8083, 2018.
[301] A. Pasquazi et al., "Micro-combs: A novel generation of optical sources," *Physics Reports,* vol. 729, pp. 1-81, 2018.
[302] J. Wu et al., "RF photonics: an optical microcombs' perspective," *IEEE Journal of Selected Topics in Quantum Electronics,* vol. 24, no. 4, pp. 1-20, 2018.
[303] L. Razzari et al., "CMOS-compatible integrated optical hyper-parametric oscillator," *Nature Photonics,* vol. 4, no. 1, pp. 41-45, 2010.
[304] X. Xu et al., "11 TOPS photonic convolutional accelerator for optical neural networks," *Nature,* vol. 589, no. 7840, pp. 44-51, 2021.
[305] C. Y. Wang et al., "Mid-infrared optical frequency combs at 2.5 μm based on crystalline microresonators," *Nat Commun,* vol. 4, no. 1, pp. 1-7, 2013.
[306] W. Liang et al., "Miniature multioctave light source based on a monolithic microcavity," *Optica* vol. 2, no. 1, pp. 40-47, 2015.
[307] H. Jung, C. Xiong, K. Y. Fong, X. Zhang, and H. X. Tang, "Optical frequency comb generation from aluminum nitride microring resonator," *Opt Lett,* vol. 38, no. 15, pp. 2810-3, Aug 1 2013.
[308] X. Liu, Z. Gong, A. W. Bruch, J. B. Surya, J. Lu, and H. X. Tang, "Aluminum nitride nanophotonics for beyond-octave soliton microcomb generation and self-referencing," *Nat Commun,* vol. 12, no. 1, p. 5428, Sep 14 2021.
[309] B. Hausmann, I. Bulu, V. Venkataraman, P. Deotare, and M. Lončar, "Diamond nonlinear photonics," *Nature Photon,* vol. 8, no. 5, pp. 369-374, 2014.
[310] A. Boes, B. Corcoran, L. Chang, J. Bowers, and A. Mitchell, "Status and potential of lithium niobate on insulator (LNOI) for photonic integrated circuits," *Laser & Photonics Reviews,* vol. 12, no. 4, p. 1700256, 2018.
[311] X. Han et al., "Mode and Polarization-Division Multiplexing Based on Silicon Nitride Loaded Lithium Niobate on Insulator Platform," *Laser & Photonics Reviews,* vol. 16, no. 1, p. 2100529, 2022.
[312] M. Pu, L. Ottaviano, E. Semenova, and K. Yvind, "Efficient frequency comb generation in AlGaAs-on-insulator," *Optica,* vol. 3, no. 8, pp. 823-826, 2016.
[313] L. Chang et al., "Ultra-efficient frequency comb generation in AlGaAs-on-insulator microresonators," *Nat Commun,* vol. 11, no. 1, p. 1331, Mar 12 2020.
[314] C. Wang et al., "High-Q microresonators on 4H-silicon-carbide-on-insulator platform for nonlinear photonics," *Light Sci Appl,* vol. 10, no. 1, p. 139, Jul 5 2021.
[315] H. Jung, S.-P. Yu, D. R. Carlson, T. E. Drake, T. C. Briles, and S. B. Papp, "Tantala Kerr nonlinear integrated photonics," *Optica,* vol. 8, no. 6, pp. 811-817, 2021.

[316] D. J. Wilson *et al.*, "Integrated gallium phosphide nonlinear photonics," *Nat. Photonics,* vol. 14, no. 1, pp. 57-62, 2020.

[317] H. Lee, M. G. Suh, T. Chen, J. Li, S. A. Diddams, and K. J. Vahala, "Spiral resonators for on-chip laser frequency stabilization," *Nat Commun,* vol. 4, no. 1, p. 2468, 2013.

[318] C. Ciminelli, F. Dell'Olio, and M. N. Armenise, "High-Q spiral resonator for optical gyroscope applications: numerical and experimental investigation," *IEEE Photonics Journal,* vol. 4, no. 5, pp. 1844-1854, 2012.

[319] X. Ji, S. Roberts, M. Corato-Zanarella, and M. Lipson, "Methods to achieve ultra-high quality factor silicon nitride resonators," *APL Photonics,* vol. 6, no. 7, p. 071101, 2021.

[320] Y. Xuan *et al.*, "High-Q silicon nitride microresonators exhibiting low-power frequency comb initiation," *Optica,* vol. 3, no. 11, pp. 1171-1180, 2016.

[321] H. Ma, X. Li, G. Zhang, and Z. Jin, "Reduction of optical Kerr-effect induced error in a resonant micro-optic gyro by light-intensity feedback technique," *Applied Optics,* vol. 53, no. 16, pp. 3465-3472, 2014/06/01 2014.

[322] C. Xie *et al.*, "Resonant microsphere gyroscope based on a double Faraday rotator system," *Optics Letters,* vol. 41, no. 20, pp. 4783-4786, 2016/10/15 2016.

[323] H. Ma, J. Zhang, Z. Chen, and Z. Jin, "Tilted Waveguide Gratings and Implications for Optical Waveguide-Ring Resonator," *Journal of Lightwave Technology,* vol. 33, no. 19, pp. 4176-4183, 2015.

[324] H. Mao, H. Ma, and Z. Jin, "Polarization maintaining silica waveguide resonator optic gyro using double phase modulation technique," *Optics Express,* vol. 19, no. 5, pp. 4632-4643, 2011/02/28 2011.

[325] J. Geng, L. Yang, S. Zhao, and Y. Zhang, "Resonant micro-optical gyro based on self-injection locking," *Optics Express,* vol. 28, no. 22, pp. 32907-32915, 2020/10/26 2020.

[326] H. Li, C. Wen, C. Feng, C. Qing, D. Zhang, and L. Feng, "Frequency Spectrum Separation Method of Suppressing Backward-Light-Related Errors for Resonant Integrated Optical Gyroscope," *Journal of Lightwave Technology,* vol. 40, no. 4, pp. 1188-1194, 2022.

[327] C. Feng *et al.*, "Resonant integrated optical gyroscope based on Si3N4 waveguide ring resonator," *Optics Express,* vol. 29, no. 26, pp. 43875-43884, 2021/12/20 2021.

[328] X. Kuai, L. Wei, F. Yang, W. Yan, Z. Li, and X. Wang, "Suppression Method of Optical Noises in Resonator-Integrated Optic Gyroscopes," (in eng), *Sensors (Basel),* vol. 22, no. 8, Apr 9 2022.

[329] H. Li, P. Ni, Q. Wang, C. Feng, and L. Feng, "Analysis and Optimization of Dynamic Performance for Resonant Integrated Optical Gyroscope," *Journal of Lightwave Technology,* vol. 39, no. 6, pp. 1858-1866, 2021.

[330] G. Matthaei, "Microwave filters, impedance-matching networks and coupling structures," *Artech House Book,* pp. 775-809, 1980.

[331] S. H. Linkwitz, "Active crossover networks for noncoincident drivers," *Journal of the Audio Engineering Society,* vol. 24, no. 1, pp. 2-8, 1976.

[332] F. Harris, E. Venosa, X. Chen, P. Muthyala, and C. Dick, "An extension of the Linkwitz-Riley crossover filters for audio systems and their sampled data implementation," in *2013 20th International Conference on Systems, Signals and Image Processing (IWSSIP)*, 2013, pp. 175-178.

[333] A. Papoulis, "Optimum Filters with Monotonic Response," *Proceedings of the IRE,* vol. 46, no. 3, pp. 606-609, 1958.

[334] A. Papoulis, "On monotonic response filters," *PROCEEDINGS OF THE INSTITUTE OF RADIO ENGINEERS,* vol. 47, no. 2, pp. 332-333, 1959.

[335] K. Ito and K. Xiong, "Gaussian filters for nonlinear filtering problems," *IEEE Transactions on Automatic Control,* vol. 45, no. 5, pp. 910-927, 2000.

[336] Y. Wu, D. Hu, M. Wu, and X. Hu, "A Numerical-Integration Perspective on Gaussian Filters," *IEEE Transactions on Signal Processing,* vol. 54, no. 8, pp. 2910-2921, 2006.

[337] B. Llc, *Image Impedance Filters: Propagation Constant, Zobel Network, Composite Image Filter, Mm'-Type Filter, Prototype Filter, M-Derived Filter*. General Books LLC, 2010.

[338] W. Jiang *et al.*, "Optical All-Pass Filter in Silicon-on-Insulator," *ACS Photonics,* vol. 7, no. 9, pp. 2539-2546, 2020.

[339] M. Fleischhauer, A. Imamoglu, and J. P. Marangos, "Electromagnetically induced transparency: Optics in coherent media," *Reviews of Modern Physics,* vol. 77, no. 2, pp. 633-673, 07/12/ 2005.

[340] M. D. Lukin and A. Imamoğlu, "Controlling photons using electromagnetically induced transparency," *Nature* vol. 413, no. 6853, pp. 273–276, Sep 20 2001.

[341] P. M. Anisimov, J. P. Dowling, and B. C. Sanders, "Objectively discerning Autler-Townes splitting from electromagnetically induced transparency," *Phys Rev Lett,* vol. 107, no. 16, p. 163604, Oct 14 2011.

[342] B. Walker, M. Kaluza, B. Sheehy, P. Agostini, and L. F. DiMauro, "Observation of continuum-continuum Autler-Townes splitting," *Phys Rev Lett,* vol. 75, no. 4, pp. 633-636, Jul 24 1995.

[343] B. Peng, Ş. K. Özdemir, W. Chen, F. Nori, and L. Yang, "What is and what is not electromagnetically induced transparency in whispering-gallery microcavities," *Nature Communications,* vol. 5, no. 1, p. 5082, 2014/10/24 2014.

[344] C. W. Hsu, B. Zhen, A. D. Stone, J. D. Joannopoulos, and M. Soljačić, "Bound states in the continuum," *Nature Reviews Materials,* vol. 1, no. 9, p. 16048, 2016/07/19 2016.

[345] Y. Liang *et al.*, "Bound States in the Continuum in Anisotropic Plasmonic Metasurfaces," *Nano Letters,* vol. 20, no. 9, pp. 6351-6356, 2020/09/09 2020.

[346] Y. Yu, Y. Chen, H. Hu, W. Xue, K. Yvind, and J. Mork, "Nonreciprocal transmission in a nonlinear photonic-crystal Fano structure with broken symmetry," *Laser & Photonics Reviews,* vol. 9, no. 2, pp. 241-247, 2015.

[347] C. Fan, F. Shi, H. Wu, and Y. Chen, "Tunable all-optical plasmonic diode based on Fano resonance in nonlinear waveguide coupled with cavities," *Optics Letters,* vol. 40, no. 11, pp. 2449-2452, 2015/06/01 2015.

[348] G. Cao *et al.*, "Fano Resonance in Artificial Photonic Molecules," *Advanced Optical Materials,* vol. 8, no. 10, p. 1902153, 2020.

[349] G. Khitrova, H. M. Gibbs, M. Kira, S. W. Koch, and A. Scherer, "Vacuum Rabi splitting in semiconductors," *Nature Physics,* vol. 2, no. 2, pp. 81-90, 2006/02/01 2006.

[350] T. Yoshie *et al.*, "Vacuum Rabi splitting with a single quantum dot in a photonic crystal nanocavity," *Nature,* vol. 432, no. 7014, pp. 200-203, 2004/11/01 2004.

[351] A. E. Schlather, N. Large, A. S. Urban, P. Nordlander, and N. J. Halas, "Near-field mediated plexcitonic coupling and giant Rabi splitting in individual metallic dimers," *Nano Lett,* vol. 13, no. 7, pp. 3281-6, Jul 10 2013.

[352] C. E. Rüter, K. G. Makris, R. El-Ganainy, D. N. Christodoulides, M. Segev, and D. Kip, "Observation of parity–time symmetry in optics," *Nature Physics,* vol. 6, no. 3, pp. 192-195, 2010.

[353] L. Feng, R. El-Ganainy, and L. Ge, "Non-Hermitian photonics based on parity–time symmetry," *Nature Photonics,* vol. 11, no. 12, pp. 752-762, 2017/12/01 2017.

[354] Ş. K. Özdemir, S. Rotter, F. Nori, and L. Yang, "Parity–time symmetry and exceptional points in photonics," *Nature Materials,* vol. 18, no. 8, pp. 783-798, 2019/08/01 2019.

[355] C. Koos *et al.*, "All-optical high-speed signal processing with silicon–organic hybrid slot waveguides," *Nature Photonics,* vol. 3, no. 4, pp. 216-219, 2009.

[356] A. Melikyan *et al.*, "High-speed plasmonic phase modulators," *Nature Photonics,* vol. 8, no. 3, pp. 229-233, 2014.

[357] C. Kieninger *et al.*, "Ultra-high electro-optic activity demonstrated in a silicon-organic hybrid modulator," *Optica,* vol. 5, no. 6, p. 739, 2018.

[358] S. Koeber *et al.*, "Femtojoule electro-optic modulation using a silicon–organic hybrid device," *Light: Science & Applications,* vol. 4, no. 2, pp. e255-e255, 2015.

[359] L. Alloatti *et al.*, "100 GHz silicon–organic hybrid modulator," *Light: Science & Applications,* vol. 3, no. 5, pp. e173-e173, 2014.

[360] Z. Zhang, Z. You, and D. Chu, "Fundamentals of phase-only liquid crystal on silicon (LCOS) devices," *Light: Science & Applications,* vol. 3, no. 10, pp. e213-e213, 2014.

[361] D. Vettese, "Liquid crystal on silicon," *Nature Photonics,* vol. 4, no. 11, pp. 752-754, 2010.

[362] T.-J. Wang, C.-K. Chaung, W.-J. Li, T.-J. Chen, and B.-Y. Chen, "Electrically Tunable Liquid-Crystal-Core Optical Channel Waveguide," *Journal of Lightwave Technology,* vol. 31, no. 22, pp. 3570-3574, 2013.

[363] F. Bonaccorso, Z. Sun, T. Hasan, and A. C. Ferrari, "Graphene photonics and optoelectronics," *Nature Photonics,* vol. 4, no. 9, pp. 611-622, 2010.

[364] J. Wu, L. Jia, Y. Zhang, Y. Qu, B. Jia, and D. J. Moss, "Graphene Oxide for Integrated Photonics and Flat Optics," *Adv Mater,* vol. 33, no. 3, p. e2006415, Jan 2021.

[365] J. Wu *et al.*, "2D Layered Graphene Oxide Films Integrated with Micro-Ring Resonators for Enhanced Nonlinear Optics," *Small,* vol. 16, no. 16, p. e1906563, Apr 2020.

[366] Y. Zhang *et al.*, "Enhanced Kerr Nonlinearity and Nonlinear Figure of Merit in Silicon Nanowires Integrated with 2D Graphene Oxide Films," *ACS Applied Materials & Interfaces,* vol. 12, no. 29, pp. 33094-33103, 2020/07/22 2020.

[367] J. Wu *et al.*, "Graphene Oxide Waveguide and Micro-Ring Resonator Polarizers," *Laser & Photonics Reviews,* vol. 13, no. 9, p. 1900056, 2019.

[368] J. An *et al.*, "Perspectives of 2D Materials for Optoelectronic Integration," *Advanced Functional Materials,* vol. 32, no. 14, p. 2110119, 2021.

[369] Z. Cheng *et al.*, "2D Materials Enabled Next-Generation Integrated Optoelectronics: from Fabrication to Applications," *Adv Sci* vol. 8, no. 11, p. e2003834, Jun 2021.

[370] T. J. Kippenberg, R. Holzwarth, and S. A. Diddams, "Microresonator-based optical frequency combs," *Science,* vol. 332, no. 6029, pp. 555-9, Apr 29 2011.

[371] L. Chang, S. Liu, and J. E. Bowers, "Integrated optical frequency comb technologies," *Nature Photonics,* vol. 16, no. 2, pp. 95-108, 2022.
[372] B. Shen *et al.*, "Integrated turnkey soliton microcombs," *Nature,* vol. 582, no. 7812, pp. 365-369, 2020/06/01 2020.
[373] S. Mahajan, V. Trivedi, P. Vora, V. Chhaniwal, B. Javidi, and A. Anand, "Highly stable digital holographic microscope using Sagnac interferometer," *Opt Lett,* vol. 40, no. 16, pp. 3743-6, Aug 15 2015.
[374] C. Ma *et al.*, "Lateral shearing common-path digital holographic microscopy based on a slightly trapezoid Sagnac interferometer," *Optics Express,* vol. 25, no. 12, pp. 13659-13667, 2017/06/12 2017.
[375] W. Zhang, C. Shi, and B. Li, "Sagnac interferometer-based transmission grating super-resolution digital microholography," *Journal of Optics,* vol. 51, no. 1, pp. 203-210, 2022/03/01 2021.
[376] L. Lu *et al.*, "16× 16 non-blocking silicon optical switch based on electro-optic Mach-Zehnder interferometers," *Opt. Express,* vol. 24, no. 9, pp. 9295-9307, 2016.
[377] A. Biberman, B. G. Lee, N. Sherwood-Droz, M. Lipson, and K. Bergman, "Broadband operation of nanophotonic router for silicon photonic networks-on-chip," *IEEE Photonics Technology Letters,* vol. 22, no. 12, pp. 926-928, 2010.
[378] T. Hu *et al.*, "Wavelength-selective 4× 4 nonblocking silicon optical router for networks-on-chip," *Optics letters,* vol. 36, no. 23, pp. 4710-4712, 2011.
[379] J. Sun, E. Timurdogan, A. Yaacobi, E. S. Hosseini, and M. R. Watts, "Large-scale nanophotonic phased array," *Nature,* vol. 493, no. 7431, pp. 195-9, Jan 10 2013.
[380] X. Xu *et al.*, "Photonic microwave true time delays for phased array antennas using a 49 GHz FSR integrated optical micro-comb source," *Photon. Res.,* vol. 6, no. 5, pp. B30-B36, 2018.
[381] J. Yang, K. Zhou, Y. Liu, and S. C. Tjin, "Photonics true-time-delay unit for phased-array antenna using multiwavelength fiber laser based on a Sagnac interferemetric filter," *International Journal of Infrared Millimeter Waves* vol. 23, no. 6, pp. 891-897, 2002.
[382] C. Zhu *et al.*, "Silicon integrated microwave photonic beamformer," *Optica,* vol. 7, no. 9, pp. 1162-1170, 2020.
[383] L. Zhuang *et al.*, "On-chip microwave photonic beamformer circuits operating with phase modulation and direct detection," *Opt Express,* vol. 22, no. 14, pp. 17079-91, Jul 14 2014.
[384] P. Moslemi, L. R. Chen, and M. Rochette, "Simultaneously generating multiple chirped microwave waveforms using an arrayed waveguide Sagnac interferometer," *Electronics Letters,* vol. 53, no. 23, pp. 1534-1535, 2017.
[385] J. Feldmann *et al.*, "Parallel convolutional processing using an integrated photonic tensor core," *Nature,* vol. 589, no. 7840, pp. 52-58, 2021.
[386] Y. Shen *et al.*, "Deep learning with coherent nanophotonic circuits," *Nature Photon,* vol. 11, no. 7, pp. 441-446, 2017.
[387] F. Ashtiani, A. J. Geers, and F. Aflatouni, "An on-chip photonic deep neural network for image classification," *Nature,* vol. 606, no. 7914, pp. 501-506, Jun 2022.
[388] Y. Zhou, H. Zheng, I. I. Kravchenko, and J. Valentine, "Flat optics for image differentiation," *Nat. Photonics,* vol. 14, no. 5, pp. 316-323, 2020.
[389] X. Zhang *et al.*, "Reconfigurable Metasurface for Image Processing," *Nano Lett,* vol. 21, no. 20, pp. 8715-8722, Oct 27 2021.
[390] T. Zhu *et al.*, "Plasmonic computing of spatial differentiation," *Nat Commun,* vol. 8, no. 1, p. 15391, May 19 2017.
[391] J. Xiang, A. Torchy, X. Guo, and Y. Su, "All-optical spiking neuron based on passive microresonator," *Journal of Lightwave Technology,* vol. 38, no. 15, pp. 4019-4029, 2020.
[392] X. Guo, J. Xiang, Y. Zhang, and Y. Su, "Integrated Neuromorphic Photonics: Synapses, Neurons, and Neural Networks," *Advanced Photonics Research,* vol. 2, no. 6, p. 2000212, 2021.
[393] X. Xu *et al.*, "Self-calibrating programmable photonic integrated circuits," *Nature Photonics,* 2022.
[394] T. Kim, M. Fiorentino, and F. N. Wong, "Phase-stable source of polarization-entangled photons using a polarization Sagnac interferometer," *Physical Review A,* vol. 73, no. 1, p. 012316, 2006.
[395] J. Park, H. Kim, and H. S. Moon, "Polarization-entangled photons from a warm atomic ensemble using a Sagnac interferometer," *Physical review letters,* vol. 122, no. 14, p. 143601, 2019.
[396] O. Kuzucu and F. N. Wong, "Pulsed Sagnac source of narrow-band polarization-entangled photons," *Physical Review A,* vol. 77, no. 3, p. 032314, 2008.
[397] S. Pirandola, B. R. Bardhan, T. Gehring, C. Weedbrook, and S. Lloyd, "Advances in photonic quantum sensing," *Nature Photon,* vol. 12, no. 12, pp. 724-733, 2018.
[398] M. Mehmet, T. Eberle, S. Steinlechner, H. Vahlbruch, and R. Schnabel, "Demonstration of a quantum-enhanced fiber Sagnac interferometer," *Opt Lett,* vol. 35, no. 10, pp. 1665-7, May 15 2010.
[399] G. C. Matos, R. M. E. Souza, P. A. M. Neto, and F. Impens, "Quantum Vacuum Sagnac Effect," *Phys Rev Lett,* vol. 127, no. 27, p. 270401, Dec 31 2021.

[400] P. Grangier, J. A. Levenson, and J.-P. Poizat, "Quantum non-demolition measurements in optics," *Nature,* vol. 396, no. 6711, pp. 537-542, 1998/12/01 1998.

[401] T. Eberle *et al.*, "Quantum enhancement of the zero-area Sagnac interferometer topology for gravitational wave detection," *Phys Rev Lett,* vol. 104, no. 25, p. 251102, Jun 25 2010.

[402] K. X. Sun, M. M. Fejer, E. Gustafson, and R. L. Byer, "Sagnac interferometer for gravitational-wave detection," *Phys Rev Lett,* vol. 76, no. 17, pp. 3053-3056, Apr 22 1996.

[403] L. Li *et al.*, "Integrated flexible chalcogenide glass photonic devices," *Nature Photonics,* vol. 8, no. 8, pp. 643-649, 2014/08/01 2014.

[404] N. Li, C. P. Ho, S. Zhu, Y. H. Fu, Y. Zhu, and L. Y. T. Lee, "Aluminium nitride integrated photonics: a review," *Nanophotonics,* vol. 10, no. 9, pp. 2347-2387, 2021.

[405] A. W. Bruch *et al.*, "Pockels soliton microcomb," *Nat. Photonics,* vol. 15, no. 1, pp. 21-27, 2021.

406. Pasquazi, et al., "Sub-picosecond phase-sensitive optical pulse characterization on a chip", Nature Photonics, vol. 5, no. 10, pp. 618-623 (2011).

407. Bao, C., et al., Direct soliton generation in microresonators, Opt. Lett, vol. 42, 2519 (2017).

408. M.Ferrera et al., "CMOS compatible integrated all-optical RF spectrum analyzer", Optics Express, vol. 22, no. 18, 21488 - 21498 (2014).

409. M. Kues, et al., "Passively modelocked laser with an ultra-narrow spectral width", Nature Photonics, vol. 11, no. 3, pp. 159, 2017.

410. L. Razzari, et al., "CMOS-compatible integrated optical hyper-parametric oscillator," Nature Photonics, vol. 4, no. 1, pp. 41-45, 2010.

411. M. Ferrera, et al., "Low-power continuous-wave nonlinear optics in doped silica glass integrated waveguide structures," Nature Photonics, vol. 2, no. 12, pp. 737-740, 2008.

412. M.Ferrera et al."On-Chip ultra-fast 1st and 2nd order CMOS compatible all-optical integration", Opt. Express, vol. 19, (23)pp. 23153-23161 (2011).

413. D. Duchesne, M. Peccianti, M. R. E. Lamont, et al., "Supercontinuum generation in a high index doped silica glass spiral waveguide," Optics Express, vol. 18, no, 2, pp. 923-930, 2010.

414. H Bao, L Olivieri, M Rowley, ST Chu, BE Little, R Morandotti, DJ Moss, ... "Turing patterns in a fiber laser with a nested microresonator: Robust and controllable microcomb generation", Physical Review Research 2 (2), 023395 (2020).

415. M. Ferrera, et al., "On-chip CMOS-compatible all-optical integrator", Nature Communications, vol. 1, Article 29, 2010.

416. A. Pasquazi, et al., "All-optical wavelength conversion in an integrated ring resonator," Optics Express, vol. 18, no. 4, pp. 3858-3863, 2010.

417. A.Pasquazi, Y. Park, J. Azana, et al., "Efficient wavelength conversion and net parametric gain via Four Wave Mixing in a high index doped silica waveguide," Optics Express, vol. 18, no. 8, pp. 7634-7641, 2010.

418. M. Peccianti, M. Ferrera, L. Razzari, et al., "Subpicosecond optical pulse compression via an integrated nonlinear chirper," Optics Express, vol. 18, no. 8, pp. 7625-7633, 2010.

419. M. Ferrera et al., "Low Power CW Parametric Mixing in a Low Dispersion High Index Doped Silica Glass Micro-Ring Resonator with Q-factor > 1 Million", Optics Express, vol.17, no. 16, pp. 14098–14103 (2009).

420. M. Peccianti, et al., "Demonstration of an ultrafast nonlinear microcavity modelocked laser", Nature Communications, vol. 3, pp. 765, 2012.

421. A.Pasquazi, et al., "Self-locked optical parametric oscillation in a CMOS compatible microring resonator: a route to robust optical frequency comb generation on a chip," Optics Express, vol. 21, no. 11, pp. 13333-13341, 2013.

422. A.Pasquazi, et al., "Stable, dual mode, high repetition rate mode-locked laser based on a microring resonator," Optics Express, vol. 20, no. 24, pp. 27355-27362, 2012.

423. Pasquazi, A. et al. Micro-combs: a novel generation of optical sources. Physics Reports 729, 1-81 (2018).

424. Moss, D. J. et al., "New CMOS-compatible platforms based on silicon nitride and Hydex for nonlinear optics", Nature photonics 7, 597 (2013).

425. H. Bao, et al., Laser cavity-soliton microcombs, Nature Photonics, vol. 13, no. 6, pp. 384-389, Jun. 2019.

426. Antonio Cutrona, Maxwell Rowley, Debayan Das, Luana Olivieri, Luke Peters, Sai T. Chu, Brent L. Little, Roberto Morandotti, David J. Moss, Juan Sebastian Totero Gongora, Marco Peccianti, Alessia Pasquazi, "High Conversion Efficiency in Laser Cavity-Soliton Microcombs", Optics Express Vol. 30, Issue 22, pp. 39816-39825 (2022). https://doi.org/10.1364/OE.470376.

427. M.Rowley, P.Hanzard, A.Cutrona, H.Bao, S.Chu, B.Little, R.Morandotti, D. J. Moss, G. Oppo, J. Gongora, M. Peccianti and A. Pasquazi, "Self-emergence of robust solitons in a micro-cavity", Nature vol. 608 (7922) 303–309 (2022).

428. Xu, X., et al., Photonic microwave true time delays for phased array antennas using a 49 GHz FSR integrated micro-comb source, Photonics Research, vol. 6, B30-B36 (2018).

429. X. Xu, M. Tan, J. Wu, R. Morandotti, A. Mitchell, and D. J. Moss, "Microcomb-based photonic RF signal processing", IEEE Photonics Technology Letters, vol. 31 no. 23 1854-1857, 2019.

430. M. Tan et al, "Orthogonally polarized Photonic Radio Frequency single sideband generation with integrated micro-ring resonators", IOP Journal of Semiconductors, Vol. 42 (4), 041305 (2021). DOI: 10.1088/1674-4926/42/4/041305.
431. Xu, et al., "Advanced adaptive photonic RF filters with 80 taps based on an integrated optical micro-comb source," Journal of Lightwave Technology, vol. 37, no. 4, pp. 1288-1295 (2019).
432. X. Xu, et al., Broadband microwave frequency conversion based on an integrated optical micro-comb source", Journal of Lightwave Technology, vol. 38 no. 2, pp. 332-338, 2020.
433. M. Tan, et al., "Photonic RF and microwave filters based on 49GHz and 200GHz Kerr microcombs", Optics Comm. vol. 465,125563, Feb. 22. 2020.
434. X. Xu, et al., "Broadband photonic RF channelizer with 90 channels based on a soliton crystal microcomb", Journal of Lightwave Technology, Vol. 38, no. 18, pp. 5116 - 5121, 2020. doi: 10.1109/JLT.2020.2997699.
435. X. Xu, et al., "Photonic RF and microwave integrator with soliton crystal microcombs", IEEE Transactions on Circuits and Systems II: Express Briefs, vol. 67, no. 12, pp. 3582-3586, 2020. DOI:10.1109/TCSII.2020.2995682.
436. X. Xu, et al., "High performance RF filters via bandwidth scaling with Kerr micro-combs," APL Photonics, vol. 4 (2) 026102. 2019.
437. M. Tan, et al., "Microwave and RF photonic fractional Hilbert transformer based on a 50 GHz Kerr micro-comb", Journal of Lightwave Technology, vol. 37, no. 24, pp. 6097 – 6104, 2019.
438. M. Tan, et al., "RF and microwave fractional differentiator based on photonics", IEEE Transactions on Circuits and Systems: Express Briefs, vol. 67, no.11, pp. 2767-2771, 2020. DOI:10.1109/TCSII.2020.2965158.
439. M. Tan, et al., "Photonic RF arbitrary waveform generator based on a soliton crystal micro-comb source", Journal of Lightwave Technology, vol. 38, no. 22, pp. 6221-6226 (2020). DOI: 10.1109/JLT.2020.3009655.
440. M. Tan, X. Xu, J. Wu, R. Morandotti, A. Mitchell, and D. J. Moss, "RF and microwave high bandwidth signal processing based on Kerr Micro-combs", Advances in Physics X, VOL. 6, NO. 1, 1838946 (2021). DOI:10.1080/23746149.2020.1838946.
441. X. Xu, et al., "Advanced RF and microwave functions based on an integrated optical frequency comb source," Opt. Express, vol. 26 (3) 2569 (2018).
442. M. Tan, X. Xu, J. Wu, B. Corcoran, A. Boes, T. G. Nguyen, S. T. Chu, B. E. Little, R.Morandotti, A. Lowery, A. Mitchell, and D. J. Moss, ""Highly Versatile Broadband RF Photonic Fractional Hilbert Transformer Based on a Kerr Soliton Crystal Microcomb", Journal of Lightwave Technology vol. 39 (24) 7581-7587 (2021).
443. Wu, J. et al. RF Photonics: An Optical Microcombs' Perspective. IEEE Journal of Selected Topics in Quantum Electronics Vol. 24, 6101020, 1-20 (2018).
444. T. G. Nguyen et al., "Integrated frequency comb source-based Hilbert transformer for wideband microwave photonic phase analysis," Opt. Express, vol. 23, no. 17, pp. 22087-22097, Aug. 2015.
445. X. Xu, J. Wu, M. Shoeiby, T. G. Nguyen, S. T. Chu, B. E. Little, R. Morandotti, A. Mitchell, and D. J. Moss, "Reconfigurable broadband microwave photonic intensity differentiator based on an integrated optical frequency comb source," APL Photonics, vol. 2, no. 9, 096104, Sep. 2017.
446. X. Xu, et al., "Broadband RF channelizer based on an integrated optical frequency Kerr comb source," Journal of Lightwave Technology, vol. 36, no. 19, pp. 4519-4526, 2018.
447. X. Xu, et al., "Continuously tunable orthogonally polarized RF optical single sideband generator based on micro-ring resonators," Journal of Optics, vol. 20, no. 11, 115701. 2018.
448. X. Xu, et al., "Orthogonally polarized RF optical single sideband generation and dual-channel equalization based on an integrated microring resonator," Journal of Lightwave Technology, vol. 36, no. 20, pp. 4808-4818. 2018.
449. X. Xu, et al., "Photonic RF phase-encoded signal generation with a microcomb source", J. Lightwave Technology, vol. 38, no. 7, 1722-1727, 2020.
450. B. Corcoran, et al., "Ultra-dense optical data transmission over standard fiber with a single chip source", Nature Communications, vol. 11, Article:2568, 2020.
451. X. Xu et al, "Photonic perceptron based on a Kerr microcomb for scalable high speed optical neural networks", Laser and Photonics Reviews, vol. 14, no. 8, 2000070 (2020). DOI: 10.1002/lpor.202000070.
452. X. Xu, et al., "11 TOPs photonic convolutional accelerator for optical neural networks", Nature 589, 44-51 (2021).
453. Mengxi Tan, Xingyuan Xu, Jiayang Wu, Roberto Morandotti, Arnan Mitchell, and David J. Moss, "Neuromorphic computing based on wavelength-division multiplexing", IEEE Journal of Selected Topics in Quantum Electronics **29** Issue: 2, 7400112 (2023). DOI:10.1109/JSTQE.2022.3203159.
454. Yang Sun, Jiayang Wu, Mengxi Tan, Xingyuan Xu, Yang Li, Roberto Morandotti, Arnan Mitchell, and David Moss, "Applications of optical micro-combs", Advances in Optics and Photonics **15** (1) 86-175 (2023). https://doi.org/10.1364/AOP.470264.

455. Yunping Bai, Xingyuan Xu,1, Mengxi Tan, Yang Sun, Yang Li, Jiayang Wu, Roberto Morandotti, Arnan Mitchell, Kun Xu, and David J. Moss, "Photonic multiplexing techniques for neuromorphic computing", Nanophotonics *Nanophotonics* **11** (2022). DOI:10.1515/nanoph-2022-0485.
456. Chawaphon Prayoonyong, Andreas Boes, Xingyuan Xu, Mengxi Tan, Sai T. Chu, Brent E. Little, Roberto Morandotti, Arnan Mitchell, David J. Moss, and Bill Corcoran, "Frequency comb distillation for optical superchannel transmission", Journal of Lightwave Technology vol. 39 (23) 7383-7392 (2021). DOI: 10.1109/JLT.2021.3116614.
457. Mengxi Tan, Xingyuan Xu, Jiayang Wu, Bill Corcoran, Andreas Boes, Thach G. Nguyen, Sai T. Chu, Brent E. Little, Roberto Morandotti, Arnan Mitchell, and David J. Moss, "Integral order photonic RF signal processors based on a soliton crystal micro-comb source", IOP Journal of Optics vol. 23 (11) 125701 (2021). https://doi.org/10.1088/2040-8986/ac2eab
458. Kues, M. et al. Quantum optical microcombs. Nature Photonics vol. 13, (3) 170-179 (2019). doi:10.1038/s41566-019-0363-0
459. C.Reimer, L. Caspani, M. Clerici, et al., "Integrated frequency comb source of heralded single photons," Optics Express, vol. 22, no. 6, pp. 6535-6546, 2014.
460. C.Reimer, et al., "Cross-polarized photon-pair generation and bi-chromatically pumped optical parametric oscillation on a chip", Nature Communications, vol. 6, Article 8236, 2015. DOI: 10.1038/ncomms9236.
461. L. Caspani, C. Reimer, M. Kues, et al., "Multifrequency sources of quantum correlated photon pairs on-chip: a path toward integrated Quantum Frequency Combs," Nanophotonics, vol. 5, no. 2, pp. 351-362, 2016.
462. C. Reimer et al., "Generation of multiphoton entangled quantum states by means of integrated frequency combs," Science, vol. 351, no. 6278, pp. 1176-1180, 2016.
463. M. Kues, et al., "On-chip generation of high-dimensional entangled quantum states and their coherent control", Nature, vol. 546, no. 7660, pp. 622-626, 2017.
464. P. Roztocki et al., "Practical system for the generation of pulsed quantum frequency combs," Optics Express, vol. 25, no. 16, pp. 18940-18949, 2017.
465. Y. Zhang, et al., "Induced photon correlations through superposition of two four-wave mixing processes in integrated cavities", Laser and Photonics Reviews, vol. 14, no. 7, pp. 2000128, 2020. DOI: 10.1002/lpor.202000128
466. C. Reimer, et al.,"High-dimensional one-way quantum processing implemented on d-level cluster states", Nature Physics, vol. 15, no.2, pp. 148–153, 2019.
467. P.Roztocki et al., "Complex quantum state generation and coherent control based on integrated frequency combs", Journal of Lightwave Technology **37** (2) 338-347 (2019).
468. S. Sciara et al., "Generation and Processing of Complex Photon States with Quantum Frequency Combs", IEEE Photonics Technology Letters **31** (23) 1862-1865 (2019). DOI: 10.1109/LPT.2019.2944564.
469. Stefania Sciara, Piotr Roztocki, Bennet Fisher, Christian Reimer, Luis Romero Cortez, William J. Munro, David J. Moss, Alfonso C. Cino, Lucia Caspani, Michael Kues, J. Azana, and Roberto Morandotti, "Scalable and effective multilevel entangled photon states: A promising tool to boost quantum technologies", Nanophotonics **10** (18), 4447–4465 (2021). DOI:10.1515/nanoph-2021-0510.
470. L. Caspani, C. Reimer, M. Kues, et al., "Multifrequency sources of quantum correlated photon pairs on-chip: a path toward integrated Quantum Frequency Combs," Nanophotonics, vol. 5, no. 2, pp. 351-362, 2016.
471. Yuning Zhang, Jiayang Wu, Yang Qu, Yunyi Yang, Linnan Jia, Baohua Jia, and David J. Moss, "Enhanced supercontinuum generated in SiN waveguides coated with GO films", Advanced Material Technologies **8** (1) 2201796 (2023). DOI:10.1002/admt.202201796.
472. Yuning Zhang, Jiayang Wu, Linnan Jia, Yang Qu, Baohua Jia, and David J. Moss, "Graphene oxide for nonlinear integrated photonics", Laser and Photonics Reviews **17** 2200512 (2023). DOI:10.1002/lpor.202200512.
473. Jiayang Wu, H.Lin, D. J. Moss, T.K. Loh, Baohua Jia, "Graphene oxide: new opportunities for electronics, photonics, and optoelectronics", Nature Reviews Chemistry **7** (2023). doi.org/10.1038/s41570-022-00458-7.
474. Yang Qu, Jiayang Wu, Yuning Zhang, Yunyi Yang, Linnan Jia, Baohua Jia, and David J. Moss, "Photo thermal tuning in GO-coated integrated waveguides", Micromachines vol. 13 1194 (2022). doi.org/10.3390/mi13081194
475. Yuning Zhang, Jiayang Wu, Yunyi Yang, Yang Qu, Houssein El Dirani, Romain Crochemore, Corrado Sciancalepore, Pierre Demongodin, Christian Grillet, Christelle Monat, Baohua Jia, and David J. Moss, "Enhanced self-phase modulation in silicon nitride waveguides integrated with 2D graphene oxide films", Journal of Selected Topics in Quantum Electronics **29** (1) 5100413 (2023). DOI: 10.1109/JSTQE.2022.3177385
476. Yuning Zhang, Jiayang Wu, Yunyi Yang, Yang Qu, Linnan Jia, Baohua Jia, and David J. Moss, "Enhanced spectral broadening of femtosecond optical pulses in silicon nanowires integrated with 2D graphene oxide films", Micromachines vol. 13 756 (2022). DOI:10.3390/mi13050756 (2022).

477. Linnan Jia, Jiayang Wu, Yuning Zhang, Yang Qu, Baohua Jia, Zhigang Chen, and David J. Moss, "Fabrication Technologies for the On-Chip Integration of 2D Materials", Small: Methods vol. 6, 2101435 (2022). DOI:10.1002/smtd.202101435.
478. Yuning Zhang, Jiayang Wu, Yang Qu, Linnan Jia, Baohua Jia, and David J. Moss, "Design and optimization of four-wave mixing in microring resonators integrated with 2D graphene oxide films", Journal of Lightwave Technology vol.39 (20) 6553-6562 (2021). DOI:10.1109/JLT.2021.3101292.
479. Yuning Zhang, Jiayang Wu, Yang Qu, Linnan Jia, Baohua Jia, and David J. Moss, "Optimizing the Kerr nonlinear optical performance of silicon waveguides integrated with 2D graphene oxide films", Journal of Lightwave Technology vol. 39 (14) 4671-4683 (2021). DOI: 10.1109/JLT.2021.3069733.
480. Yang Qu, Jiayang Wu, Yuning Zhang, Yao Liang, Baohua Jia, and David J. Moss, "Analysis of four-wave mixing in silicon nitride waveguides integrated with 2D layered graphene oxide films", Journal of Lightwave Technology vol. 39 (9) 2902-2910 (2021). DOI: 10.1109/JLT.2021.3059721.
481. Jiayang Wu, Linnan Jia, Yuning Zhang, Yang Qu, Baohua Jia, and David J. Moss," Graphene oxide: versatile films for flat optics to nonlinear photonic chips", Advanced Materials vol. 33 (3) 2006415, pp.1-29 (2021). DOI:10.1002/adma.202006415.
482. Y. Qu, J. Wu, Y. Zhang, L. Jia, Y. Yang, X. Xu, S. T. Chu, B. E. Little, R. Morandotti, B. Jia, and D. J. Moss, "Graphene oxide for enhanced optical nonlinear performance in CMOS compatible integrated devices", Paper No. 11688-30, PW21O-OE109-36, 2D Photonic Materials and Devices IV, SPIE Photonics West, San Francisco CA March 6-11 (2021). doi.org/10.1117/12.2583978
483. Yang Qu, Jiayang Wu, Yunyi Yang, Yuning Zhang, Yao Liang, Houssein El Dirani, Romain Crochemore, Pierre Demongodin, Corrado Sciancalepore, Christian Grillet, Christelle Monat, Baohua Jia, and David J. Moss, "Enhanced nonlinear four-wave mixing in silicon nitride waveguides integrated with 2D layered graphene oxide films", Advanced Optical Materials vol. 8 (21) 2001048 (2020). DOI: 10.1002/adom.202001048.
484. Yuning Zhang, Yang Qu, Jiayang Wu, Linnan Jia, Yunyi Yang, Xingyuan Xu, Baohua Jia, and David J. Moss, "Enhanced Kerr nonlinearity and nonlinear figure of merit in silicon nanowires integrated with 2D graphene oxide films", ACS Applied Materials and Interfaces vol. 12 (29) 33094−33103 (2020). DOI:10.1021/acsami.0c07852
485. Jiayang Wu, Yunyi Yang, Yang Qu, Yuning Zhang, Linnan Jia, Xingyuan Xu, Sai T. Chu, Brent E. Little, Roberto Morandotti, Baohua Jia,* and David J. Moss*, "Enhanced nonlinear four-wave mixing in microring resonators integrated with layered graphene oxide films", Small vol. 16 (16) 1906563 (2020). DOI: 10.1002/smll.201906563
486. Jiayang Wu, Yunyi Yang, Yang Qu, Xingyuan Xu, Yao Liang, Sai T. Chu, Brent E. Little, Roberto Morandotti, Baohua Jia, and David J. Moss, "Graphene oxide waveguide polarizers and polarization selective micro-ring resonators", Paper 11282-29, SPIE Photonics West, San Francisco, CA, 4 - 7 February (2020). doi: 10.1117/12.2544584
487. Jiayang Wu, Yunyi Yang, Yang Qu, Xingyuan Xu, Yao Liang, Sai T. Chu, Brent E. Little, Roberto Morandotti, Baohua Jia, and David J. Moss, "Graphene oxide waveguide polarizers and polarization selective micro-ring resonators", Laser and Photonics Reviews vol. 13 (9) 1900056 (2019). DOI:10.1002/lpor.201900056.
488. Yunyi Yang, Jiayang Wu, Xingyuan Xu, Sai T. Chu, Brent E. Little, Roberto Morandotti, Baohua Jia, and David J. Moss, "Enhanced four-wave mixing in graphene oxide coated waveguides", Applied Physics Letters Photonics vol. 3 120803 (2018); doi: 10.1063/1.5045509.
489. O. Darrigol, "Georges Sagnac: A life for optics," *Comptes Rendus Physique,* vol. 15, no. 10, pp. 789-840, 2014.